\documentclass{article}

\usepackage{PRIMEarxiv}

\usepackage[utf8]{inputenc} 
\usepackage[T1]{fontenc}    
\usepackage{hyperref}       
\usepackage{url}            
\usepackage{booktabs}       
\usepackage{amsfonts}       
\usepackage{nicefrac}       
\usepackage{microtype}      
\usepackage{lipsum}
\usepackage{fancyhdr}       
\usepackage{graphicx}       
\graphicspath{{media}}     

\usepackage{comment}
\usepackage{xcolor}
\usepackage{amsmath, amssymb}
\usepackage{bm}
\usepackage{enumitem}

\usepackage{graphicx}%
\usepackage{multirow}%
\usepackage{amsmath,amssymb,amsfonts}%
\usepackage{mathrsfs}%
\usepackage{textcomp}%
\usepackage{manyfoot}%
\usepackage{booktabs}%
\usepackage{algorithm}%
\usepackage{algorithmic}
\usepackage{listings}%
\usepackage[numbers]{natbib}

\usepackage{comment}
\usepackage{caption}
\usepackage{subcaption}
\usepackage{tikz}
\usepackage{stmaryrd}

\newcommand{\Ehi}{\mathcal{E}_{h}^I}

\newcommand{\ave}[1]{\{ #1 \} }

\newcommand{\bq}{{\bf q}}

\newcommand{\be}{{\bf e}}
\newcommand{\bx}{{\bf x}}
\newcommand{\bn}{{\bf n}}
\newcommand{\bu}{{\bf u}}
\newcommand{\bv}{{\bf v}}

\newcommand{\bw}{{\bf w}}

\newcommand{\bbf}{{\bf f}}

\newcommand{\bbeta}{\bm{\beta}}

\def\ljump{{[\![}}
\def\rjump{{]\!]}}

\newcommand{\jump}[1]{\left\llbracket #1\right\rrbracket}
\newcommand{\avg}[1]{\{ #1\}}

\newcommand{\Th}{\mathcal{T}_h}

\newcommand{\Eh}{\mathcal{E}_h}
\newcommand{\Eho}{\mathcal{E}_h^o}

\newcommand{\trinorm}[1]{{\left\vert\kern-0.25ex\left\vert\kern-0.25ex\left\vert #1 \right\vert\kern-0.25ex\right\vert\kern-0.25ex\right\vert}}

\newcommand{\sumin}{\sum_{T \in \mathcal{T}_h}}
\newcommand{\sumi}{\sum_{e \in \mathcal{E}_h^I}}
\newcommand{\sumd}{\sum_{e \in \mathcal{E}_{h, \bu}^{\partial, D}}}
\newcommand{\sumid}{\sum_{e \in \mathcal{E}_h^I \cup \mathcal{E}_{h, \bu}^{\partial, D}}}
\newcommand{\sumnT}{\sum_{e \in \mathcal{E}_{h, \theta}^{\partial, N}}}
\newcommand{\sumnU}{\sum_{e \in \mathcal{E}_{h, \bu}^{\partial, N}}}

\newcommand{\Ra}{\mathrm{Ra}}
\newcommand{\Rey}{\text{Re}}
\newcommand{\Ri}{\text{Ri}}
\newcommand{\Prr}{\text{Pr}}

\usepackage{amsthm}
\newtheorem{theorem}{Theorem}
\newtheorem{remark}{Remark}%

\theoremstyle{thmstylethree}%
\title{Pressure-robust enriched Galerkin finite element methods for coupled Navier-Stokes and heat equations}

\author{
  Sanjeeb Poudel \\
  Department of Scientific Computing\\
  Florida State University \\
  Tallahassee, FL\\
  \texttt{spoudel@fsu.edu} \\
   \And
  Sanghyun Lee \\
  Department of Mathematics\\
  Florida State University \\
  Tallahassee, FL\\
  \texttt{slee17@fsu.edu} \\
  \AND
  Lin Mu \\
  Department of Mathematics \\
  University of Georgia \\
  Athens, GA \\
  \texttt{linmu@uga.edu} \\
}

\begin{document}

\maketitle

\begin{abstract}
We propose a pressure-robust enriched Galerkin (EG) finite element method for the incompressible Navier-Stokes and heat equations in the Boussinesq regime.
For the Navier-Stokes equations, the EG formulation combines continuous
Lagrange elements with a discontinuous enrichment vector per element in the
velocity space and a piecewise constant pressure space, and it can be implemented
efficiently within standard finite element frameworks. To enforce pressure
robustness, we construct velocity reconstruction operators that map the
discrete EG velocity field into exactly divergence-free,
$H(\mathrm{div})$-conforming fields. In particular, we develop reconstructions
based on Arbogast-Correa (AC) mixed finite element spaces on quadrilateral
meshes and demonstrate that the resulting schemes remain stable and accurate
even on highly distorted grids.  
The nonlinearity of the coupled
Navier-Stokes-Boussinesq system is treated with several iterative strategies,
including Picard iterations and Anderson-accelerated iterations; our numerical
study shows that Anderson acceleration yields robust and efficient
convergence for high Rayleigh number flows within the proposed framework. The
performance of the method is assessed on a set of benchmark problems and
application-driven test cases. These numerical experiments highlight the
potential of pressure-robust EG methods as flexible and accurate tools for
coupled flow and heat transport in complex geometries.
\end{abstract}


\section{Introduction} \label{sec:intro}
The coupling of incompressible fluid flow and heat transport arises in a wide
range of scientific and engineering applications, including the design of heat
exchangers~\cite{sha_multidimensional_1982,stamou2006verification},
temperature control in turbine blades of a jet
engines~\cite{sondak2000simulation}, enhanced geothermal energy
systems~\cite{olasolo2016enhanced}, among many others.
A standard framework for modeling buoyancy-driven flows is the Boussinesq
approximation~\cite{boussinesq_theorie_1897, pollock2021acceleration}, in
which all variations of the fluid properties except for density are neglected,
and density variations appear only in the gravitational body force. For
sufficiently small density contrasts, this approximation provides an accurate
and computationally tractable model and underpins simulations of natural
convection~\cite{de1983natural}, mantle
convection~\cite{kronbichler2012high}, oceanic general
circulation~\cite{mcwilliams1996modeling}, and related geophysical and
engineering flows.

The numerical solution of the coupled Navier-Stokes-Boussinesq system
presents several challenges. For the incompressible Navier-Stokes equations,
the velocity and pressure finite element spaces must satisfy the inf-sup
stability condition for saddle-point
problems~\cite{babuska_finite_1973, brezzi_existence_1974}. Beyond inf-sup
stability, it is now well understood that many classical mixed finite element
pairs produce velocity errors that depend explicitly on the continuous
pressure gradient and the Reynolds number. As a consequence, the discrete
velocity may be severely polluted by irrotational pressure modes in the
high Reynolds number regime unless one employs a pressure-robust
discretization~\cite{su2025pressure}.
These difficulties are further exacerbated on strongly distorted meshes, which
are often unavoidable in complex geometries and porous media applications.

In this work, we employ and extend the enriched Galerkin (EG) finite element
method~\cite{yi2022enriched,hu2024pressure,yi2022locking,lee2023locking}
for the coupled incompressible Navier-Stokes and heat equations in the
Boussinesq setting. The EG velocity space consists of continuous Lagrange
finite elements enriched by suitable discontinuous functions, paired with a piecewise constant pressure space. The resulting formulation retains a variational structure closely related to interior-penalty discontinuous Galerkin methods and can be implemented efficiently in standard finite element codes, with significantly fewer degrees of freedom than many classical inf-sup stable elements.

However, as with other standard mixed methods, the basic EG discretization is
not pressure-robust: in the high-Reynolds-number regime, the velocity error
bound typically depends on the pressure error and scales directly with the
Reynolds number. To overcome this limitation, we design pressure-robust EG
schemes based on a velocity reconstruction operator in the spirit
of~\cite{hu2024pressure}. The reconstruction maps the discrete EG
velocity into an exactly divergence-free, $H(\mathrm{div})$-conforming field,
thereby removing the spurious influence of the irrotational component of the
pressure gradient on the discrete velocity. We consider such reconstructions
for both the pure Navier-Stokes system and the coupled Navier-Stokes-heat
equations, and demonstrate that the resulting schemes yield velocity errors
that are essentially independent of the pressure and the Reynolds number in
practice.

A particular focus of this work is the robust treatment of distorted
quadrilateral meshes. To this end, we construct pressure-robust reconstructions
for EG velocities using the Arbogast-Correa (AC) mixed finite element
spaces~\cite{arbogast2016two} on quadrilaterals. The AC elements are
constructed using vector polynomials defined directly on the quadrilaterals,
with additional functions mapped by the Piola transformation. This approach is
carefully designed to remain stable and accurate under significant mesh
distortion, which is crucial for pore-scale simulations in complex geometries
and for upscaled models in fractured media. The resulting pressure-robust EG
schemes thus combine local mass
conservation~\cite{lee2016locally,choo2018enriched}, inf-sup stability, and
robustness with respect to both pressure and mesh quality.

In addition to the spatial discretization, the nonlinearity of the coupled
Navier-Stokes-Boussinesq system requires efficient and robust nonlinear
solvers. We systematically investigate several iterative strategies, including
the Picard fixed-point iteration and Anderson-accelerated Picard
iteration~\cite{anderson1965iterative, evans2020proof}. The Picard method is
known to be globally convergent under suitable smallness conditions on the
Reynolds number and related parameters~\cite{li2023accelerating}, but may
converge slowly in strongly nonlinear regimes. Anderson acceleration has been
shown to significantly enhance the convergence of fixed-point schemes, in
particular for Navier-Stokes flows and natural convection
problems~\cite{pollock2019anderson, pollock2021acceleration}.
Alternatively, Newton’s method offers quadratic convergence but requires an
initial guess that is sufficiently close to the solution and involves solving
the fully coupled fluid-heat system at each iteration. Therefore, in this
work, we primarily focus on the Anderson-accelerated Picard method, which
decouples the fluid and heat equations for efficient numerical solution. Our
numerical experiments indicate that Anderson acceleration applied to fixed-point
iterations provides a robust and efficient solver for high-Rayleigh-number
flows within the proposed pressure-robust EG framework.

The main contributions of this paper can be summarized as follows:
\begin{itemize}[leftmargin=15pt]
  \item We develop an enriched Galerkin finite element framework for the
  coupled incompressible Navier-Stokes and heat equations under the
  Boussinesq approximation.
  \item We design and analyze pressure-robust EG schemes for the Boussinesq
  approximation based on velocity reconstruction operators, and construct
  corresponding pressure-robust reconstructions on distorted quadrilateral
  meshes using AC mixed finite element spaces. The resulting methods yield
  velocity errors that are essentially independent of the pressure and
  Reynolds number, and enable accurate simulations on highly distorted grids.
  \item We investigate nonlinear iterative solvers for the coupled system and
  demonstrate that Anderson-accelerated schemes are particularly effective for
  high Rayleigh number flows.
\end{itemize}

These developments are validated through a series of benchmark and
application-driven numerical experiments. We first study cavity flow and
natural convection benchmarks to verify the accuracy of the proposed method.
We then examine pressure robustness and convergence at high Reynolds numbers.
Finally, we apply the method to pore-scale heat transfer in porous media,
evaluating the heat extraction rate for different input data.

\section{Governing System}
\label{sec:governing}

The Boussinesq approximation~\cite{boussinesq_theorie_1897, pollock2021acceleration}
is a widely used model for the numerical simulation of natural convection
problems. It assumes that all fluid properties, including viscosity, specific
heat capacity, and thermal conductivity, remain constant, and that the fluid
density can be approximated by a constant reference value in all terms except
for the gravitational body force term in the momentum equation, where its
temperature dependence is retained. For sufficiently small relative density
variations, the Boussinesq approximation provides an accurate and simplified
description of many buoyancy-driven flows.

Let $\Omega \subset \mathbb{R}^2$, be a bounded, simply connected
Lipschitz domain with boundary $\partial \Omega$, and let
$I := (0,t_f]$ denote the time interval of interest, where $t_f > 0$ is the
final time. We seek the vector-valued fluid velocity
$\bu : \Omega \times I \rightarrow \mathbb{R}^2$, the scalar-valued pressure
$p : \Omega \times I \rightarrow \mathbb{R}$, and the scalar-valued temperature
$\theta : \Omega \times I \rightarrow \mathbb{R}$ such that
\begin{subequations}\label{eqn:governing}
\begin{alignat}{2}
\frac{\partial \bu}{\partial t} + (\bu\cdot\nabla)\bu
- 2 \Rey^{-1}\nabla\cdot\bm{\varepsilon}(\bu)  + \nabla p
&= \Ri\,\theta\,\hat{\mathbf{e}} + \bbf,
&& \quad \text{in } \Omega \times I, \label{eqn:be_momentum} \\
\nabla\cdot \bu &= 0,
&& \quad \text{in } \Omega \times I,  \label{eqn:be_continuity} \\
\frac{\partial \theta}{\partial t} + \bu \cdot \nabla \theta
- \kappa \Delta \theta &= \gamma,
&& \quad \text{in } \Omega \times I. \label{eqn:be_temperature}
\end{alignat}
\end{subequations}
Here, $\bm{\varepsilon}(\bu) := \tfrac{1}{2}(\nabla \bu + (\nabla \bu)^{\!T})$
denotes the symmetric part of the velocity gradient. The Reynolds number $\Rey$
is the ratio of inertial to viscous forces, and the Richardson number $\Ri$
measures the ratio of buoyancy to shear in the flow. The unit vector
$\hat{\mathbf{e}}$ is taken to be opposite to the direction of the
gravitational acceleration. The dimensionless thermal diffusion coefficient is
defined as $\kappa := \Rey^{-1}\Prr^{-1}$, where $\Prr$ is the Prandtl number,
given by the ratio of kinematic viscosity to thermal diffusivity. The function
$\bbf : \Omega \times I \rightarrow \mathbb{R}^2$ denotes an external body
force in the momentum equation, and $\gamma : \Omega \times I \rightarrow
\mathbb{R}$ denotes a volumetric thermal source term. We assume 
$\bbf \in L^2(0,t_f;\,[L^2(\Omega)]^2)$,
and
$\gamma \in L^2(0,t_f;\,L^2(\Omega)),$

For the above system, the boundary $\partial \Omega$ is assumed to be suitably
decomposed into Dirichlet and Neumann parts for both the fluid and temperature
boundary conditions. For the fluid flow, we impose a Dirichlet boundary
condition for the velocity on $\partial \Omega_D^{\bu}$, and prescribe the
total stress on the Neumann part of the boundary, denoted by
$\partial \Omega_N^{\bu} := \partial \Omega \setminus \partial \Omega_D^{\bu}$.
Similarly, we impose Dirichlet and Neumann boundary conditions for the heat
equation on $\partial \Omega_D^\theta$ and
$\partial \Omega_N^\theta := \partial \Omega \setminus \partial \Omega_D^\theta$,
respectively. On the Dirichlet part of the boundary, we prescribe the
temperature values, whereas on the Neumann part we prescribe the normal heat
flux. The boundary conditions are summarized as
\begin{subequations}
\begin{alignat}{2}
\bu & = \mathbf{u}_D
&& \quad \text{on } \partial\Omega_D^\bu \times I, \\
\bigl(2\Rey^{-1} \bm{\varepsilon}(\bu) - p \mathbf{I} \bigr)\,\bn
& = \mathbf{t}_N
&& \quad \text{on } \partial\Omega_N^\bu \times I, \\
\theta &= \theta_D
&& \quad \text{on } \partial \Omega_D^\theta \times I, \\
\kappa \nabla \theta \cdot \bn &= q_N
&& \quad \text{on } \partial \Omega_N^\theta \times I,
\end{alignat}
\end{subequations}
where $\mathbf{u}_D : \partial\Omega_D^\bu \times I \to \mathbb{R}^2$ and
$\mathbf{t}_N : \partial\Omega_N^\bu \times I \to \mathbb{R}^2$ denote
prescribed velocity and traction data, respectively,
$\theta_D : \partial\Omega_D^\theta \times I \to \mathbb{R}$ and
$q_N : \partial\Omega_N^\theta \times I \to \mathbb{R}$ denote prescribed
temperature and normal heat flux, and $\bn$ denotes the
outward unit normal vector on the boundary. We assume that the given data are sufficiently regular such that
$\bu_D \in L^2(0,t_f;\,[H^{1/2}(\partial\Omega_D^\bu)]^2),
\mathbf{t}_N \in L^2(0,t_f;\,[H^{-1/2}(\partial\Omega_N^\bu)]^2),
\theta_D \in L^2(0,t_f;\,H^{1/2}(\partial\Omega_D^\theta)),$ 
and 
$q_N \in L^2(0,t_f;\,H^{-1/2}(\partial\Omega_N^\theta)).$
Moreover, we specify the initial conditions on the whole domain as
\begin{equation}
\bu(\cdot, 0) = \bu^0
\ \text{ and } \
\theta(\cdot, 0) = \theta^0
\ \text{ in } \Omega,
\label{eqn:IC}
\end{equation}
where $\bu^0 : \Omega \to \mathbb{R}^2$ and $\theta^0 : \Omega \to \mathbb{R}$
are the prescribed initial velocity and temperature, respectively.

The system \eqref{eqn:governing}--\eqref{eqn:IC} is fully coupled and presents
several numerical challenges. Owing to its nonlinear character, a first step
is to linearize the equations before computing a numerical solution. In
addition, suitable finite element spaces must be chosen to ensure stability
and convergence. In the following sections, we describe the linearization
procedure and a decoupling strategy for the fluid and temperature equations that
enables efficient solution of the resulting linear systems. Furthermore, we
discuss the choice of finite element spaces with a reduced number of degrees
of freedom and pressure-robust enhancements of the finite element
discretization in order to obtain optimal convergence properties, in
particular for large Reynolds numbers.

\section{Numerical Algorithm}\label{sec:Numerical_Algorithm}
In this section, we discuss the temporal and spatial discretization of the system
\eqref{eqn:governing}--\eqref{eqn:IC}. First, we present a fully implicit temporal
discretization based on the backward Euler scheme. We then introduce iterative
schemes to solve the resulting nonlinear system and describe the spatial
discretization, which employs an enriched Galerkin (EG) space of continuous
piecewise linear functions supplemented by discontinuous enrichment functions~\cite{yi2022enriched} for the velocity, a discontinuous piecewise constant
space for the pressure, and a continuous Galerkin space of piecewise linear
functions for the temperature.

\subsection{Temporal Discretization}
Over the computational time interval $(0,t_f]$, we consider a uniform partition
$ 0 = t_0 < t_1 < \dots < t_N = t_f$,   
$\delta t := t_{n+1} - t_n$ 
for  $n = 0,1,\dots,N-1$,
where $N \in \mathbb{N}$ is the number of time steps. For each $n$, we
denote by
$\bu_n \approx \bu(\cdot,t_n), p_n \approx p(\cdot,t_n), \theta_n \approx \theta(\cdot,t_n),$
the discrete approximations of the continuous solutions at time $t_n$.
Similarly, we write
$\bbf_n \approx \bbf(\cdot,t_n)$,
and
$\gamma_n \approx \gamma(\cdot,t_n).
$

Given the initial conditions $\bu_0$ and $\theta_0$, the unknowns
$[\bu_n, p_n, \theta_n]^\top$ at each time step $t_n$ ($n=1,\dots,N$) are
computed by solving the following system of nonlinear equations:
\begin{subequations}\label{eqn:backward_euler}
    \begin{alignat}{2}
        \frac{\bu_{n} - \bu_{n-1}}{\delta t}
        + (\bu_{n} \cdot\nabla)\bu_{n}
        - 2 \Rey^{-1}\nabla\cdot\bm{\varepsilon}(\bu_{n})
        + \nabla p_{n}
        &= \Ri\,\theta_{n} \hat{\mathbf{e}} + \bbf_{n}, \\[0.3em]
        \nabla\cdot \bu_{n} &= 0, \\[0.3em]
        \frac{\theta_{n} - \theta_{n-1}}{\delta t}
        + \bu_{n} \cdot \nabla \theta_{n}
        - \kappa \Delta \theta_{n}
        &= \gamma_n.
    \end{alignat}
\end{subequations}
To solve the nonlinear system \eqref{eqn:backward_euler}, one may employ various
iterative schemes, such as Picard iteration or Newton's method. Newton's method
offers quadratic convergence provided that the initial guess is sufficiently
close to the exact solution; however, each Newton step requires solving the
fully coupled velocity--pressure--temperature system, which can be
computationally expensive. In contrast, the Picard method, as described below,
decouples the fluid and heat equations at each iteration, enabling a more efficient solution strategy for the coupled problem. For this reason, our primary focus in this work is on Picard iteration, enhanced by Anderson acceleration to improve its convergence behavior; see, e.g., \cite{anderson1965iterative, walker2011anderson}. In the following subsections, we provide the precise formulation of the Picard scheme, its Anderson-accelerated variant~\cite{pollock2019anderson, pollock2021acceleration}, and their implementation within the proposed pressure-robust EG framework~\cite{yi2022enriched, hu2024pressure}.

\subsubsection{Picard Iteration}
The Picard method is a fixed-point iteration for finding solutions of equations
of the form $\bx = g(\bx)$. Starting from an initial guess $\bx_0$, it generates
a sequence of approximations $\{\bx_k\}_{k=1}^{N_k}$ by repeatedly applying the
mapping $g$: $\bx_k = g(\bx_{k-1}),  k=1,\dots,N_k,$
where $k$ denotes the iteration index and $N_k$ is the prescribed maximum number
of iterations. If the method converges, the iterates approach a fixed point
$\bx^*$ satisfying $\bx^* = g(\bx^*)$. In our case,
$\bx = [\bu_n, p_n, \theta_n]^\top$, and $g(\cdot)$ represents one Picard update
of the linearized system~\eqref{eqn:backward_euler} at time level $t_n$.

At a fixed time step $t_n$, we linearize the nonlinear advection terms by
freezing the velocity at the previous Picard iterate. More precisely, given
$\bu_{n,k-1}$, we replace
\[
  (\bu_n \cdot \nabla)\bu_n \approx (\bu_{n,k-1} \cdot \nabla)\bu_{n,k},
  \qquad
  \bu_n \cdot \nabla \theta_n \approx \bu_{n,k-1} \cdot \nabla \theta_{n,k}.
\]
Starting with the initial guess taken from the previous time step,
\[
  \bx_{n,0} := [\bu_{n-1}, p_{n-1}, \theta_{n-1}]^\top,
\]
the Picard iteration produces a sequence $\{\bx_{n,k}\}_{k \ge 1}$ with
$\bx_{n,k} = [\bu_{n,k}, p_{n,k}, \theta_{n,k}]^\top$. At each iteration
$k \ge 1$, we solve the following linearized system:
\begin{subequations}
\label{eqn:picard}
\begin{align}
        \frac{1}{\delta t} \bu_{n,k}
        + (\bu_{n,k-1} \cdot\nabla)\bu_{n,k}
        - 2 \Rey^{-1} \nabla\cdot \bm{\varepsilon}(\bu_{n,k})
        + \nabla p_{n,k}
        &= \frac{1}{\delta t} \bu_{n-1}
        + \Ri\, \theta_{n,k} \hat{\mathbf{e}} + \bbf_n, \\
        \nabla\cdot \bu_{n,k} &= 0, \\
        \frac{1}{\delta t} \theta_{n,k}
        + \bu_{n,k-1} \cdot \nabla \theta_{n,k}
        - \kappa \Delta \theta_{n,k}
        &= \frac{1}{\delta t} \theta_{n-1} + \gamma_n.
\end{align}    
\end{subequations}

In practice, this linear system is solved in a decoupled manner. Given the
velocity iterate $\bu_{n,k-1}$, we first update the temperature by solving the
advection–diffusion equation,
\[
  \frac{1}{\delta t} \theta_{n,k}
  + \bu_{n,k-1} \cdot \nabla \theta_{n,k}
  - \kappa \Delta \theta_{n,k}
  = \frac{1}{\delta t} \theta_{n-1} + \gamma_n,
\]
and then update the velocity and pressure by solving the linearized
fluid system,
\begin{align*}
\frac{1}{\delta t} \bu_{n,k}
      + (\bu_{n,k-1} \cdot\nabla)\bu_{n,k}
      - 2 \Rey^{-1} \nabla\cdot \bm{\varepsilon}(\bu_{n,k})
      + \nabla p_{n,k}
      &= \frac{1}{\delta t} \bu_{n-1}
      + \Ri\, \theta_{n,k} \hat{\mathbf{e}} + \bbf_n, \\
      \nabla\cdot \bu_{n,k} &= 0.
\end{align*}
This procedure is repeated until a suitable convergence criterion is satisfied,
for example,
$$
  \|\bu_{n,k} - \bu_{n,k-1}\| 
  +  ||p_{n, k} - p_{n, k-1}|| + \|\theta_{n,k} - \theta_{n,k-1}\|
  \le \varepsilon_{\text{Picard}},
$$
for a prescribed tolerance $\varepsilon_{\text{Picard}} > 0$, where
$\|\cdot\|$ denotes an appropriate norm. 
Here, we employ the discrete $L^2(\Omega)$ norm, applied componentwise to the velocity, pressure, and temperature fields. In Section~\ref{sec:AA}, we combine this Picard scheme with Anderson acceleration
to further improve its convergence properties.

\subsubsection{Anderson-accelerated Picard iteration (AA--Picard)}
\label{sec:AA}
Anderson acceleration is an extrapolation technique that is effective in
accelerating and enhancing the convergence of fixed-point methods. In
particular, it has proven useful for Picard iterations applied to the
Navier--Stokes and Boussinesq
equations~\cite{pollock2019anderson, pollock2021acceleration}.

Consider a generic fixed-point iteration $\bx_k = g(\bx_{k-1})$. We define the
differences between successive iterates $(\be_k)$ and the fixed-point residuals
$(\bw_k)$ by
$$
  \be_k := \bx_k - \bx_{k-1}, 
  \qquad
  \bw_k := g(\bx_{k-1}) - \bx_{k-1},
$$
for $k \ge 1$. Thus, $\bw_k$ denotes the residual associated with the iterate
$\bx_{k-1}$. The Anderson acceleration algorithm starts with a (possibly
damped) fixed-point update at $k=1$, and the extrapolation (acceleration) is
activated for $k \ge 2$. We denote by $m > 0$ the maximum allowable
algorithmic depth, by $m_k$ the depth used at iteration~$k$, and by $\beta_k$
the relaxation (damping or mixing) parameter.

At each iteration $k \ge 2$, we form matrices $\mathbf{E}_k$ and $\mathbf{F}_k$
whose columns collect recent differences between iterates and residuals,
respectively. More precisely, we set
$m_k := \min\{k-1, m\}$,
and define
$\Delta \bw_j := \bw_{j+1} - \bw_j, 
j \ge 1,$
so that
$$
  \mathbf{F}_k 
    := \bigl[\,\Delta \bw_{k-1},\,\Delta \bw_{k-2},\,\dots,\,\Delta \bw_{k-m_k}\,\bigr],
  \
  \mathbf{E}_k 
    := \bigl[\,\be_{k-1},\,\be_{k-2},\,\dots,\,\be_{k-m_k}\,\bigr].
$$
The Anderson coefficients $\gamma_k \in \mathbb{R}^{m_k}$ provide an optimal (in
the Euclidean norm) linear combination of the residual differences that
approximates the current residual. They are obtained by solving the linear
least-squares problem:
$$
  \gamma_k 
  = \operatorname*{argmin}_{\gamma \in \mathbb{R}^{m_k}}
    \bigl\| \mathbf{F}_k \gamma - \bw_k \bigr\|_2,
$$
which can be efficiently solved, for instance, by a QR factorization. We then
define the affine combinations
$$
  \bx_{k-1}^\alpha := \bx_{k-1} - \mathbf{E}_k \gamma_k,
  \qquad
  \bw_k^\alpha := \bw_k - \mathbf{F}_k \gamma_k,
$$
and perform the Anderson-accelerated update
$$
  \bx_k 
  = \bx_{k-1}^\alpha + \beta_k \bw_k^\alpha
  = \bx_{k-1} + \beta_k \bw_k - (\mathbf{E}_k + \beta_k \mathbf{F}_k)\gamma_k.
$$
A detailed description of the algorithm is given in Algorithm~\ref{alg:AA}.

\begin{algorithm}
    \caption{Anderson acceleration~\cite{pollock_filtering_2023, li2023accelerating}}
    \label{alg:AA}
    \begin{algorithmic}[1]    
        \STATE{Choose initial iterate $\bx_0$ and maximum allowable algorithmic depth parameter $m$.}
        \STATE{Compute $\bw_1 = g(\bx_0) - \bx_0$, choose relaxation parameter $\beta_1$, and set $\bx_1 = \bx_0 + \beta_1 \bw_1$ and $\be_1 = \bx_1 - \bx_0$.}        
        \FOR{$k = 2, 3, \dots$}
            \STATE{Compute residual $\bw_{k} = g(\bx_{k-1}) - \bx_{k-1}$}
            \STATE{Set $m_k = \min\{k-1, m\}$}
            \STATE{Form $\mathbf{F}_k = [\,\Delta \bw_{k-1},\dots,\Delta \bw_{k-m_k}\,]$ with $\Delta \bw_j = \bw_{j+1} - \bw_j$}
            \STATE {Form $\mathbf{E}_k = [\,\be_{k-1},\dots,\be_{k-m_k}\,]$ with $\be_j = \bx_j - \bx_{j-1}$}
            \STATE{Compute $\displaystyle \gamma_k = \operatorname*{argmin}_{\gamma \in \mathbb{R}^{m_k}} \bigl\|\mathbf{F}_k \gamma - \bw_k \bigr\|_2$}
            \STATE{Choose relaxation parameter $\beta_k$}
            \STATE{Update $\displaystyle \bx_{k} = \bx_{k-1} + \beta_k \bw_{k} - (\mathbf{E}_k + \beta_k \mathbf{F}_k) \gamma_{k}$}
            \STATE{Set $\be_k = \bx_k - \bx_{k-1}$}     
        \ENDFOR
    \end{algorithmic}
\end{algorithm}

\subsection{Spatial Discretization}
We perform the spatial discretization of the domain using the finite element method (FEM). 
In this section, we present the weak formulation, the enriched Galerkin finite element method (EG--FEM), and the pressure-robust enhancement of the EG method. 
The formulation is given for the Picard iteration; the Anderson-accelerated Picard method uses the same discrete system at each iteration. 

We use the standard notation for the Sobolev space $H^s(\Omega)$ for a domain $\Omega \subset \mathbb{R}^2$ and an integer $s \ge 0$, where $H^s(\Omega)$ consists of functions with square-integrable weak derivatives up to order $s$. 
We also define
$ L^2_0(\Omega) := \bigl\{ q \in L^2(\Omega) : \int_\Omega q \, dx = 0 \bigr\}$.

The weak solution of \eqref{eqn:picard} at time step $t_n$ and Picard iteration $k$ is a triple 
$$
  (\bu_{n,k}, p_{n,k}, \theta_{n,k}) 
  \in 
  \begin{cases}
    [H^1(\Omega)]^2 \times L^2(\Omega) \times H^1(\Omega), & \text{if } |\partial\Omega_N^\bu| > 0,\\[0.2em]
    [H^1(\Omega)]^2 \times L^2_0(\Omega) \times H^1(\Omega), & \text{if } |\partial\Omega_N^\bu| = 0,
  \end{cases}
$$
such that $\bu_{n,k}\vert_{\partial\Omega_D^\bu} = \mathbf{u}_D$, $\theta_{n,k}\vert_{\partial\Omega_D^\theta} = \theta_D$ and
\begin{multline*}
\left(\frac{1}{\delta t} \bu_{n,k}, \bv \right) \;+\; \bigl( (\bu_{n,k-1} \cdot \nabla) \bu_{n,k}, \bv \bigr)  \;+\; 2\Rey^{-1}\bigl(\bm{\varepsilon}(\bu_{n,k}), \bm{\varepsilon}(\bv)\bigr) -\; (p_{n,k}, \nabla \cdot \bv) \\
= \left(\frac{1}{\delta t} \bu_{n-1}, \bv \right)  + \Ri(\theta_{n,k} \hat{\mathbf{e}}, \bv) + (\bbf_n, \bv) + (\mathbf{t}_N, \bv)_{\partial\Omega_N^\bu},
\quad \forall \bv \in [H^1_{0,\bu}(\Omega)]^2,
\end{multline*}
\begin{equation*}
    (\nabla \cdot \bu_{n,k}, w) = 0, \quad \forall w \in 
    \begin{cases}
          L^2(\Omega), & \text{if } |\partial\Omega_N^\bu| > 0,\\
          L^2_0(\Omega), & \text{if } |\partial\Omega_N^\bu| = 0,
    \end{cases}
\end{equation*}
\begin{multline*}
    \left(\frac{1}{\delta t} \theta_{n,k}, \tau \right) \;+\;(\bu_{n,k-1} \cdot \nabla \theta_{n,k}, \tau)\;+\; \kappa \left(\nabla \theta_{n,k}, \nabla \tau \right) \\ 
    = \left(\frac{1}{\delta t} \theta_{n-1}, \tau \right) + (\gamma_n, \tau) + (q_N, \tau)_{\partial\Omega_N^\theta}, \quad \forall \tau \in H^1_{0,\theta}(\Omega).
\end{multline*}
Here, $H^1_{0,\bu}(\Omega)$ and $H^1_{0,\theta}(\Omega)$ denote the spaces
\begin{align*}
  &H^1_{0,\bu}(\Omega) := \{ \bv \in [H^1(\Omega)]^2 : \bv = \mathbf{0} \text{ on } \partial\Omega_D^\bu \},\\
  &H^1_{0,\theta}(\Omega) := \{ \tau \in H^1(\Omega) : \tau = 0 \text{ on } \partial\Omega_D^\theta \},
\end{align*}
and $(\cdot,\cdot)$ denotes the $L^2(\Omega)$ inner product, with $(\cdot,\cdot)_{\partial\Omega}$ the $L^2$ inner product on the boundary.

For a saddle-point problem to have a unique solution, the finite element spaces in the mixed formulation must satisfy an inf--sup stability (LBB) condition \cite{babuska_finite_1973, brezzi_existence_1974}. 
By enriching the continuous finite element space for the velocity with discontinuous mean-zero vector functions, the EG--FEM scheme yields an inf--sup stable velocity--pressure pair for the Stokes problem with a minimal number of degrees of freedom \cite{yi2022enriched}. 
Therefore, we use the EG--FEM scheme for the fluid part of the governing system to ensure an efficient and stable numerical solution. 
For the temperature, we use a continuous Galerkin space with piecewise linear basis functions.

\subsubsection{Enriched Galerkin Finite Element Method}
We consider a shape regular partition of the computational domain, $\overline{\Omega} = \cup_{T \in \mathcal{T}_h} \overline{T},$ where $T \in \mathcal{T}_h$ are quadrilaterals. We denote the set of all edges by $\Eh$, which has a partition $\Eh = \Ehi \cup \Eho$, where $\Ehi$ is the set of all interior edges while $\Eho$ is the set of all boundary edges. 

The EG finite-element space for the velocity is obtained by extending the continuous space with a discontinuous function. First, we define a vector-valued linear continuous Galerkin (CG) finite-element space:
$$\mathcal{CG}_1 := \{\psi \in [H^1(\Omega)]^2 \; | \; \psi\vert_T \in [\mathbb{Q}_1(T)]^2, \; \forall T \in \mathcal{T}_h \},$$
where $\mathbb{Q}_1(T)$ is the space of polynomials with each variable degree 1. 
Then, the EG space for velocity, defined as $\mathcal{V}_h$, is obtained by extending $\mathcal{CG}_1$ with the discontinuous enrichment space, 
$$ \mathcal{D} := \{ \psi \in [L^2(\Omega)]^2 \; | \; \psi\vert_T = c_T(\bx - \bx_T), \; c_T \in \mathbb{R}, \; \forall T \in \mathcal{T}_h \},$$
where $\bx$ is the position vector and $\bx_T$ is the centroid of $T \in \mathcal{T}_h$. 
Then, our EG finite element space for the velocity is defined by
\begin{equation*}
\mathcal{V}_h := \mathcal{CG}_1 \oplus \mathcal{D} \subset [L^2(\Omega)]^2.
\label{eqn:eg_def}
\end{equation*}
For the pressure $p$, we use the piecewise constant function space:
$$\mathcal{W}_h := \begin{cases}
                \{\psi \in L^2(\Omega) \; | \; \psi|_T \in \mathbb{Q}_0(T), \forall T \in \mathcal{T}_h \} \cap L_0^2(\Omega) \quad &\text{if } |\partial\Omega_N^\bu| = 0, \\
                \{\psi \in L^2(\Omega) \; | \; \psi|_T \in \mathbb{Q}_0(T), \forall T \in \mathcal{T}_h \} \quad &\text{if } |\partial\Omega_N^\bu| > 0.
                \end{cases}$$
For the temperature $\theta$, we use a linear CG finite-element space:
$$\Theta_h := \{\psi \in H^1(\Omega), \; | \; \psi \vert_T \in \mathbb{Q}_1(T), \; \forall T \in \mathcal{T}_h\}.$$
As mentioned earlier, the Picard iteration method does not require solving the entire fluid and heat system simultaneously. Instead, we can first solve the heat equation and then use the resulting temperature field to solve the fluid dynamics system. To solve the heat equation, we define the following linear and bilinear forms:
\begin{align*}
a_\theta(\bbeta, \theta, \tau) &:= \sumin \Big( \frac{1}{\delta t}(\theta, \tau)_T + (\bbeta \cdot \nabla \theta, \tau)_T + \kappa (\nabla \theta, \nabla \tau)_T \Big), \\
{\bf F}_\theta(\theta, \tau) &:= \sumin \Big( \frac{1}{\delta t} (\theta, \tau)_T + (\gamma, \tau)_T \Big) + \sumnT(q_N, \tau)_e,
\end{align*}
where $e \in \mathcal{E}_{h, \theta}^{\partial, N}$ is the boundary edge where the Neumann condition is specified for the temperature. With the above definitions, the Picard iteration involves solving the following linear system for temperature to find $\theta^h_{n,k} \in \Theta_h$ such that
\begin{equation}
    a_\theta(\bu_{n,k-1}^h, \theta_{n, k}^h, \tau) = {\bf F}_\theta(\theta_{n-1}^h, \tau), \quad \forall \tau \in \Theta_h  
    \label{eq:linear_system_temp}
\end{equation}
Similarly, for the fluid system, we define the following linear and bilinear forms:
\begin{multline*}
a_\bu(\bu, \bv) := \sumin \left( \frac{1}{\delta t} (\bu, \bv)_T + 2 \Rey^{-1}  (\epsilon(\bu), \epsilon(\bv))_T \right) \\
- 2\Rey^{-1}\sumid \Bigl( (\avg{\epsilon(\bu)}\bn, \jump{\bv})_e  - \zeta \left(\jump{\bu}, \avg{\epsilon(\bv)}\bn \right)_e - \frac{\alpha}{h_e} \bigl(\jump{\bu}, \jump{\bv} \bigr)_e \Bigr),
\end{multline*}
\begin{multline*}
b(\bbeta, \bu, \bv) := \sumin \left( (\bbeta \cdot \nabla)\bu, \bv \right)_T + \sumi \left( |\avg{\bbeta} \cdot \bn|, (\bu^+ - \bu^-)\cdot \bv^+ \right)_e + \sum_{e\in\mathcal{E}_h^{o-}} \left( |\bbeta \cdot \bn |, \bu^+ \cdot \bv^+\right)_e
\end{multline*}
\begin{equation*}
c({\bf v}, w) := -\sumin(w, \nabla \cdot \bv)_T + \sumid (\{w\}, \jump{\bv} \cdot \bn)_e, 
\end{equation*}
\begin{multline*}
{\bf F}_\bu(\bu, \bv, \theta) := \sumin \Big( \frac{1}{\delta t} (\bu, \bv)_T  + \Ri(\theta \hat{\be}, \bv)_T  + (f, \bv)_T \Big) + \sumnU({\bf t}_N, \bv)_e \\
+ 2\Rey^{-1} \sumd \Bigl( \zeta (\bu_D, \epsilon(\bv)\bn)_e + \frac{\alpha}{h_e}(\bu_D, \bv)_e \Bigr) + \sum_{e\in\mathcal{E}_h^{o-}} \left( |\bbeta \cdot \bn |, \bu^+_D \cdot \bv \right)_e,
\end{multline*}
where the constant $\zeta$ is a symmetrization parameter chosen from $\{-1, 0, 1\}$, $\alpha > 0$ is the penalty parameter, and $h_e = \vert e \vert ^{\frac{1}{d-1}}$ is the mesh size. 
The boundary edges or faces have two subsets: $\mathcal{E}_{h, \bu}^{\partial, D}$ and $\mathcal{E}_{h, \bu}^{\partial, N}$ for Dirichlet and Neumann boundary conditions for the fluid system, respectively.

For each interior edge $e\in \mathcal{E}_h^I$, let $T^{+}$ and $T^{-}$ be the neighboring elements in $\Th$ such that $e =\partial T^{+}\cap \partial T^{-}$ and let $\bn^{+}$ and $\bn^{-}$ be the unit outward normal vector to $\partial T^{+}$ and $\partial T^{-}$, respectively. Then, for a given vector function $\mathbf{q}$, we define the average operator $\{\cdot\}$ and the jump operator $\left\llbracket \cdot\right\rrbracket $ on $e$ by
$$
\{\bq\} = \frac{\bq|_{T^{+}} + \bq|_{T^{-}}}{2}, \quad \quad 
\left\llbracket \bq \right\rrbracket = \bq|_{T^{+}} - \bq|_{T^{-}}.
$$
On the other hand, for a boundary edge $e \in \Eho \cap \partial T$ for some $T \in \Th$, we define
$$
\ave{\bq}=(\bq \vert_T) \vert_{e}, \quad \jump{\bq}  =  (\bq \vert_{T} )\vert_e \bn,
$$
where $\bn$ is the unit outward normal vector to $\partial \Omega$.

For the upwind scheme on an interior edge $e \in \Ehi$, we denote $\bu^+$ as the velocity $\bu$ approaching from inside a cell $T^+$ when $\avg{\bbeta} (\bx)\cdot \bn^+ < 0$ (inflow to $T^+$) and $\bu^-$ is the limit of $\bu$ approaching from the neighboring element $T^-$ when $\avg{\bbeta} (\bx) \cdot \bn^- > 0$. For boundary edges, let $\mathcal{E}_h^{o-}$ denote inflow boundary edges, i.e., $\bbeta(\bx)\cdot \bn < 0$. At these inflow boundaries, we define $\bu_D^+$ as the boundary velocity. If the inflow is the Dirichlet boundary, we use the prescribed value, i.e., $\bu_D^+ = \bu_D$. For the Neumann part of the boundary, refer to Remark~\ref{remark:upwind_for_Neumann} for further details.

With the above definitions, each Picard iteration involves solving the following linear system for fluid velocity and pressure to find $(\bu_{n,k}^h, p_{n,k}^h) \in \mathcal{V}_h \times \mathcal{W}_h$ such that
\begin{subequations}
    \begin{alignat}{3}
        a_\bu(\bu_{n,k}^h, \bv) + b(\bu_{n,k-1}^h, \bu_{n,k}^h, \bv) + c(\bv, p_{n, k}^h) &= {\bf F}_\bu(\bu_{n-1}^h, \bv, \theta_{n, k}^h), \; &&\forall {\bf v} \in \mathcal{V}_h,  \\
        c(\bu_{n, k}^h, w) &= \sum_{e\in \mathcal{E}_h^{\partial, D}} ({w}, \bu_D \cdot {\bf n_e})_e, \; &&\forall w \in \mathcal{W}_h.
    \end{alignat}    
\label{eq:linear_system_fluid}
\end{subequations}
\begin{remark}
    \label{remark:upwind_for_Neumann}
    {To compute the upwind flux at the the Neumann part of the boundary ($e \in \mathcal{E}_{h, \bu}^{\partial, N}$), where the flow is directed into the domain (inflow), the boundary velocity is taken from the solution at the previous iterative step; i.e., the upwind flux is computed as:
    $$
    \left( |\bbeta \cdot \bn |, \bu^+_D \cdot \bv \right)_e = \left( |\bbeta \cdot \bn |, \bbeta \cdot \bv \right)_e 
    $$
    where, in the Picard iteration~\eqref{eq:linear_system_fluid}, we set $\bbeta = \bu^h_{n, k}$.}
\end{remark}

\begin{remark}
For all the numerical experiments, when imposing the Dirichlet condition on velocity, we enforce the prescribed boundary condition strongly on the continuous component of the velocity, i.e., the continuous part of the velocity exactly matches the boundary condition. For the discontinuous part, we impose the boundary condition weakly to zero through the penalty term.
\end{remark}

At each time step $n$, the Picard iteration proceeds by solving the linear system in Eqn. \eqref{eq:linear_system_temp} to update the temperature field. This updated temperature is then used in Eqn. \eqref{eq:linear_system_fluid} to update the velocity and pressure. This process continues until the Picard method achieves the desired level of convergence (in terms of relative norm or residual). Once it converges, the solution is advanced to the next time step. Algorithm \ref{alg:Picard_EG} shows the details of Picard iteration for the proposed finite element scheme. 
\begin{algorithm}
    \caption{Picard Iteration with EG FEM}\label{alg:Picard_EG}
    \begin{algorithmic}[1]    
        \STATE{\textbf{Input:} Initial and boundary conditions: $\bu_0, \theta_0, {\bu_D}, {\bu_N}, \theta_D, \theta_N$}      
        \FOR{n = 1, \dots}
            \STATE{$\bu_{n, 0}^h = \bu_{n-1}^h$}
            \FOR{k = 1, \dots}                
                \STATE {Solve Eqn. \eqref{eq:linear_system_temp} to obtain $\theta_{n,k}^h$}
                \STATE {Solve Eqn. \eqref{eq:linear_system_fluid} to obtain $\bu_{n, k}^h$ and $p_{n,k}^h$}
            \ENDFOR
            \STATE{$\bu_{n}^h = \bu_{n, k}^h, p_{n}^h = p_{n, k}^h, \theta_{n}^h = \theta_{n, k}^h$}            
        \ENDFOR
    \end{algorithmic}
\end{algorithm}
\subsubsection{Arbogast-Correa (AC) elements}
First, we briefly recapitulate the AC$_0$ space established in \cite{arbogast2016two}, which will be used to perform velocity reconstruction. Unlike Raviart-Thomas (RT) \cite{raviart2006mixed} and Brezzi-Douglas-Marini (BDM) \cite{brezzi1985two} elements that are defined on rectangles and extended to quadrilaterals using the Piola transform, the elements for AC$_0$ are constructed from vector polynomials defined directly on the quadrilaterals, which maintains optimal approximation of $\nabla \cdot \bu$. \par
 
The following example illustrates the construction of basis functions in the 
 AC$_0$ space on a quadrilateral element.The physical quadrilateral element $T$ is defined with four vertices as $(x_1,y_1)$, $(x_2,y_2)$, $(x_3,y_3)$, $(x_4,y_4)$ with coordinate ($x,y$) and the reference rectangle $\hat{T}$ is denoted as $[-1,1]^2$ with coordinate $(\hat{x},\hat{y})$. 
\begin{figure} [!h]
\centering
\begin{tabular}{cc}
    \includegraphics[width=0.25\linewidth]{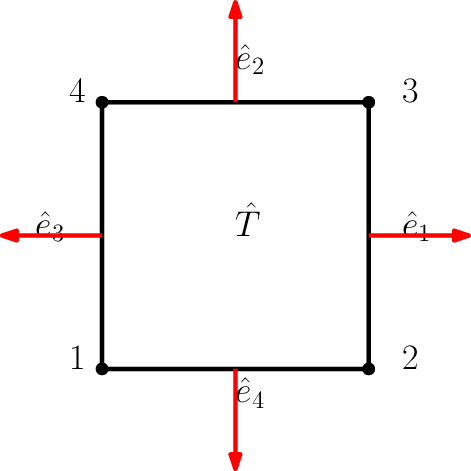}
    &\includegraphics[width=0.20\linewidth]{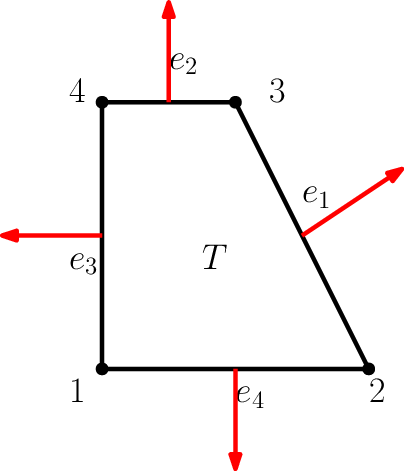}
    \end{tabular}
    \caption{Reference rectangle element $\hat{T}$ and physical trapezoid element $T$.}
    \label{fig:placeholder}
\end{figure}
\begin{figure}
    \centering
    \includegraphics[width=0.2\linewidth]{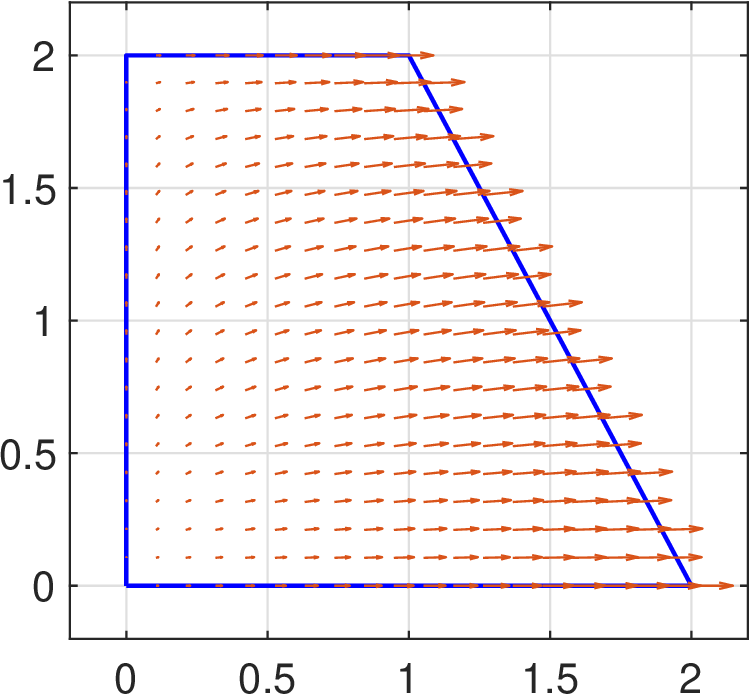}
    \includegraphics[width=0.2\linewidth]{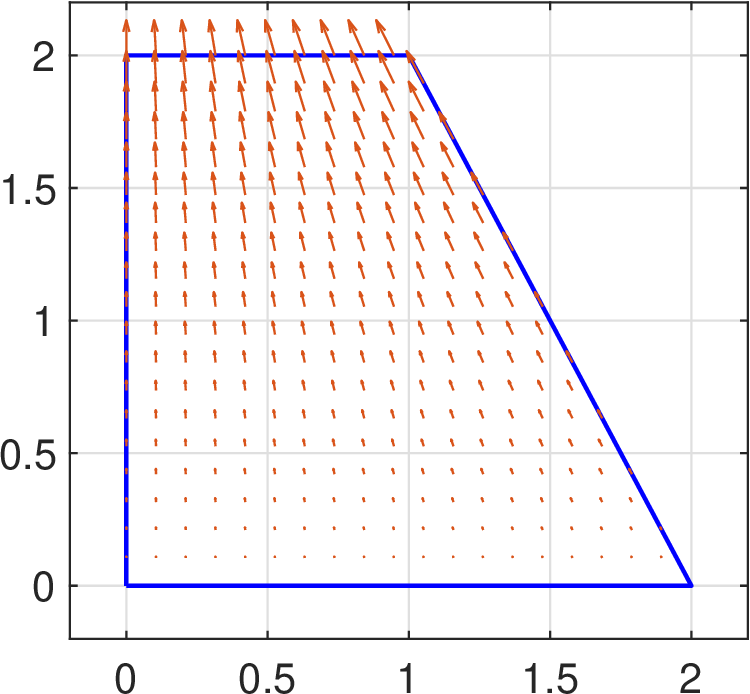}
    \includegraphics[width=0.2\linewidth]{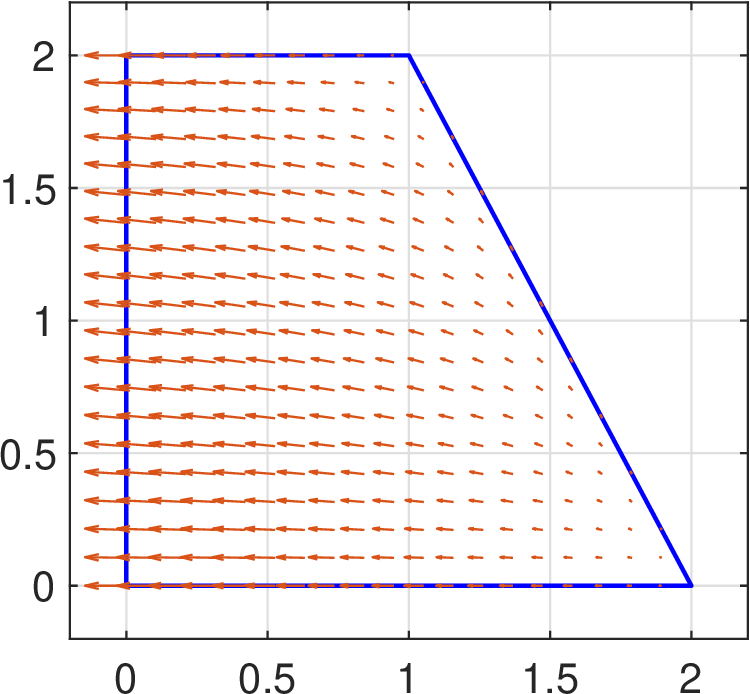}
    \includegraphics[width=0.2\linewidth]{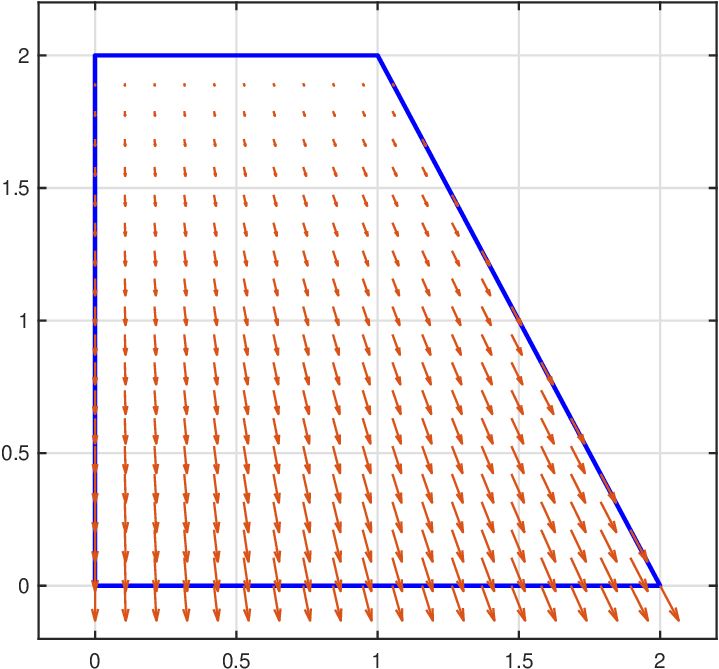}
    \caption{An example of the AC$_0$ basis functions: vector plot in a trapezoid element with four vertices $\{(0,0),(2,0),(1,2),(0,2)\}$.}
    \label{fig:AC0-basis} 
\end{figure}
Let $J_T$ denote the Jacobian matrix of the mapping from the physical quadrilateral element $T$ to the reference rectangle $\hat{T}$. For the basis construction, we use the following vector-valued functions:
\begin{eqnarray*}
    \text{span}\{\Psi_k\}_{k=1}^4 = \left\{\begin{bmatrix}
       x-x_2\\ y-y_2
    \end{bmatrix}, \begin{bmatrix}
       x-x_4\\ y-y_4
    \end{bmatrix},
    \begin{bmatrix}
       x-x_1\\ y-y_1 
    \end{bmatrix}, \frac{J_T}{\text{det}(J_T)}\begin{bmatrix}
        \hat{x}\\-\hat{y}
    \end{bmatrix}
    \right\}.
\end{eqnarray*}
It is easy to verify that $\nabla\cdot\Psi_k = $constant ($k=1,4)$.
Therefore, for each edge $e_j\in\partial T$, we define the degrees of freedom (DOFs) used to compute the AC$_0$ basis functions by: 
$$\int_{e_j} \psi_i\cdot\bn ds = \delta_{ij} \text{ with } \psi_i = \sum_k \alpha_k\Psi_k,\ i,j,k = 1,\cdots,4.$$ 
Thus, the coefficients are determined by using the following inverted matrix:
\begin{eqnarray*}
\mathcal{M} = 
\begin{bmatrix}
    0 & 2|\Delta_{234}| & 2|\Delta_{123}| & 2\\
    2|\Delta_{234}|  &0 & 2|\Delta_{134}| &-2\\
    2|\Delta_{124}| & 0 & 0 & 2\\
    0 &2|\Delta_{124}| & 0 & -2 &
\end{bmatrix}^{-1},
\end{eqnarray*}
where $|\Delta_{234}|$ denotes the area of the triangle formed in the physical domain with vertex 2, 3, and 4, and similarly $|\Delta_{123}|$, $|\Delta_{134}|$, and $|\Delta_{124}|$ are the areas of the triangles with corresponding vertices. 
Finally, for a particular example of a quadrilateral element with four vertices $\{(0,0),(2,0),(1,2),(0,2)\}$, we invert the following matrix to determine the coefficients 
\begin{eqnarray*}
\mathcal{M} = 
\begin{bmatrix}
0&2&4&2\\
2&0&2&-2\\
4&0&0&2\\
0&4&0&-2
\end{bmatrix}^{-1} = 
\begin{bmatrix}
    -1/18& 1/9& 7/36& 1/36\\
    1/18& -1/9& 1/18& 2/9\\
    1/6& 1/6& -1/12& -1/12\\
    1/9& -2/9& 1/9& -1/18
\end{bmatrix}.
\end{eqnarray*}
The resulting basis functions (also shown in Fig. \ref{fig:AC0-basis}) are:
\begin{eqnarray*}
    \psi_1 &=& 
\frac{-1}{18}\Psi_1 + \frac{1}{9}\Psi_2 + \frac{7}{36}\Psi_3 + \frac{1}{36}\Psi_4,\\
\psi_2 &=&\frac{1}{18}\Psi_1 + \frac{-1}{9}\Psi_2 + \frac{1}{18}\Psi_3 + \frac{2}{9}\Psi_4,\\ 
\psi_3 &=& \frac{1}{6}\Psi_1 + \frac{1}{6}\Psi_2 + \frac{-1}{12}\Psi_3 + \frac{-1}{12}\Psi_4,\\
\psi_4 &=& \frac{1}{9}\Psi_1 + \frac{-2}{9}\Psi_2 + \frac{1}{9}\Psi_3 + \frac{-1}{18}\Psi_4.\\
\end{eqnarray*}

\subsubsection{Pressure-Robust Enhancement with AC element}
In this section, we will enhance the linear system for the fluid part given in Eqn. \eqref{eq:linear_system_fluid}  with the desired pressure robustness. We introduce the velocity reconstruction operation $R:\mathcal{V}_h\to \bar{\mathcal{V}}_h\subset H(\text{div},\Omega)$. 
Here we employ the piecewise AC$_0$ basis for $\bar{\mathcal{V}}_h$ and define:
\begin{equation}\label{eq:vel-reconstruction}
    \int_{e}\{ \bv \} \cdot \bn ds = \int_{e}R\bv\cdot\bn ds,\ \forall e\in \mathcal{E}_h^I.
\end{equation}
It is noted that the reconstruction is done locally on the edge of element $T\in\mathcal{T}_h$. 
Then the enhanced algorithm is to find $(\bu^h_{n,k},p^h_{n,k})\in \mathcal{V}_h\times \mathcal{W}_h$ such that
\begin{subequations}
\begin{alignat}{3} 
        a_R(\bu_{n,k}^h, \bv) + b_R(\bu_{n,k-1}^h, \bu_{n,k}^h, \bv) + c_R(\bv, p_{n, k}^h) &= {\bf F}_R(\bu_{n-1}^h, \bv, \theta_{n, k}^h), \; &&\forall {\bf v} \in \mathcal{V}_h,  \\
        c_R(\bu_{n, k}^h, w) &= \sum_{e\in \mathcal{E}_h^{\partial, D}} (w, \bu_D\cdot {\bf n_e})_e, \; &&\forall w \in \mathcal{W}_h,
\end{alignat}    
\label{eq:pr-ac-scheme}
\end{subequations}
where
\begin{multline*}
a_R(\bu, \bv) := \sumin \left( \frac{1}{\delta t} (R\bu, R\bv)_T + 2 \Rey^{-1}  (\epsilon(\bu), \epsilon(\bv))_T \right) \\
- 2\Rey^{-1}\sumid \Bigl( (\avg{\epsilon(\bu)}\bn, \jump{\bv})_e  - \zeta \left(\jump{\bu}, \avg{\epsilon(\bv)}\bn \right)_e - \frac{\alpha}{h_e} \bigl(\jump{\bu}, \jump{\bv} \bigr)_e \Bigr),
\end{multline*}
\begin{multline*}
b_R(\bbeta, \bu, \bv) := \sumin \left( (R\bbeta \cdot \nabla)\bu, R\bv \right)_T + \sumi \left( |\avg{R\bbeta} \cdot \bn|, (\bu^+ - \bu^-)\cdot \bv^+ \right)_e + \sum_{e\in\mathcal{E}_h^{o-}} \left( |R\bbeta \cdot \bn |, \bu^+ \cdot \bv^+\right)_e
\end{multline*}
\begin{equation*}
c_R({\bf v}, w) := -\sumin(w, \nabla \cdot \bv)_T + \sumid (\{w\}, \jump{\bv} \cdot \bn)_e, 
\end{equation*}
\begin{multline*}
{\bf F}_R(\bu, \bv, \theta) := \sumin \Big( \frac{1}{\delta t} (R\bu, R\bv)_T  + \Ri(\theta \hat{\be}, R\bv)_T  + (f, R\bv)_T \Big) + \sumnU({\bf t}_N, \bv)_e \\
+ 2\Rey^{-1} \sumd \Bigl( \zeta (\bu_D, \epsilon(\bv)\bn)_e + \frac{\alpha}{h_e}(\bu_D, \bv)_e \Bigr) + \sum_{e\in\mathcal{E}_h^{o-}} \left( |R\bbeta \cdot \bn |, \bu^+_D \cdot \bv \right)_e,
\end{multline*}

We note that any EG basis functions $\bv \in \mathcal{V}_h$ has a unique decomposition $\bv = \bv_{CG} + \bv_{DG}$ such that 
$\bv_{CG} \in \mathcal{CG}_1$ and $\bv_{DG} \in \mathcal{D}$.
Then, while performing velocity reconstruction of the DG component of the velocity ($\bv_{DG} $) at the Dirichlet boundary condition, we set the normal flux to zero, i.e.,
$$
\int_{\partial \Omega^\bu_D} R\bv_{DG} \cdot \bn \; d\bx = 0.
$$
In contrast, for the CG component ($\bv_{CG}$), the normal flux is preserved, 
$$
\int_{\partial \Omega^\bu_D} R\bv_{CG} \cdot \bn \; d\bx = \int_{\partial \Omega^\bu_D} \bv_{CG} \cdot \bn \; d\bx.
$$
For the Neumann part of the boundary, we preserve the normal flux for both the CG and DG components of the velocity:
$$
\int_{\partial \Omega^\bu_N} R\bv_{CG} \cdot \bn \; d\bx = \int_{\partial \Omega^\bu_N} \bv_{CG} \cdot \bn \; d\bx,
$$
$$
\int_{\partial \Omega^\bu_N} R\bv_{DG} \cdot \bn \; d\bx = \int_{\partial \Omega^\bu_N} \bv_{DG} \cdot \bn \; d\bx.
$$

\subsection{Mass Conservation}
In this section, we check the conservation law preserved by our proposed scheme. By the definition of $c_R(\cdot,\cdot)$, we have
\begin{eqnarray*}
   \sum_{e\in\mathcal{E}_{h,\bu}^{\partial,D}}(q_h,\bu_D\cdot\bn)_e &=& c_R(\bu_h,q_h) 
   = -\sum_{T\in\mathcal{T}_h}(\nabla\cdot\bu_h,q_h)_T + \sum_{e\in\mathcal{E}_h^I\cup\mathcal{E}_{h,\bu}^{\partial,D}}(\{q_h\},\ljump\bu_h\rjump\cdot\bn)_e \\
   &=& \sum_{T\in\mathcal{T}_h}-( \bu_h\cdot\bn,q_h)_{\partial T} + \sum_{e\in\mathcal{E}_h^I\cup\mathcal{E}_{h,\bu}^{\partial,D}}(\{q_h\},\ljump\bu_h\rjump\cdot\bn)_e\\
   &=& \sum_{e\in\mathcal{E}_h^I\cup\mathcal{E}_{h,\bu}^{\partial,N}} - (\ljump q_h\rjump,\{\bu_h\}\cdot\bn)_e.
\end{eqnarray*}
Reorganizing the above terms and use the velocity reconstruction in \eqref{eq:vel-reconstruction}, it implies
\begin{eqnarray*}
  0 = \sum_{e\in\mathcal{E}_h^I\cup\mathcal{E}_{h}^o} - (\ljump q_h\rjump,\{\bu_h\}\cdot\bn)_e &=& \sum_{T\in\mathcal{T}_h}- (q_h,R\bu_h\cdot\bn)_{\partial T} \\
   &=& \sum_{T\in\mathcal{T}_h}- (\nabla\cdot R\bu_h,q_h)_T.
\end{eqnarray*}
Taking $q_h = \nabla\cdot R\bu_h\in P_0(T)$ in the above equation, we can prove $\nabla\cdot R\bu_h\vert_T = 0$ and $\ljump R\bu_h\rjump_e\cdot\bn_e = 0$ for $T\in\mathcal{T}_h,e\in\mathcal{E}_h^I$ and derive the following theorem.

\begin{theorem}
    For the numerical solution $(\bu_n,p_h)$ obtained from Eqn. \eqref{eq:pr-ac-scheme}, we have the following mass conservation 
    $$
    \nabla\cdot R\bu_h = 0.
    $$
\end{theorem}

\section{Numerical Experiments}\label{sec13}
In this section, we present a series of numerical experiments designed to
validate and illustrate the capabilities of the proposed algorithm. We begin
with the classical natural convection benchmark in a square cavity in
Example~\ref{sub:ex1}, and then verify the convergence of the algorithm for
large Reynolds numbers in Example~\ref{sub:ex2}. Finally, we investigate heat
transfer in a porous medium under varying input data in
Example~\ref{sub:ex3}. All computations are performed using the deal.II finite
element library~\cite{deal9402022}.

\subsection{Example 1. Natural convection in a square cavity}
\label{sub:ex1}

In this example, we compare the benchmark solution for natural convection in a
square cavity obtained with the proposed EG
method against reference results from the
literature~\cite{de1983natural, kuznik2007double,
choi2011comparative}. 
In the computational domain
$\Omega = [0,1]^2$, following boundary conditions are given:
\begin{align*}
\bu_D &= \bm{0} && \text{on } \partial \Omega,  \\
\theta_D &= 1 && \text{on } \partial \Omega \cap \{x = 0\}, \\
\theta_D &= 0 && \text{on } \partial \Omega \cap \{x = 1\}, \\
q_N     &= 0 && \text{on } \partial \Omega \cap \bigl( \{y = 0\}  \cup  \{y = 1\} \bigr).
\end{align*}
The initial conditions are set to $\bu^0=\bm{0}$ and $\theta^0=0$. We consider four different Rayleigh numbers, defined by $\Ra := \Ri\,\Rey^2\,\Prr$,
with
$$
  \Ra \in \{10^3, 10^4, 10^5, 10^6\},
$$
by setting $\Prr=0.71$ and $\Rey=1.408$, and varying $\Ri$ accordingly. It is
well known that larger $\Ra$ values lead to more complex flow and make the system more challenging to solve
numerically~\cite{pollock2021acceleration}. In particular, classical Picard
iterations exhibit poor convergence properties at large $\Ra$. This example
highlights the performance of the AA-Picard scheme, which provides consistent
convergence behavior across a wide range of Rayleigh numbers. For the numerical
experiments, we use the pressure-robust EG algorithm and set the discretization parameters to
$\delta t=0.01$, $h=1/128$, and $t_f=1$.

\begin{figure}[!h]
     \centering
     \begin{subfigure}[t]{0.24\textwidth}
         \centering
         \includegraphics[trim={2.2cm 2.2cm 1.5cm 2cm},clip,width=\textwidth]{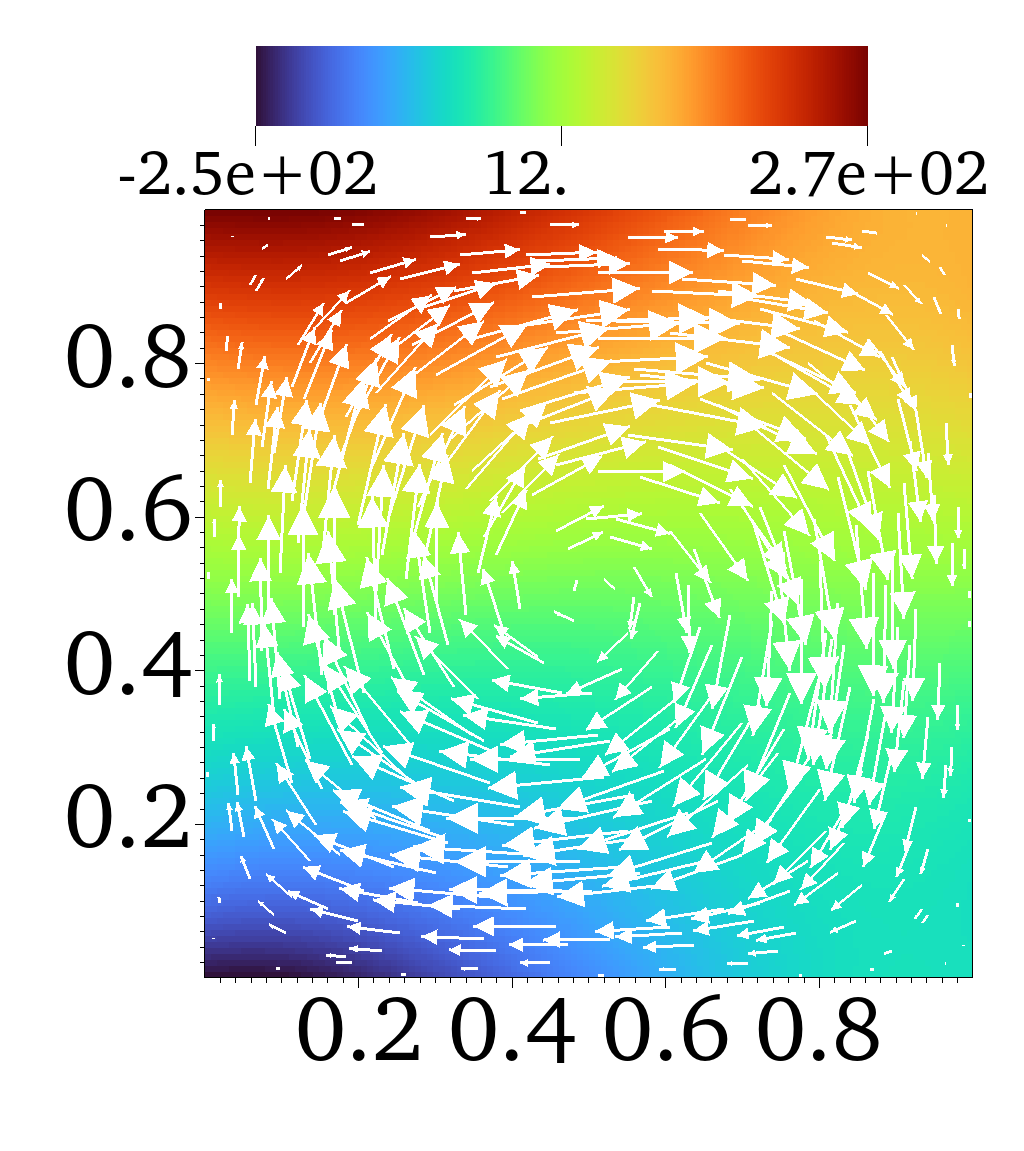}
         \subcaption{$\text{Ra} = 10^3$}
     \end{subfigure}
     \begin{subfigure}[t]{0.24\textwidth}
         \centering
         \includegraphics[trim={2.2cm 2.2cm 1.5cm 2cm},clip,width=\textwidth]{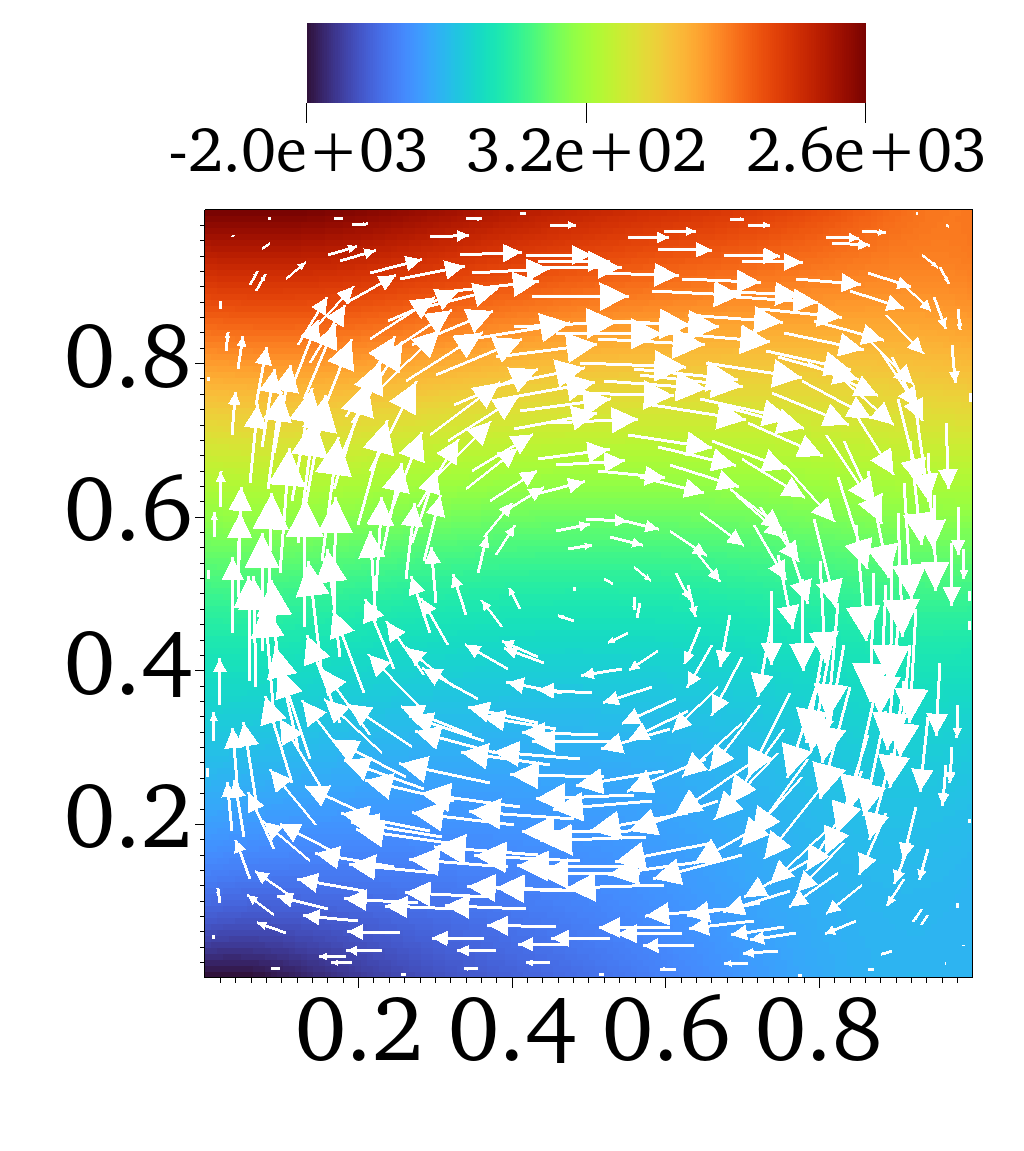}
         \subcaption{$\text{Ra} = 10^4$}
     \end{subfigure}
     \begin{subfigure}[t]{0.24\textwidth}
         \centering
         \includegraphics[trim={2.2cm 2.2cm 1.5cm 2cm},clip,width=\textwidth]{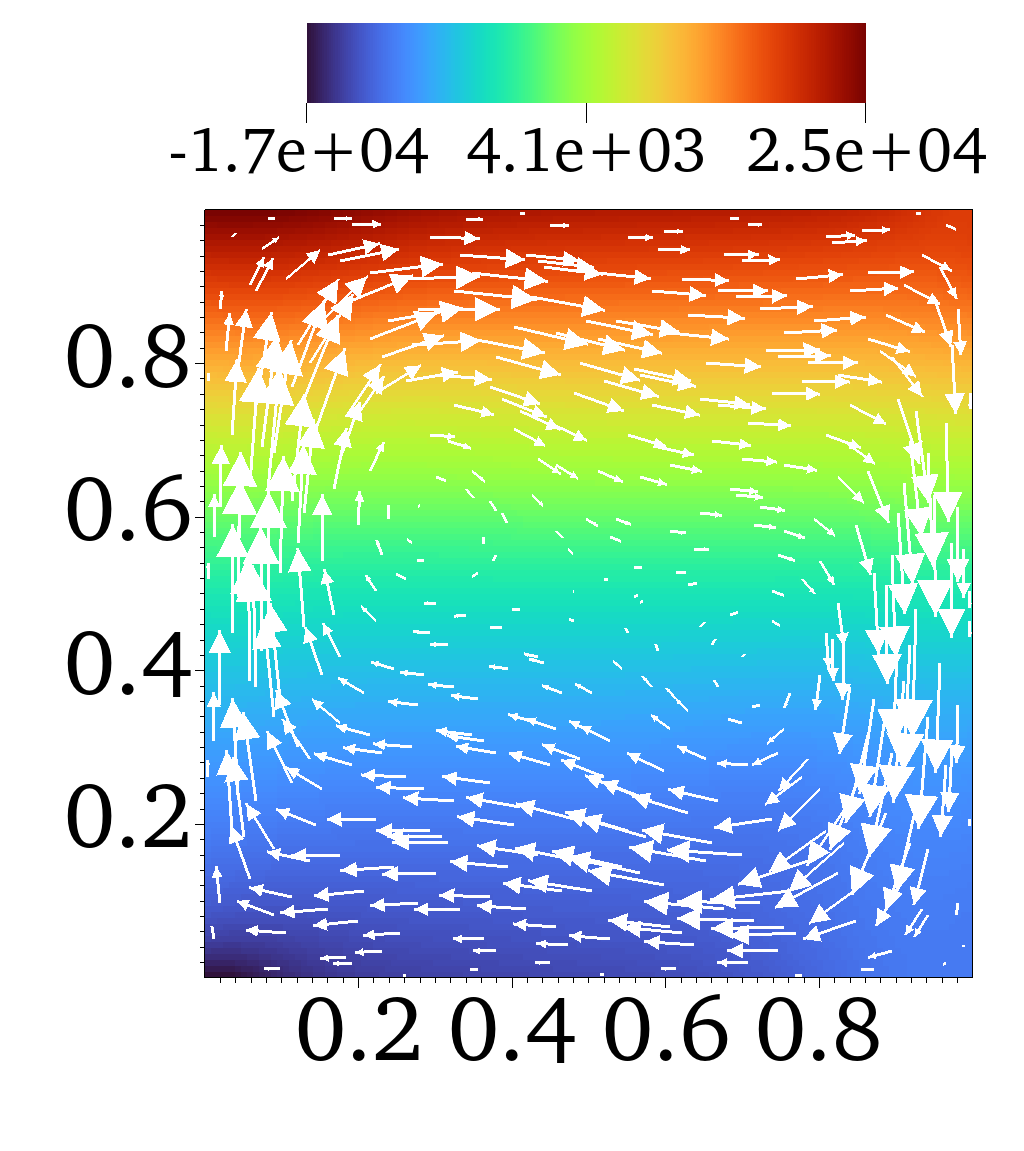}
         \subcaption{$\text{Ra} = 10^5$}
     \end{subfigure}
     \begin{subfigure}[t]{0.24\textwidth}
         \centering
         \includegraphics[trim={2.2cm 2.2cm 1.5cm 2cm},clip,width=\textwidth]{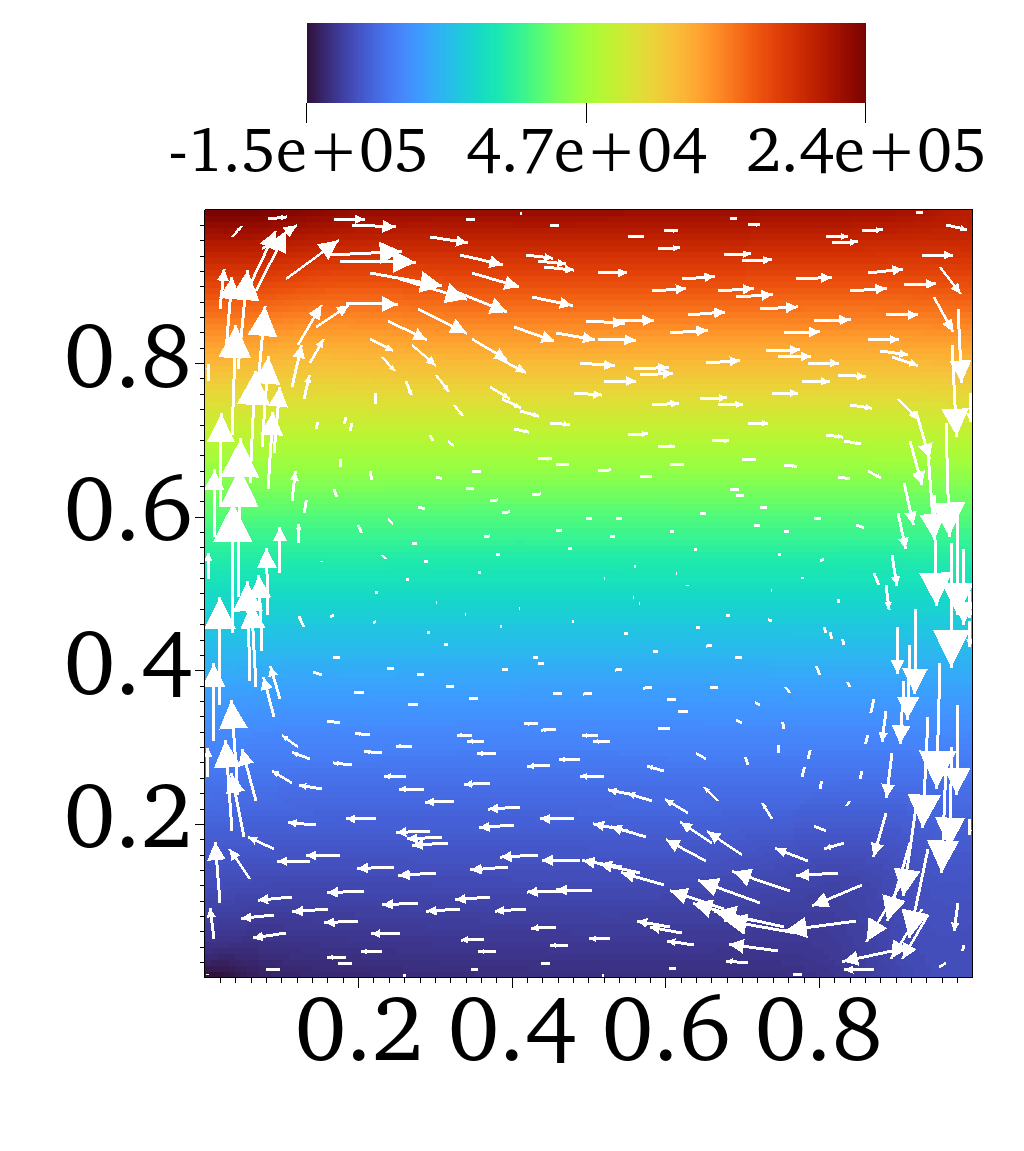}
         \subcaption{$\text{Ra} = 10^6$}
     \end{subfigure}
        \caption{Pressure distribution and velocity field at $t_f = 1$ for various Ra numbers ($\text{Ra}$).}
        \label{fig:streamlines}
\end{figure} 
First, Figure~\ref{fig:streamlines} shows the pressure and velocity field for the four
Rayleigh numbers. For all cases, we observe a circulation of fluid within the
cavity: warm fluid near the hot surface (left boundary at $x=0$) rises, and
cold fluid near the cold surface (right boundary at $x=1$) sinks, creating a
clockwise convective circulation. As $\Ra$ increases, the thermal buoyancy on the left
side generates a stronger upward motion, and the circulation becomes more
vigorous and increasingly confined to thin boundary layers near the vertical
walls.

\begin{figure}[!h]
     \centering
     \begin{subfigure}[t]{0.24\textwidth}
         \centering
         \includegraphics[trim={2.2cm 2.2cm 1.5cm 1cm},clip,width=\textwidth]{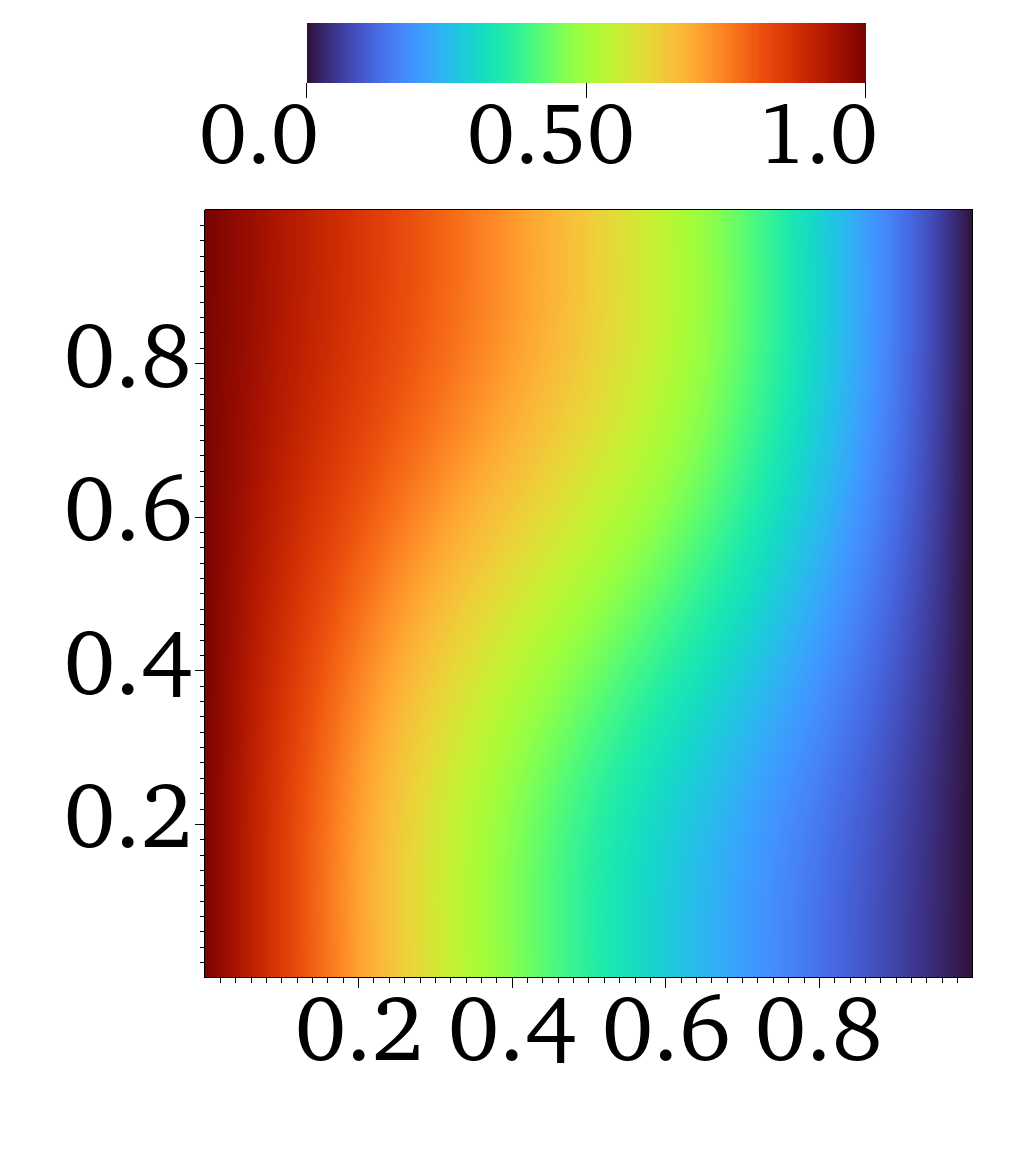}
         \subcaption{$\text{Ra} = 10^3$}
     \end{subfigure}
     \begin{subfigure}[t]{0.24\textwidth}
         \centering
         \includegraphics[trim={2.2cm 2.2cm 1.5cm 1cm},clip,width=\textwidth]{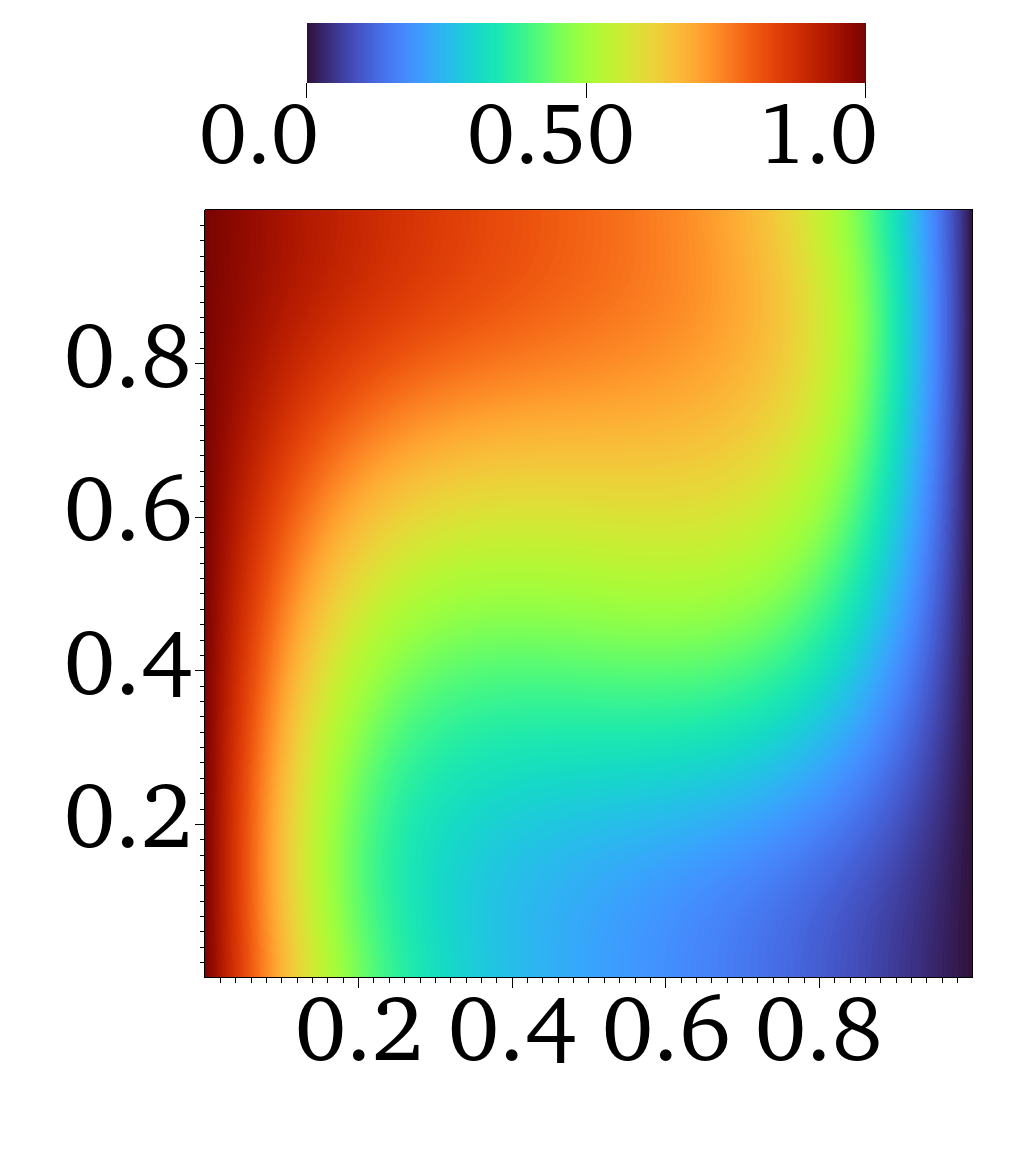}
         \subcaption{$\text{Ra} = 10^4$}
     \end{subfigure}
     \begin{subfigure}[t]{0.24\textwidth}
         \centering
         \includegraphics[trim={2.2cm 2.2cm 1.5cm 1cm},clip,width=\textwidth]{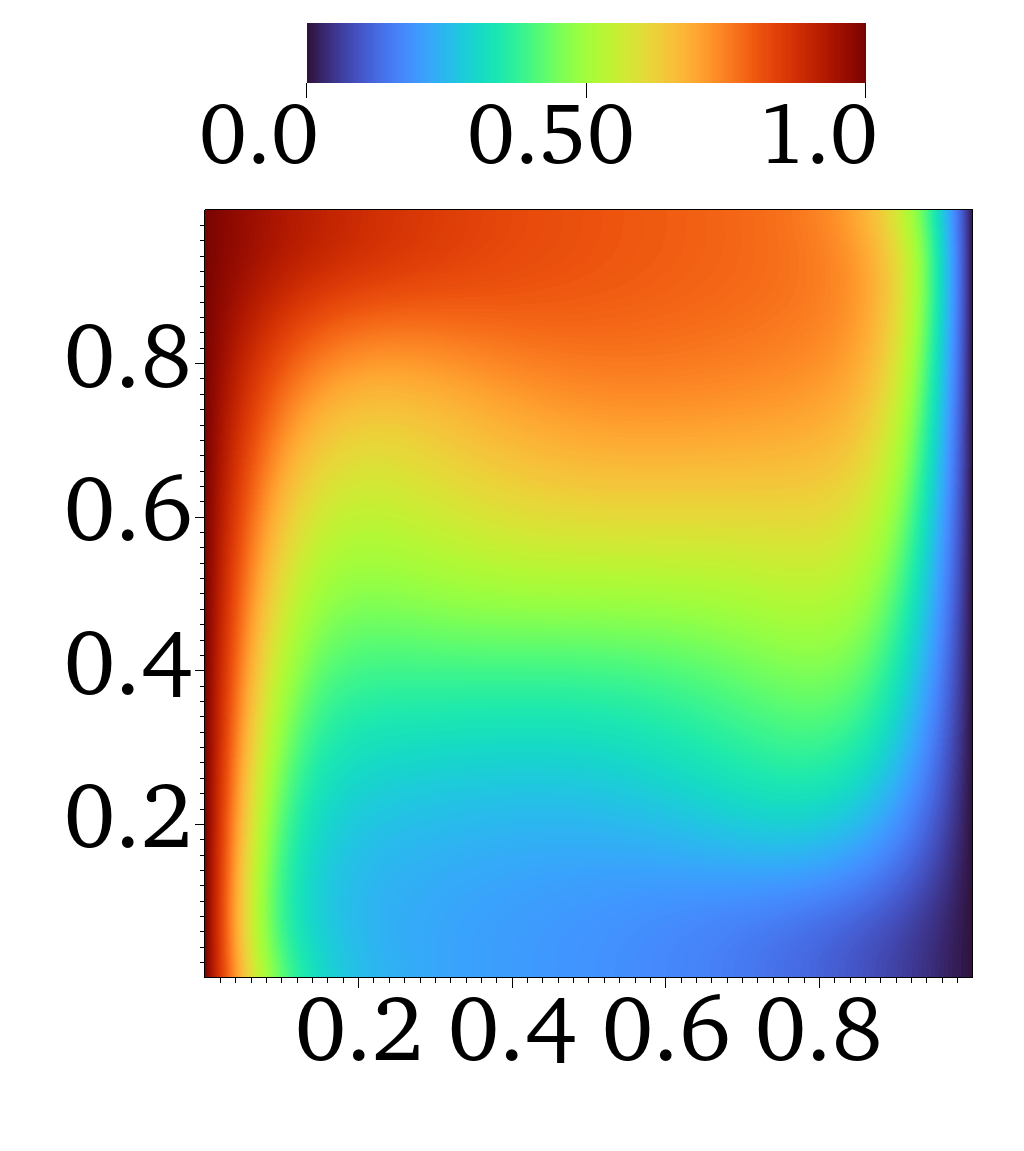}
         \subcaption{$\text{Ra} = 10^5$}
     \end{subfigure}
     \begin{subfigure}[t]{0.24\textwidth}
         \centering
         \includegraphics[trim={2.2cm 2.2cm 1.5cm 1cm},clip,width=\textwidth]{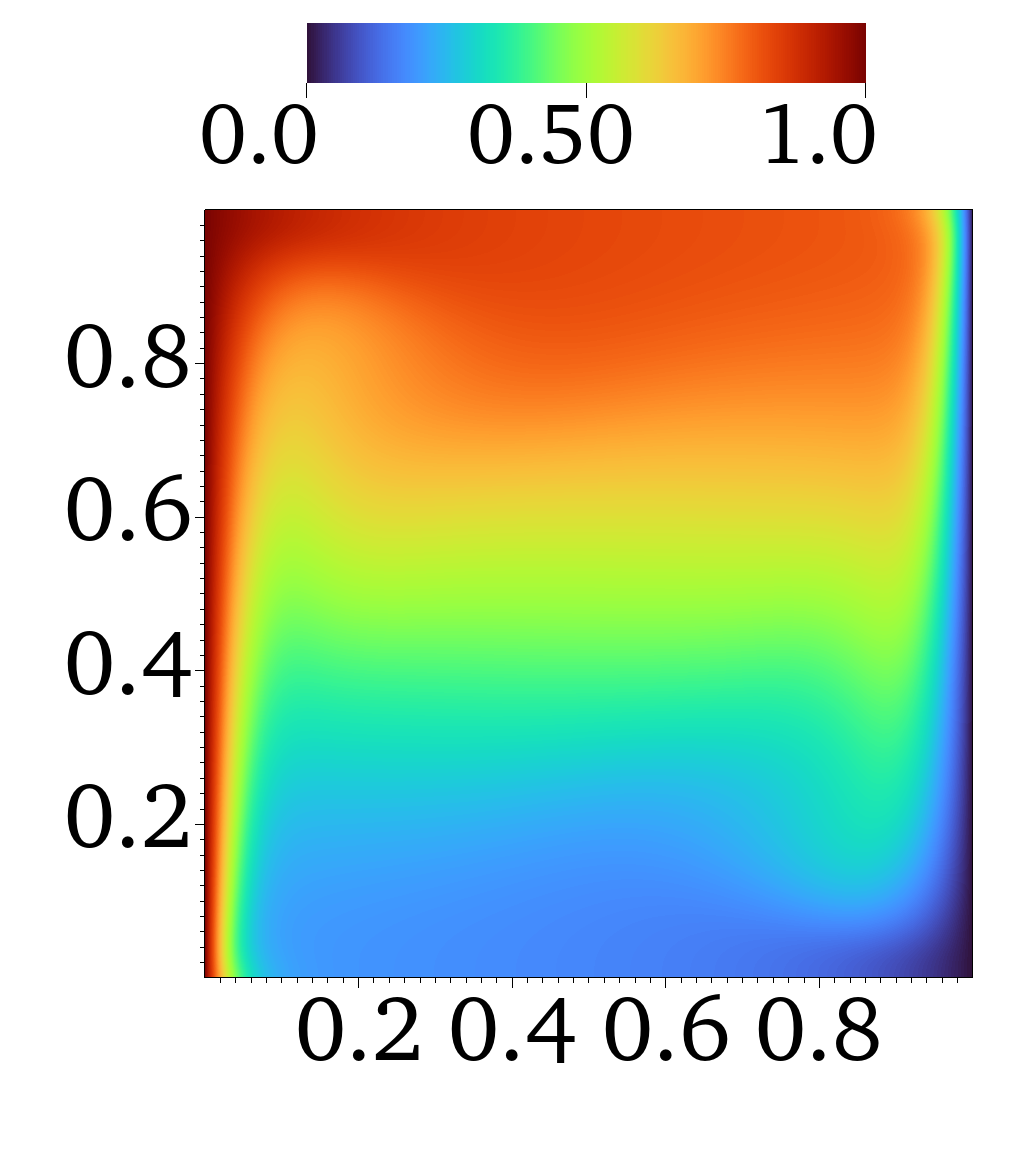}
         \subcaption{$\text{Ra} = 10^6$}
     \end{subfigure}
        \caption{Temperature distribution at $t_f = 1$ for various Rayleigh number.}
        \label{fig:temp_contour}
\end{figure}

Similarly, Figure~\ref{fig:temp_contour} shows temperature contours at $t_f = 1$. Due to the fluid circulation, the isothermal lines are no longer vertical but instead wrap around and follow the clockwise flow pattern. As $\Ra$ increases, thermal boundary layers near the hot and cold walls sharpen, in agreement with classical cavity benchmarks.

\begin{figure}[!h]
     \centering
     \begin{subfigure}[b]{0.35\textwidth}
         \centering
         \includegraphics[width=\textwidth]{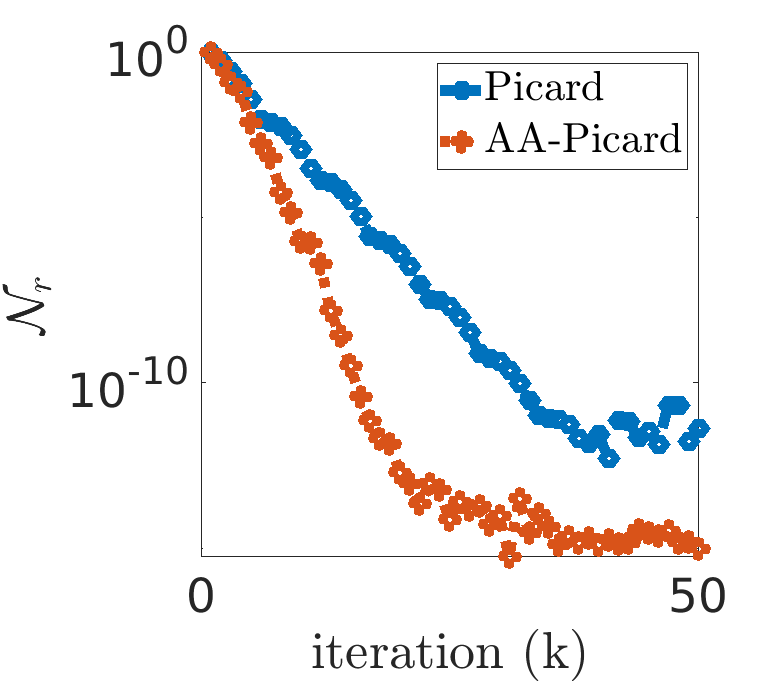}
         \subcaption{$\text{Ra} = 10^5$}
         \label{fig:benchmark_comparison_a}
     \end{subfigure}
     \begin{subfigure}[b]{0.35\textwidth}
         \centering
         \includegraphics[width=\textwidth]{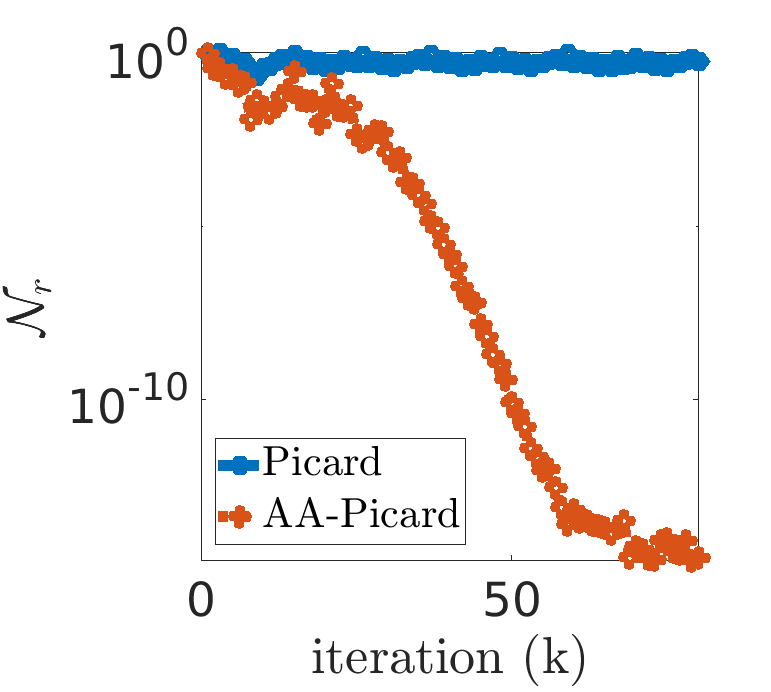}
         \subcaption{$\text{Ra} = 10^6$}
         \label{fig:benchmark_comparison_b}
     \end{subfigure}
    \caption{Comparison of the convergence of Picard and AA-Picard (with $m = 10$) for different Rayleigh numbers.}
    \label{fig:benchmark_comparison}
\end{figure}

In addition, we compare the performance of the nonlinear iteration algorithms
discussed above: Picard and AA--Picard. In this example, we observe that the
AA--Picard method offers improved convergence properties across a wide range of
Rayleigh numbers. For AA--Picard, we set the algorithmic depth parameter to
$m = 10$ and the relaxation parameter to $\beta_k = 1$, and test both algorithms
at the first time step ($n=1$). While both Picard and AA--Picard perform well
at low Rayleigh numbers, AA--Picard provides faster convergence, as illustrated
in Figure~\ref{fig:benchmark_comparison_a}.
In this figure, the $x$-axis represents the iteration number $k$, and the
$y$-axis represents the relative difference between successive discrete
solution vectors in the discrete $L^2$ norm, denoted by $\mathcal{N}_r$:
$$
   \mathcal{N}_r := 
   \frac{\|\bx_{n,k} - \bx_{n,k-1}\|_{L^2(\Omega)}}{\|\bx_{n,k}\|_{L^2(\Omega)}},
$$
where $\bx_{n,k} = [\bu_{n,k}, p_{n,k}, \theta_{n,k}]^\top$.

As the Rayleigh number is increased to $10^6$, the AA--Picard method
significantly outperforms the Picard method, as illustrated in
Figure~\ref{fig:benchmark_comparison_b}. The Picard method fails to achieve
convergence at this higher Rayleigh number, whereas Anderson acceleration
remains effective and successfully converges even under these challenging
conditions. This example highlights the superior performance of Anderson
acceleration in the high-Rayleigh-number regime.

Finally, Table~\ref{tab:comparison} compares the quantities of interest with
those reported in previous benchmark
studies~\cite{de1983natural,kuznik2007double,
choi2011comparative}. Here, $U_{\max}$ denotes the maximum horizontal
velocity (in the $x$-direction) along the vertical line $x = 0.5$, and
$y_{\max}$ denotes the vertical position at which $U_{\max}$ is attained.
Similarly, $V_{\max}$ is the maximum vertical velocity (in the $y$-direction)
along the horizontal line $y = 0.5$, and $x_{\max}$ is the horizontal position
of $V_{\max}$. The average Nusselt number at $x = 0$, denoted by $N_{u_0}$, is
computed as $N_{u_0} := \int_0^1 -\frac{\partial \theta}{\partial x} \, dy$.
The table shows excellent agreement with the established benchmark data across
all Rayleigh numbers. This close consistency confirms the accuracy and
reliability of the proposed method in capturing natural convection in a square
cavity.

 \begin{table}[!h]
     \centering
     \begin{tabular}{c c c c c c }
         \hline
         & & de Vahl Davis~\cite{de1983natural} & Kuznik et al.~\cite{kuznik2007double} & Choi et al. \cite{choi2011comparative} & Present \\
         \hline
         \hline
         $Ra = 10^3$ & $U_{\max}$ & 3.639 & 3.636 & 3.647 & 3.650 \\ 
         & $y_{\max}$ & 0.831 & 0.809 & 0.811 & 0.813 \\
         & $V_{\max}$ & 3.679 & 3.686 & 3.695 & 3.698 \\
         & $x_{\max}$ & 0.178 & 0.174 & 0.180 & 0.180 \\
         & $N_{u_0}$ & 1.117 & 1.117 & 1.117 & 1.118 \\
         \hline
         $Ra = 10^4$ & $U_{\max}$ & 16.178 & 16.167 & 16.177 & 16.188 \\ 
         & $y_{\max}$ & 0.823 & 0.821 & 0.820 & 0.820 \\
         & $V_{\max}$ & 19.617 & 19.597 & 19.614 & 19.638 \\
         & $x_{\max}$ & 0.119 & 0.120 & 0.122 & 0.117 \\
         & $N_{u_0}$ & 2.238 & 2.246 & 2.244 & 2.245 \\
         \hline
         $Ra = 10^5$ & $U_{\max}$ & 34.730 & 34.962 & 34.762 & 34.764 \\ 
         & $y_{\max}$ & 0.855 & 0.854 & 0.846 & 0.852 \\
         & $V_{\max}$ & 68.590 & 68.578 & 68.623 & 68.633 \\
         & $x_{\max}$ & 0.066 & 0.067 & 0.066 & 0.063 \\
         & $N_{u_0}$ & 4.509 & 4.518 & 4.521 & 4.519 \\
         \hline
         $Ra = 10^6$ & $U_{\max}$ & 64.630 & 64.133 & 64.815 & 64.878 \\ 
         & $y_{\max}$ & 0.850 & 0.860 & 0.846 & 0.852 \\
         & $V_{\max}$ & 219.360 & 220.537 & 220.613 & 221.639 \\
         & $x_{\max}$ & 0.038 & 0.038 & 0.038 & 0.039 \\
         & $N_{u_0}$ & 8.817 & 8.792 & 8.829 & 8.805 \\
         \hline
     \end{tabular}
     \caption{Comparison of benchmark quantities of interest with reference solutions for natural convection in a square cavity.}
     \label{tab:comparison}
\end{table}

\subsection{Example 2. Convergence of a smooth solution with high Reynolds number}
\label{sub:ex2}
\begin{figure}[!h]
     \centering
     \begin{subfigure}[b]{0.2\textwidth}
         \centering 
         \includegraphics[trim={2.8cm 5cm 1cm 2cm},clip, width=\textwidth]{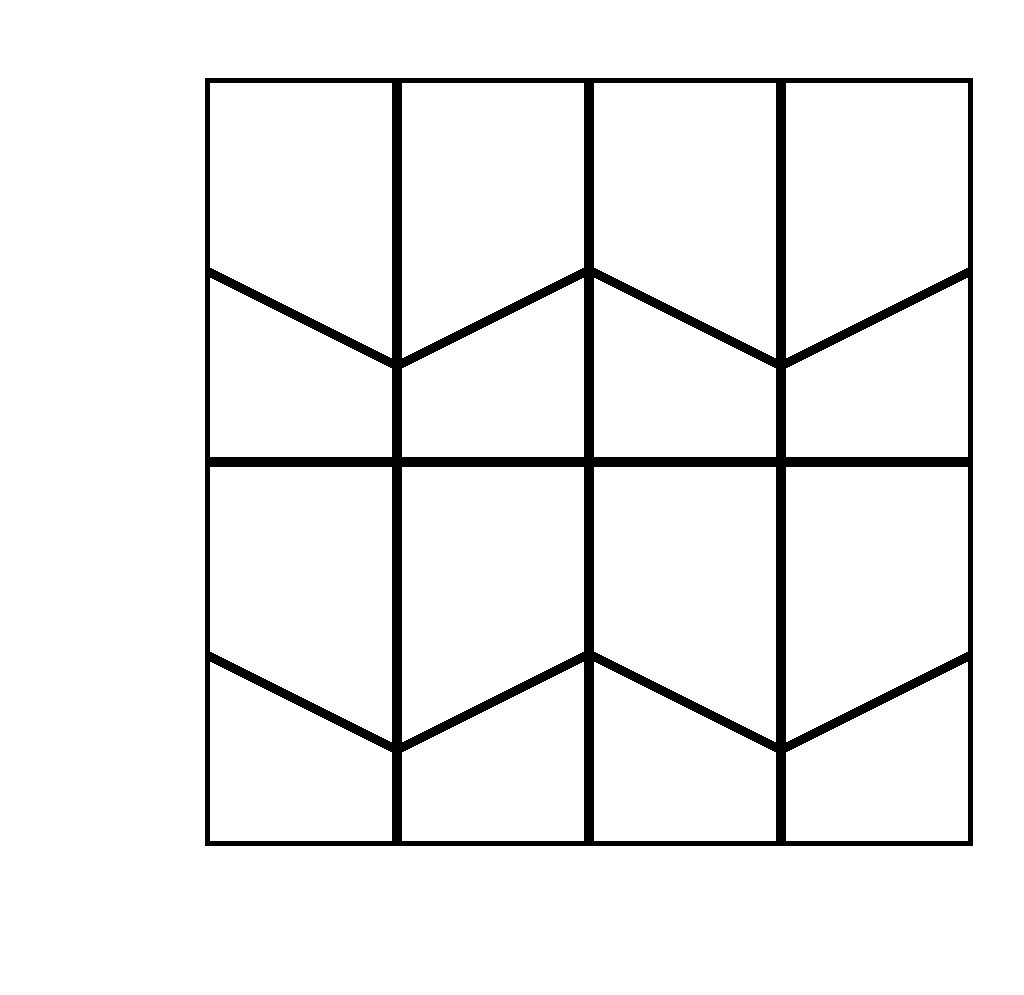}
         \caption{Level 1}
     \end{subfigure}
     \begin{subfigure}[b]{0.2\textwidth}
         \centering
         \includegraphics[trim={2.8cm 5cm 1cm 2cm},clip,width=\textwidth]{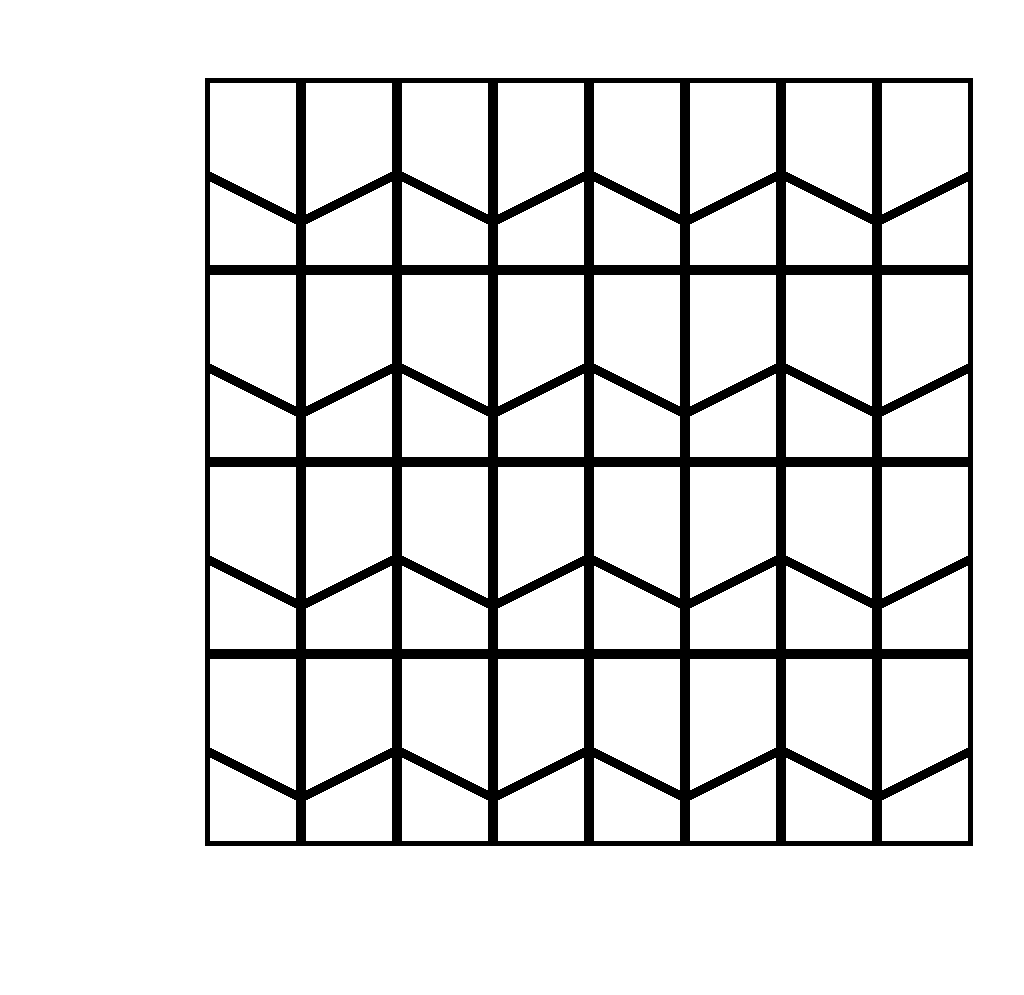}
         \caption{Level 2}
     \end{subfigure}
    \caption{Illustration of two levels of mesh distortion of $\Omega$ using trapezoidal elements.}
    \label{fig:trapezoids}
\end{figure}
In this example, we compare the convergence of the standard enriched Galerkin (ST--EG) method and the proposed pressure-robust enriched Galerkin (PR--EG) method, focusing on the incompressible Navier--Stokes equations. We employ distorted quadrilateral meshes as shown in Figure~\ref{fig:trapezoids} with two levels of mesh refinement using trapezoidal elements. We perform two experiments: one with homogeneous boundary conditions ($\bu = 0 \text{ on } \partial \Omega$) and the other with non-homogeneous boundary conditions ($\bu \neq 0 \text{ on } \partial \Omega$). For both experiments, we choose a time step size $\delta t = 0.1$ and final time $t_f = 1.0$, and we study the convergence behavior for various Reynolds numbers $\Rey$.
\subsubsection{Homogeneous boundary case}
For the homogeneous boundary case, the manufactured solution is given by
\begin{equation}
    \bu(x,y,t) =
    \begin{bmatrix}
        t x^2 (x-1)^2 y (y-1)(2y-1) \\
        -t x (x-1)(2x-1) y^2 (y-1)^2
    \end{bmatrix},
    \qquad
    p(x,y) = (x-1)(y-1),
\end{equation}
in $\Omega = [0,1]^2$.  The body force $\bbf$ in the momentum equation is chosen
so that $(\bu,p)$ satisfies the given system exactly.

We impose the following mixed boundary conditions:
\begin{alignat*}{2}
    \bu_D &= \bu &&\quad \text{on } \partial \Omega \cap \bigl(\{x = 0\} \cup \{y = 0\} \cup \{y = 1\}\bigr),\\
    \mathbf{t}_N &= \bigl(2\Rey^{-1} \bm{\varepsilon}(\bu) - p\mathbf{I}\bigr)\bn
    &&\quad \text{on } \partial \Omega \cap \{x = 1\},
\end{alignat*}
where $\bu_D$ and $\mathbf{t}_N$ are taken from the exact manufactured
solution.

\begin{figure}[!h]
     \centering
     \begin{subfigure}[b]{0.40\textwidth}
         \centering
         \includegraphics[width=\textwidth]{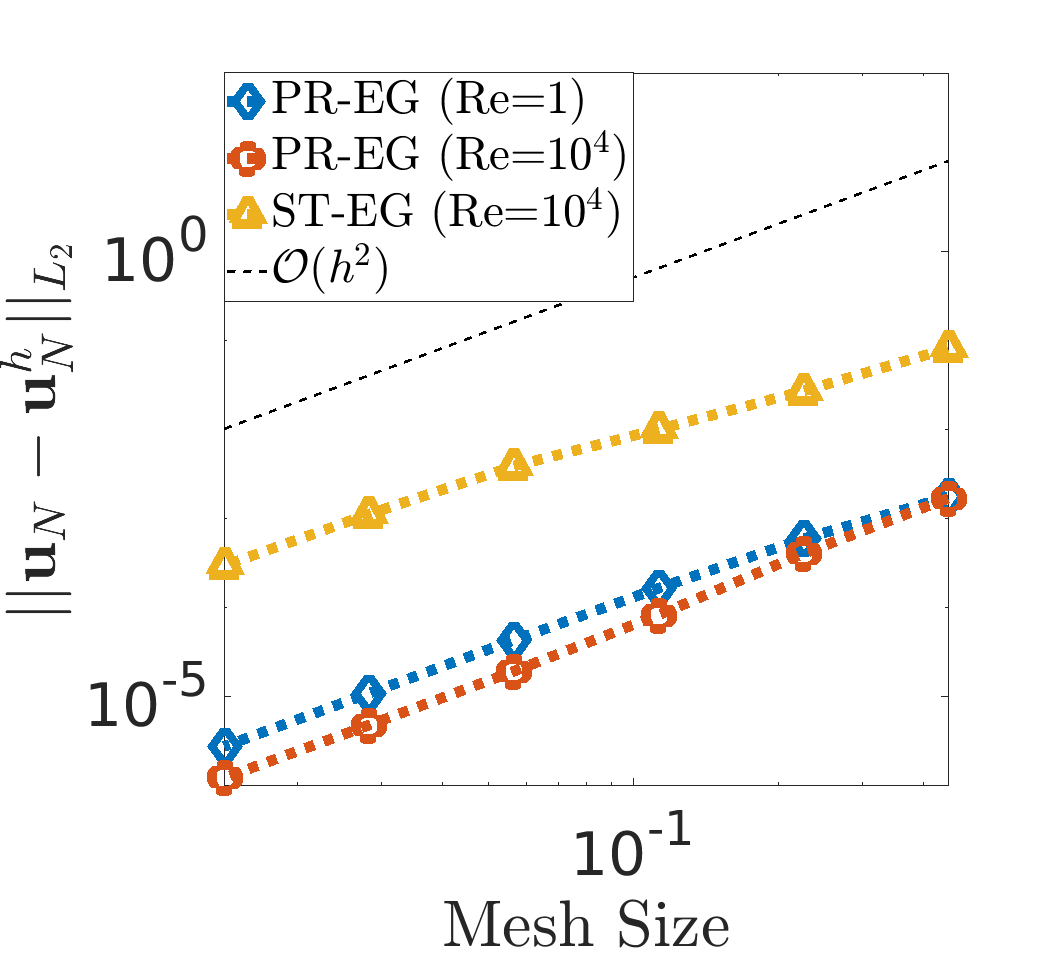}
         \caption{Velocity error}
         \label{fig:convergence_ST-EG_vs_PR-EG_velocity}
     \end{subfigure}
     \begin{subfigure}[b]{0.40\textwidth}
         \centering
         \includegraphics[width=\textwidth]{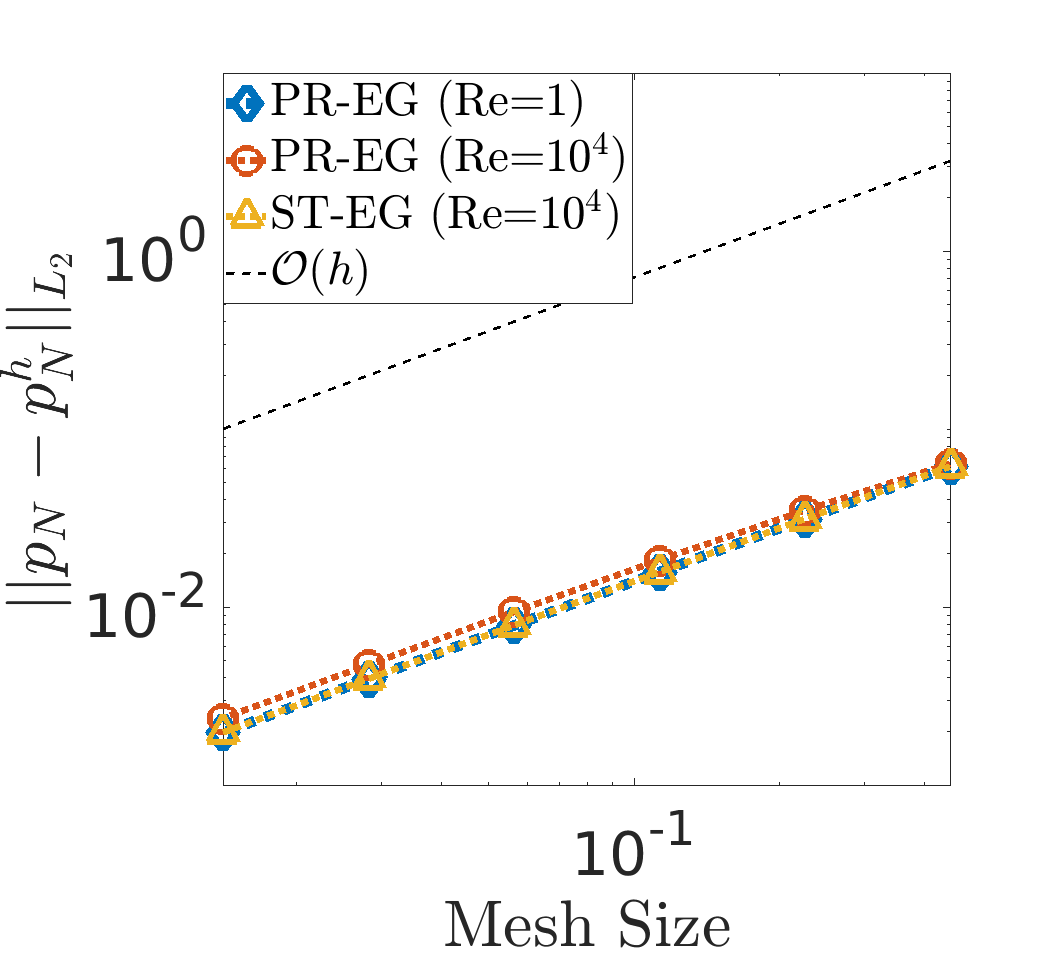}
         \caption{Pressure error}
         \label{fig:convergence_ST-EG_vs_PR-EG_pressure}
     \end{subfigure}
    \caption{Comparison of errors from the ST-EG and PR-EG methods.}
    \label{fig:convergence_ST-EG_vs_PR-EG}
\end{figure}

From Figure~\ref{fig:convergence_ST-EG_vs_PR-EG}, for a high Reynolds number ($\Rey = 10^4$), the velocity error for the ST--EG method exhibits an observed convergence rate that is below second order in the $L^2$ norm. Moreover, the ST--EG method yields velocity errors that are several orders of magnitude higher than those of the PR--EG method. With the pressure-robust enhancement, we observe from Figure~\ref{fig:convergence_ST-EG_vs_PR-EG_velocity} that the velocity errors for $\Rey = 1$ and $\Rey = 10^4$ are similar, indicating that the velocity error is independent of the Reynolds number. Furthermore, Figure~\ref{fig:convergence_ST-EG_vs_PR-EG_pressure} shows that the pressure error is also independent of the Reynolds number, and that both ST--EG and PR--EG yield similar pressure errors even for large $\Rey$. 

\begin{figure}[!h]
    \centering
    \begin{subfigure}{\textwidth}
        \centering
        \begin{subfigure}[b]{0.32\textwidth}
            \includegraphics[trim={4cm 0 1.5cm 0.4cm},clip,width=\textwidth]{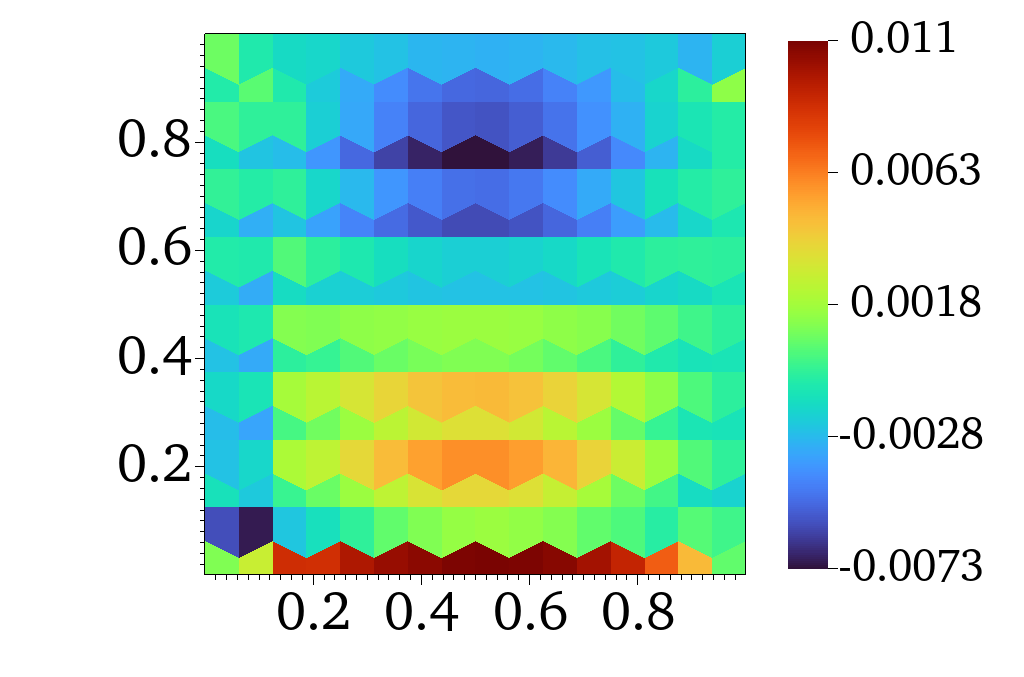}
        \end{subfigure}
        \begin{subfigure}[b]{0.32\textwidth}
             \includegraphics[trim={4cm 0 1.5cm 0.4cm},clip,width=\textwidth]{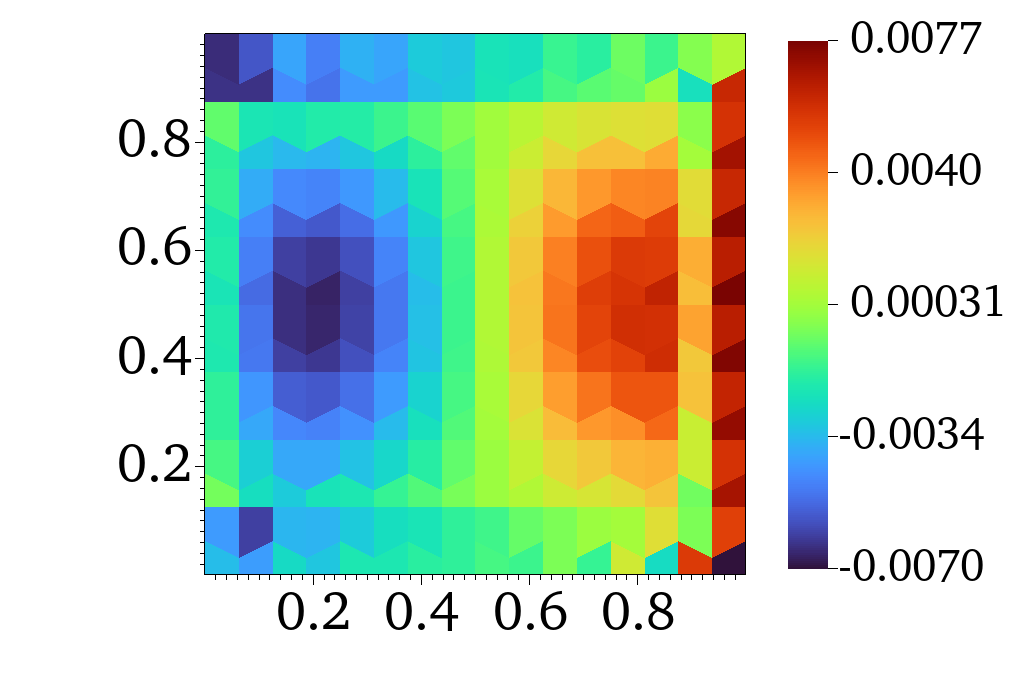}        
        \end{subfigure}
        \begin{subfigure}[b]{0.32\textwidth}
             \includegraphics[trim={4cm 0 1.5cm 0.4cm},clip,width=\textwidth]{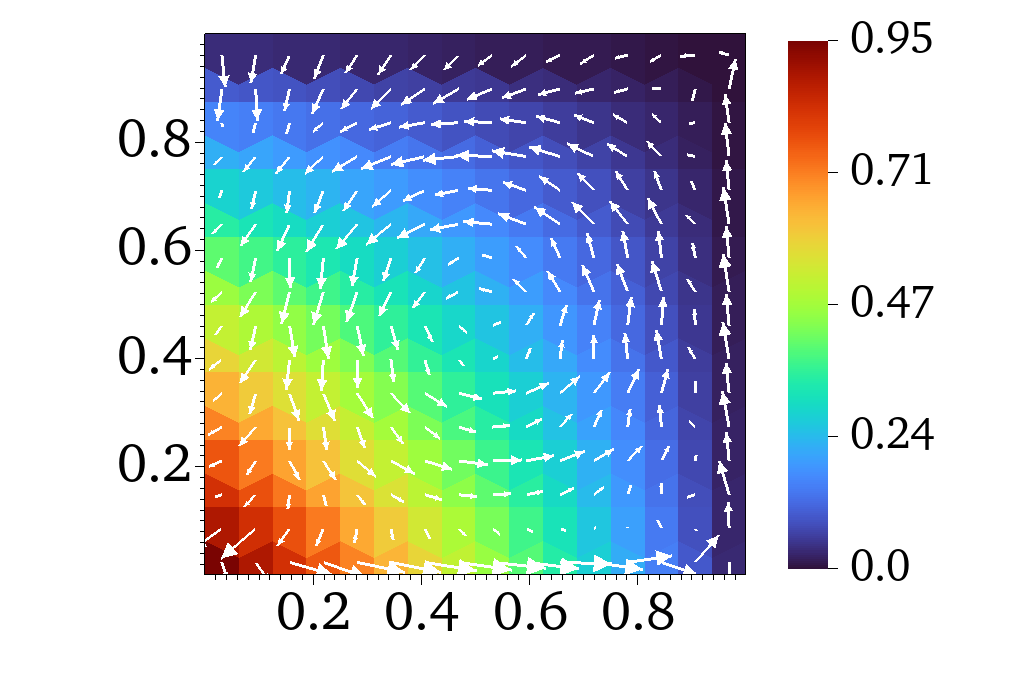}        
        \end{subfigure}
        \caption{ST-EG: x-component velocity, y-component velocity, and pressure, with the velocity vector fields, from left to right.}
    \end{subfigure}    
    \begin{subfigure}{\textwidth}
        \centering
        \begin{subfigure}[b]{0.32\textwidth}
            \includegraphics[trim={4cm 0 1.5cm 0.4cm},clip,width=\textwidth]{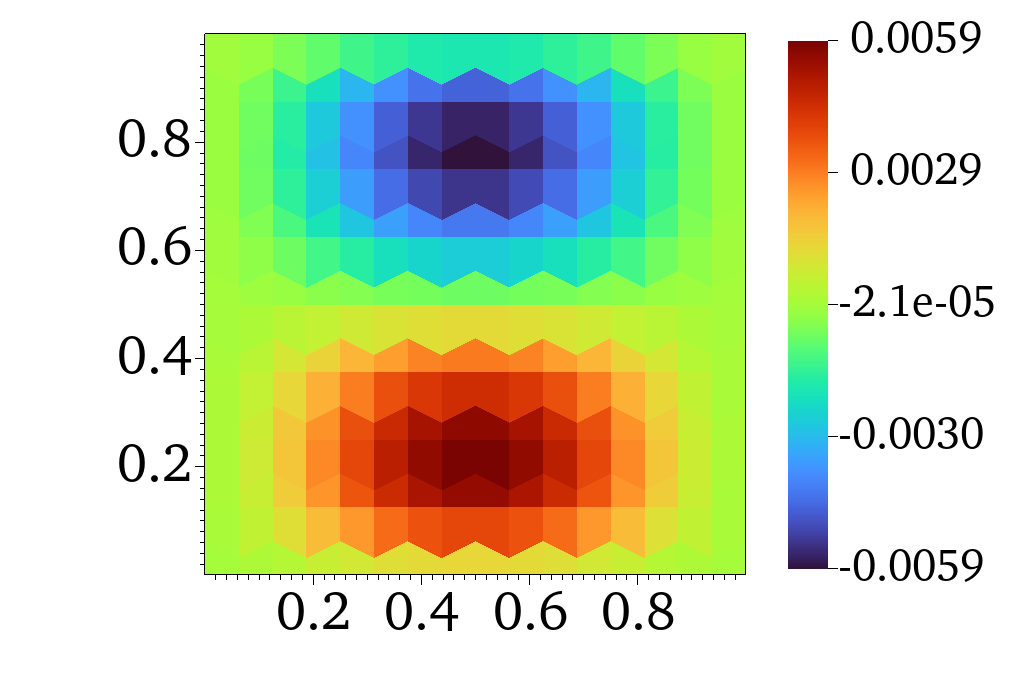}
        \end{subfigure}
        \begin{subfigure}[b]{0.32\textwidth}
             \includegraphics[trim={4cm 0 1.5cm 0.4cm},clip,width=\textwidth]{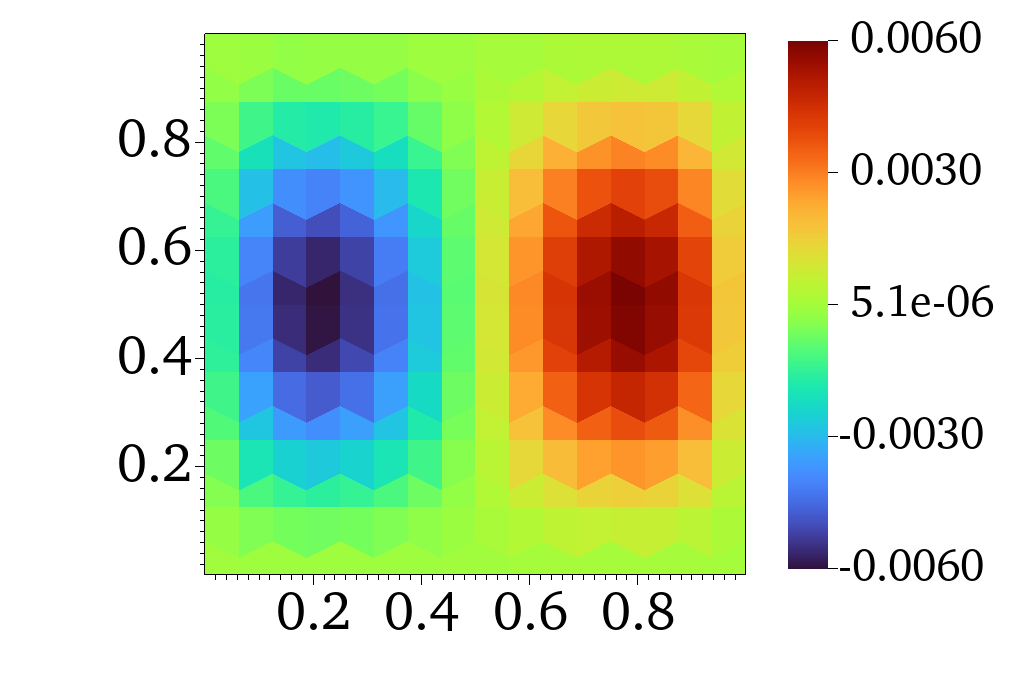}        
        \end{subfigure}
        \begin{subfigure}[b]{0.32\textwidth}
             \includegraphics[trim={4cm 0 1.5cm 0.4cm},clip,width=\textwidth]{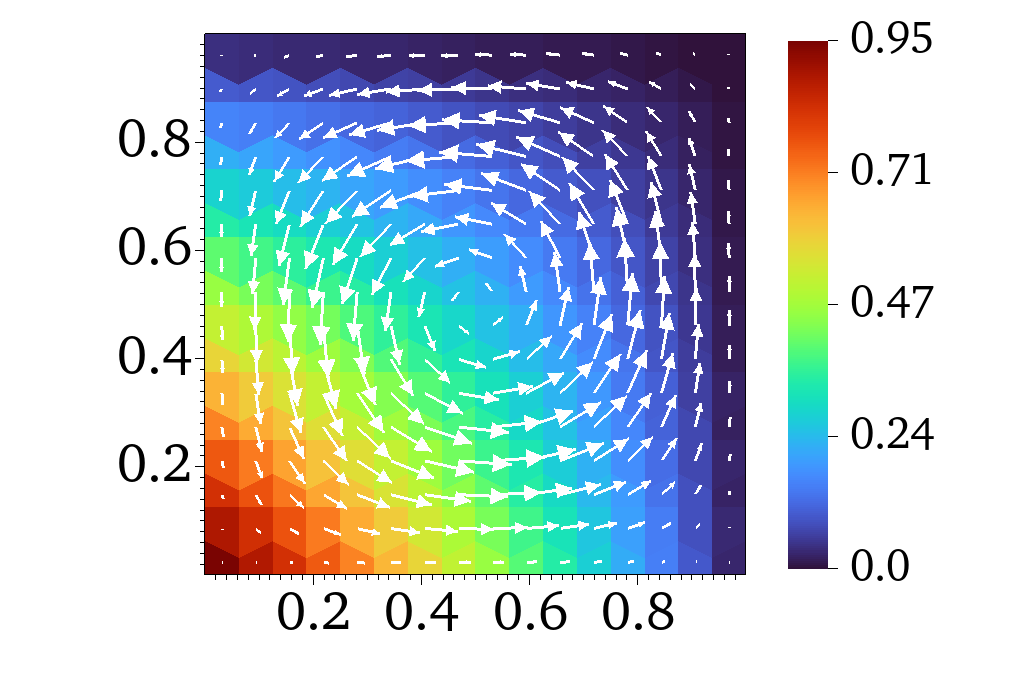}        
        \end{subfigure}
        \caption{PR-EG: x-component velocity, y-component velocity, and pressure, with the velocity vector fields, from left to right.}
    \end{subfigure}
    \caption{Comparison of solutions between ST-EG and PR-EG method for $\Rey = 10^4$.}
    \label{fig:compare_contour_PR_ST}
\end{figure}

Next, to visualize the difference between the two methods, we present the
numerical solutions obtained with $\Rey = 10^4$ and a mesh resolution of
$32 \times 32$ cells in Figure~\ref{fig:compare_contour_PR_ST}. As expected
from the pressure error results, the two methods produce nearly identical
pressure fields. For the velocity, however, the PR--EG method accurately
captures the flow pattern, whereas the ST--EG method fails to do so. Moreover, as we can observe from the vector plots in the left Figure 8(a), several unphysical vectors are appearing near the walls. This highlights the superior performance of the proposed PR--EG method over the standard ST--EG method.

\subsubsection{Non-homogeneous boundary case}
For the non-homogeneous boundary case, the manufactured solution is given by
\begin{equation}
    \bu(x,y,t) =
    \begin{bmatrix}
        0.1t\sin(x)\sin(y) \\
        0.1t\cos(x)\cos(y)
    \end{bmatrix},
    \qquad
    p(x,y) = \sin(\pi x)\cos(\pi y),
\end{equation}
in $\Omega = [0,1]^2$.  Similar to the previous example, the body force $\bbf$ in the momentum equation is chosen
so that $(\bu,p)$ satisfies the given system exactly.

We impose the following mixed boundary conditions:
\begin{alignat*}{2}
    \bu_D &= \bu &&\quad \text{on } \partial \Omega \cap \bigl(\{x = 0\} \cup \{y = 0\} \cup \{y = 1\}\bigr),\\
    \mathbf{t}_N &= \bigl(2\Rey^{-1} \bm{\varepsilon}(\bu) - p\mathbf{I}\bigr)\bn
    &&\quad \text{on } \partial \Omega \cap \{x = 1\},
\end{alignat*}
where $\bu_D$ and $\mathbf{t}_N$ are computed from the exact solution.
\begin{figure}[!h]
     \centering
     \begin{subfigure}[b]{0.4\textwidth}
         \centering
         \includegraphics[width=\textwidth]{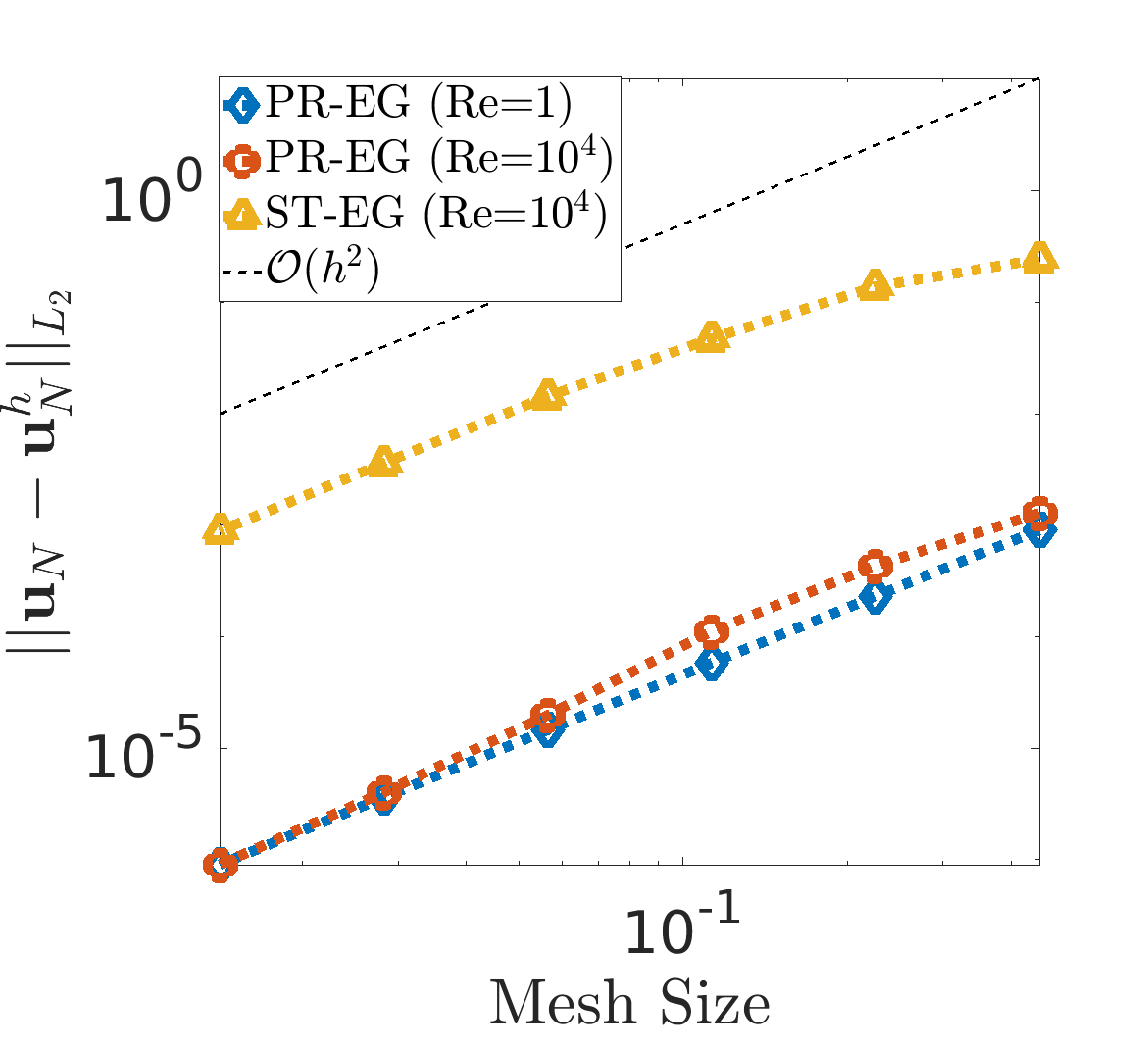}
         \caption{Velocity error}
         \label{fig:convergence_ST-EG_vs_PR-EG_velocity_non_homogeneous}
     \end{subfigure}
     \begin{subfigure}[b]{0.4\textwidth}
         \centering
         \includegraphics[width=\textwidth]{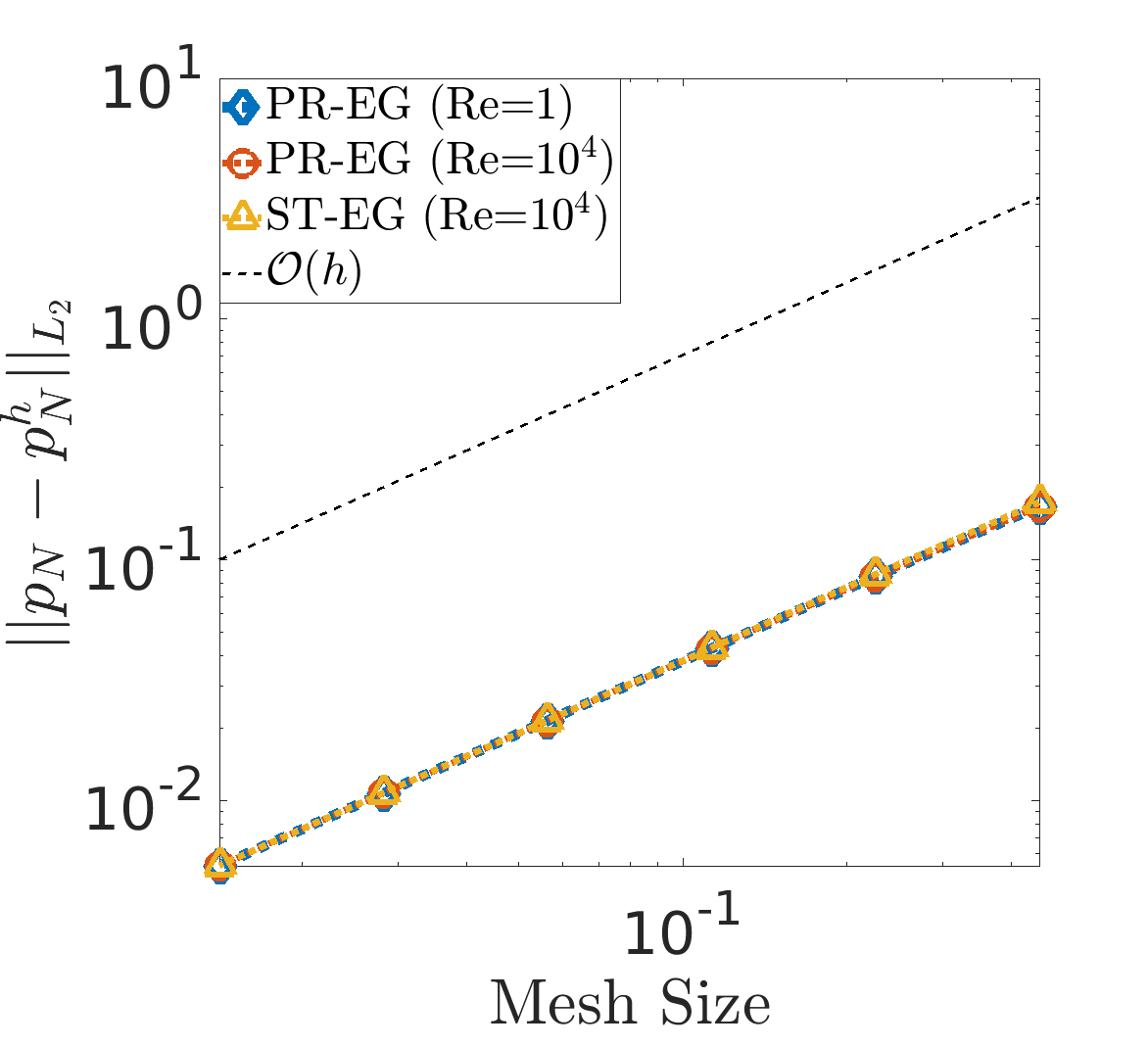}
         \caption{Pressure error}
         \label{fig:convergence_ST-EG_vs_PR-EG_pressure_non_homogeneous}
     \end{subfigure}
    \caption{Comparison of errors from the ST-EG and PR-EG methods.}
    \label{fig:convergence_ST-EG_vs_PR-EG_non_homogeneous}
\end{figure}

\begin{figure}
    \centering
    \begin{subfigure}{\textwidth}
        \centering
        \begin{subfigure}[b]{0.32\textwidth}
            \includegraphics[trim={4cm 0 2cm 0.4cm},clip,width=\textwidth]{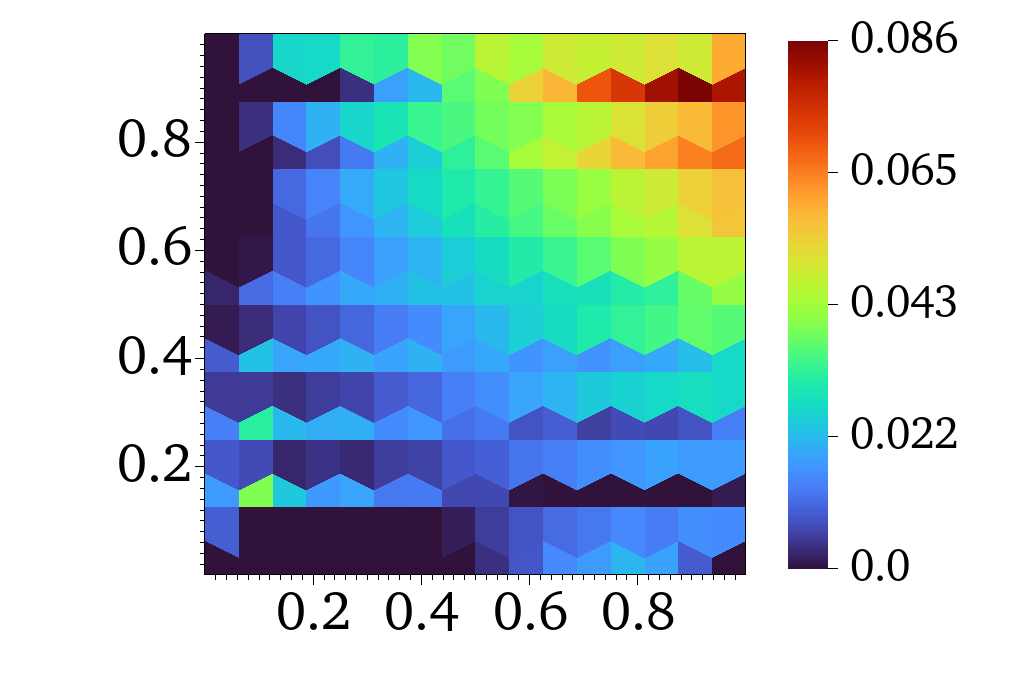}
        \end{subfigure}
        \begin{subfigure}[b]{0.32\textwidth}
             \includegraphics[trim={4cm 0 2cm 0.4cm},clip,width=\textwidth]{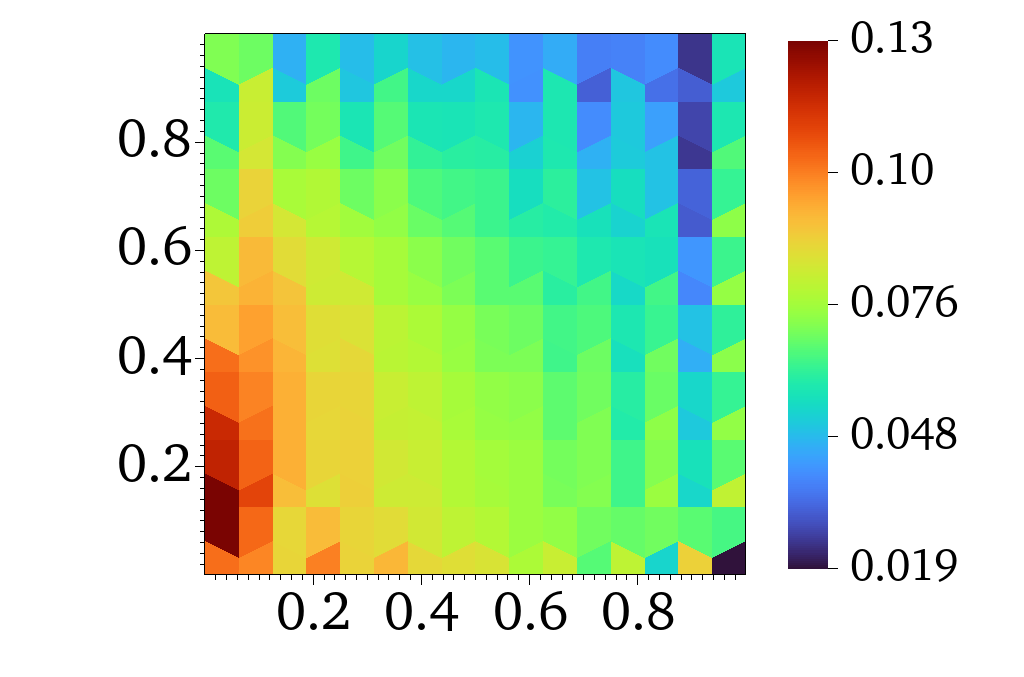}        
        \end{subfigure}
        \begin{subfigure}[b]{0.32\textwidth}
             \includegraphics[trim={4cm 0 2cm 0.4cm},clip,width=\textwidth]{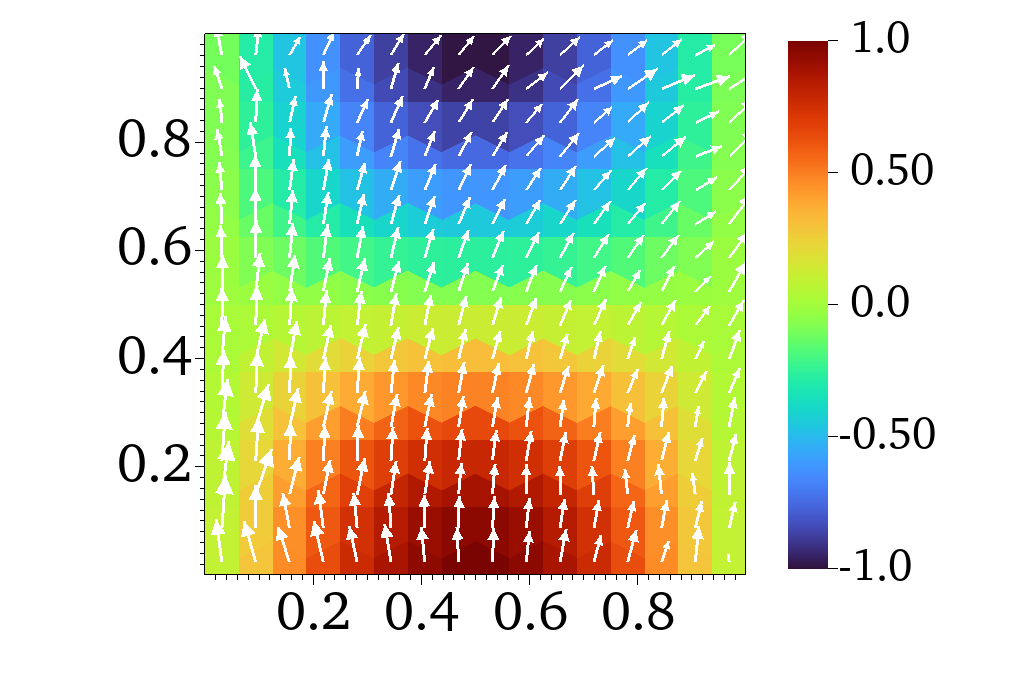}        
        \end{subfigure}
        \caption{ST-EG: x-component velocity, y-component velocity, and pressure, with the velocity vector fields, from left to right.}
    \end{subfigure}    
    \begin{subfigure}{\textwidth}
        \centering
        \begin{subfigure}[b]{0.32\textwidth}
            \includegraphics[trim={4cm 0 2cm 0.4cm},clip,width=\textwidth]{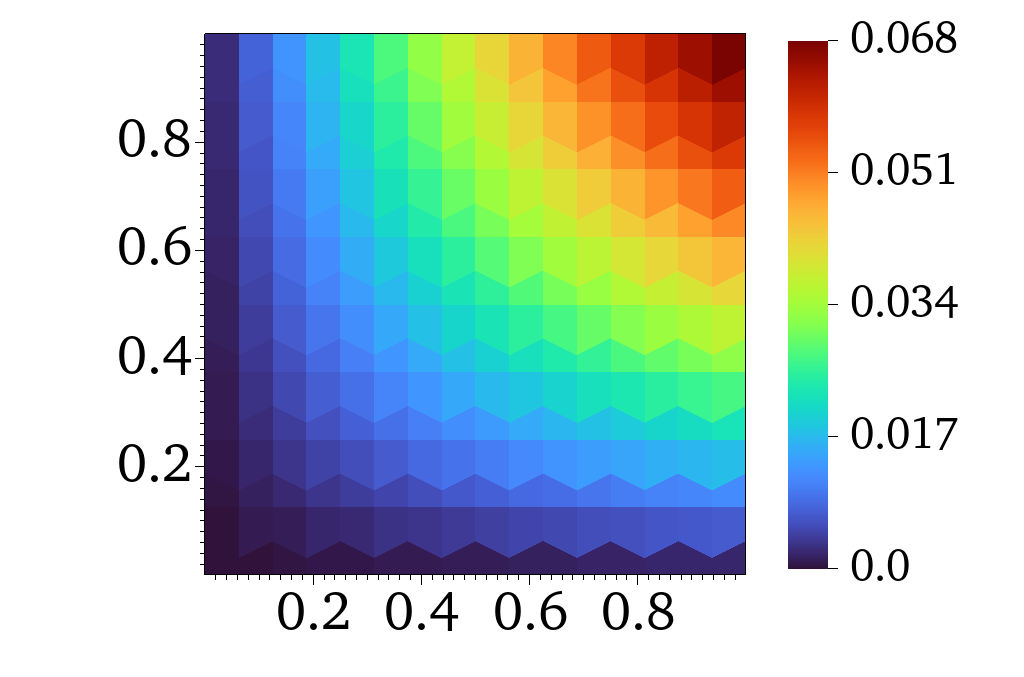}
        \end{subfigure}
        \begin{subfigure}[b]{0.32\textwidth}
             \includegraphics[trim={4cm 0 2cm 0.4cm},clip,width=\textwidth]{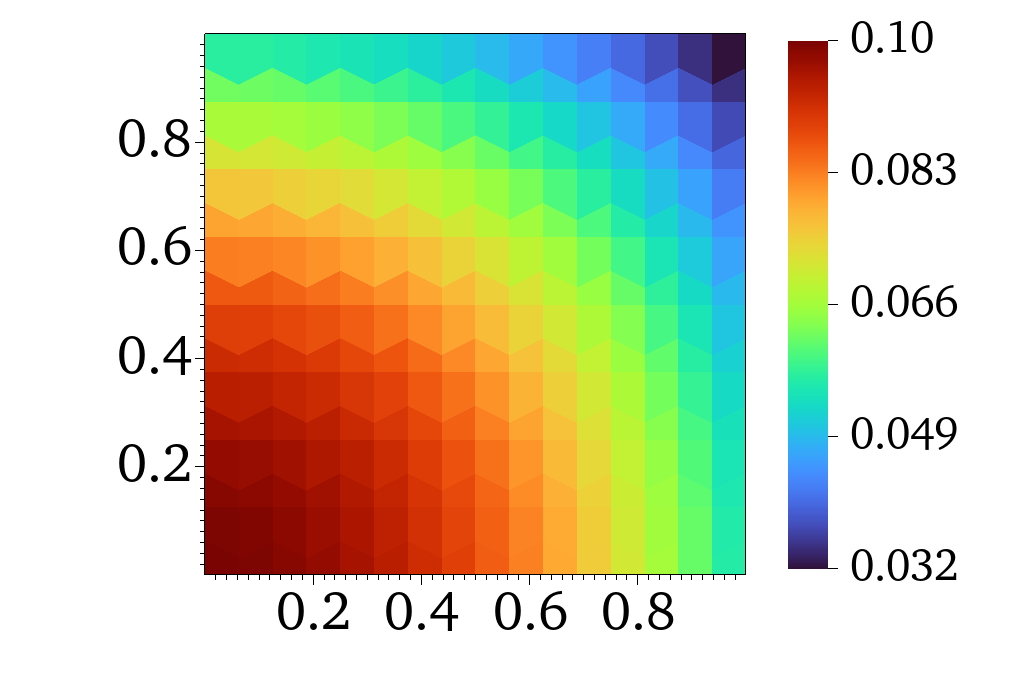} 
        \end{subfigure}
        \begin{subfigure}[b]{0.32\textwidth}
             \includegraphics[trim={4cm 0 2cm 0.4cm},clip,width=\textwidth]{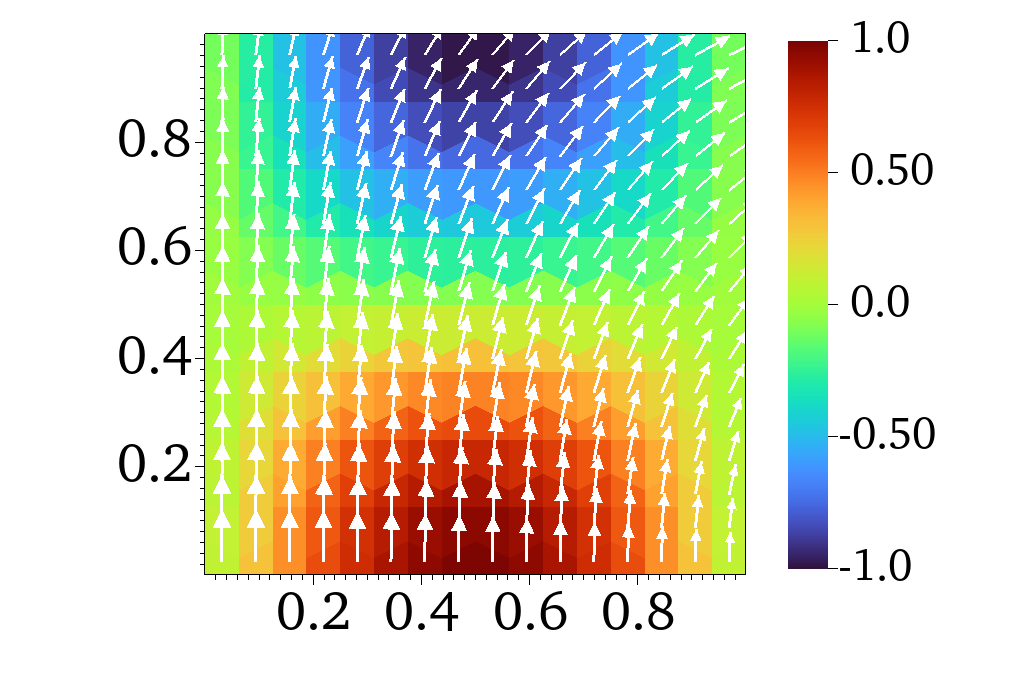}        
        \end{subfigure}
        \caption{PR-EG: x-component velocity, y-component velocity, and pressure, with the velocity vector fields, from left to right.}
    \end{subfigure}
    \caption{Comparison of numerical simulation for ST-EG and PR-EG method for $\Rey = 10^4$.}
    \label{fig:nH_compare_contour_PR_ST}
\end{figure}

As in the previous result, Figure~\ref{fig:convergence_ST-EG_vs_PR-EG_non_homogeneous} illustrates that the PR--EG method exhibits a convergence rate close to second order even at a large Reynolds number $(\Rey)$. Additionally, the velocity errors for the PR--EG method are several orders of magnitude smaller than those for the ST--EG method. With the pressure-robust enhancement, Figure~\ref{fig:convergence_ST-EG_vs_PR-EG_velocity_non_homogeneous} reveals that the velocity errors for both $\Rey = 1$ and $\Rey = 10^4$ are similar. Moreover, Figure~\ref{fig:convergence_ST-EG_vs_PR-EG_pressure} shows that the pressure error remains unaffected by the Reynolds number as well, with both ST--EG and PR--EG methods producing comparable pressure errors. Finally, Figure~\ref{fig:nH_compare_contour_PR_ST} shows the visual comparison of velocity and pressure solution for different Reynolds numbers. As previously observed, the two methods produce nearly identical pressure fields. However, for the velocity, the PR--EG method accurately captures the flow pattern, whereas the ST--EG method fails to do so. This confirms that the proposed PR–EG method maintains superior performance even in the presence of non-homogeneous boundary conditions.

\subsection{Example 3. Flow and heat transfer in a porous medium}
\label{sub:ex3}

In subsurface studies, modeling and simulation of heat transfer in pore-scale structures play a crucial role in various applications, including geothermal energy extraction. Analyzing heat transfer from the solid matrix to the flowing fluid under varying conditions enhances our understanding of the parameters
that influence heat extraction rates. In this example, we investigate
pore-scale heat transfer for different values of the dimensionless constants.

\begin{figure}[!h]
    \centering
    \includegraphics[width=0.45\linewidth]{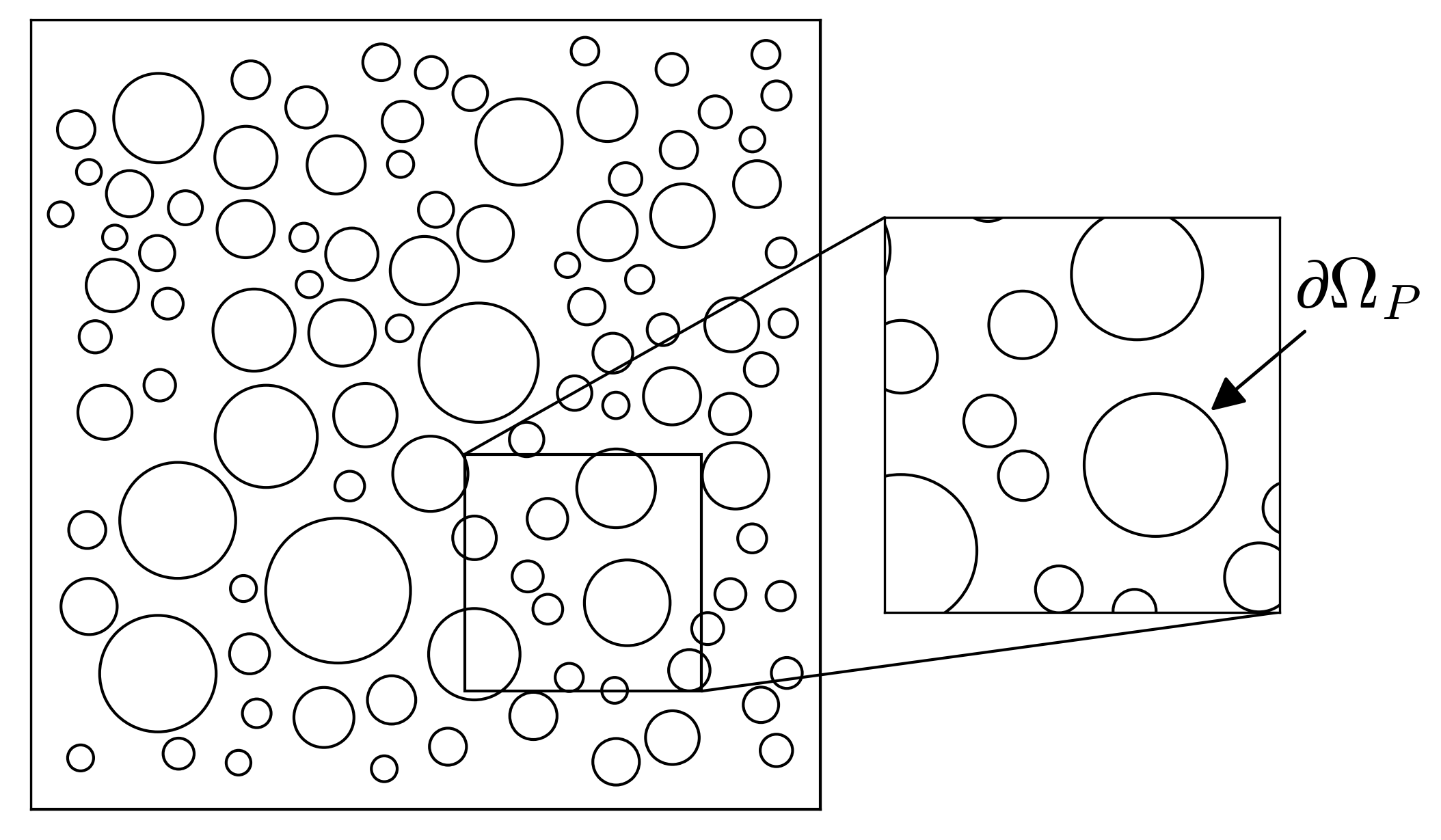}
    \caption{Geometry of the pore-scale domain $\Omega$. The circles represent
    the pore boundaries $\partial\Omega_P$, and the outer boundary is denoted
    by $\partial\Omega$.}
    \label{fig:pore_geometry}
\end{figure}

The computational domain $\Omega = [0,1]^2$ is shown in
Figure~\ref{fig:pore_geometry}, where the circles indicate the solid matrix. The
domain $\Omega$ has an outer boundary $\partial\Omega$ and an inner boundary
$\partial\Omega_P$, which represents the pore boundaries.

The boundary conditions are specified as follows:
\begin{alignat*}{2}
\bu_D &= \bigl(4y(1-y),\,0\bigr)^\top &&\text{ on } \partial\Omega \cap \{x = 0\}, \\
\bu_D &= \bm{0} &&\text{ on } \partial \Omega_P \cup \{y = 0\} \cup \{ y=1\}, \\
\mathbf{t}_N &= \bm{0} &&\text{ on } \partial\Omega \cap \{x = 1\}, \\
\theta_D &= 0 &&\text{ on } \partial \Omega  \cap \{x = 0\}, \\
q_N &= 1 &&\text{ on } \partial \Omega_P, \\
q_N &= 0 &&\text{ on } \partial \Omega \cap \bigl(\{ x = 1\} \cup \{y = 0\} \cup \{y=1\}\bigr).
\end{alignat*}
Thus, cold fluid enters the domain through the left boundary and exits through
the right boundary. As the fluid flows through the pore structure, it absorbs
heat from the solid matrix, where a constant heat flux is imposed. The initial conditions are set to $\bu^0=\bm{0}$ and $\theta^0=0$.

For all numerical experiments, we use a time step size $\delta t = 0.05$ and
compute the solution up to $t_f = 2$. The discrete system has 96,695 degrees of
freedom for the velocity (65,948 for the continuous part and 30,747 for the
discontinuous enrichment), 30,747 degrees of freedom for the pressure, and
32,974 degrees of freedom for the temperature. For AA--Picard, we fix the
algorithmic depth parameter and relaxation parameter as
$m = 10$ and $\beta_k = 1$. 

Moreover, we employ a first-order artificial diffusion stabilization for
the heat equation to prevent spurious oscillations in advection-dominated heat
transport. In the bilinear form of the heat equation, we add the following
diffusion term:
$$
  \mathcal{D}_\theta(\theta,\tau)
  := \sum_{T \in \mathcal{T}_h} \kappa_\theta \, (\nabla \theta, \nabla \tau)_T,
$$
where the artificial diffusion coefficient is defined by
$\kappa_\theta := c\,h_T \,\|\bu\|_{L^\infty(T)}$. 
Here, $h_T$ is a characteristic mesh size on element $T$,
$\|\bu\|_{L^\infty(T)}$ is the maximum magnitude of the velocity on $T$, and
$c$ is a constant, which we set to $c = 0.1$ for all experiments. Using this
setup, we study the velocity and temperature profiles within the domain for
various values of the dimensionless constants.

\begin{figure}[!h]
     \centering
     \begin{subfigure}[b]{0.4\textwidth}
         \centering
         \includegraphics[width=\textwidth]{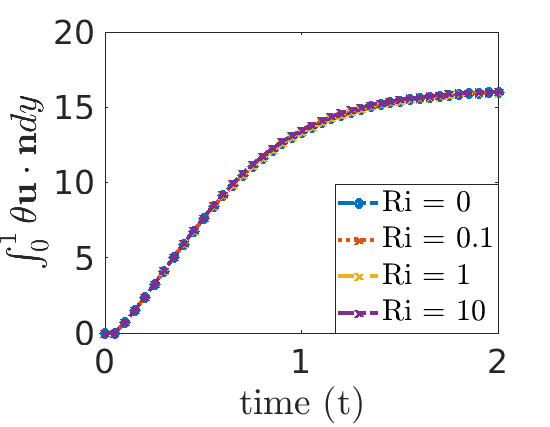}
         \caption{Reynolds number (Re) = 10}
         \label{fig:fig:heat_flux_different_Re_10}
     \end{subfigure}
     \centering
     \begin{subfigure}[b]{0.4\textwidth}
         \centering
         \includegraphics[width=\textwidth]{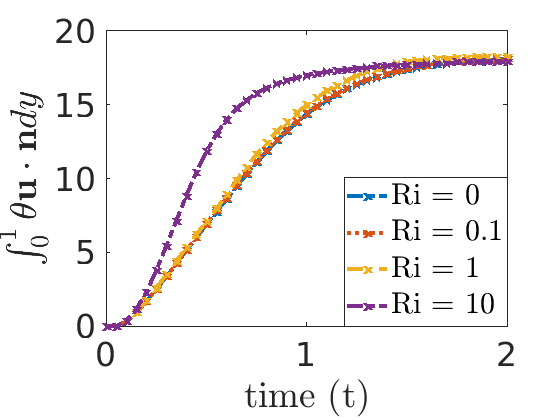}
         \caption{Reynolds number (Re) = 1000}
         \label{fig:fig:heat_flux_different_Re_1000}
     \end{subfigure}
    \caption{Convective heat flux across the outlet boundary for different Reynolds and Richardson numbers.}
    \label{fig:heat_flux_different_Re}
\end{figure}

First, we compare the convective heat flux across the right boundary for
different values of the Richardson number $\Ri$. We define the convective heat
flux as
\[
F_\theta := \int_0^1 \theta \, \bu \cdot \bn \, dy
\quad \text{on } \partial \Omega \cap \{x = 1\}.
\]
As shown in Figure~\ref{fig:heat_flux_different_Re}, for a low Reynolds number
($\Rey = 10$), the convective heat flux remains essentially constant across all
Richardson numbers. At such low $\Rey$, diffusion dominates over convection, so
increasing $\Ri$ has a negligible impact on the solution. However, for a higher
Reynolds number ($\Rey = 1000$), we observe that the heat flux varies
significantly with $\Ri$. In this regime, convection becomes the dominant mode
of heat transfer, and increasing the Richardson number strengthens the
buoyancy forcing in the momentum equation, thereby altering the velocity field.
These effects are further illustrated by the velocity, pressure, and
temperature distributions at $t_f = 2$.

\begin{figure}[!h]
     \centering
     \begin{subfigure}[t]{0.235\textwidth}
         \centering
         \includegraphics[trim={2.7cm 0 3cm 2cm},clip,width=\textwidth]{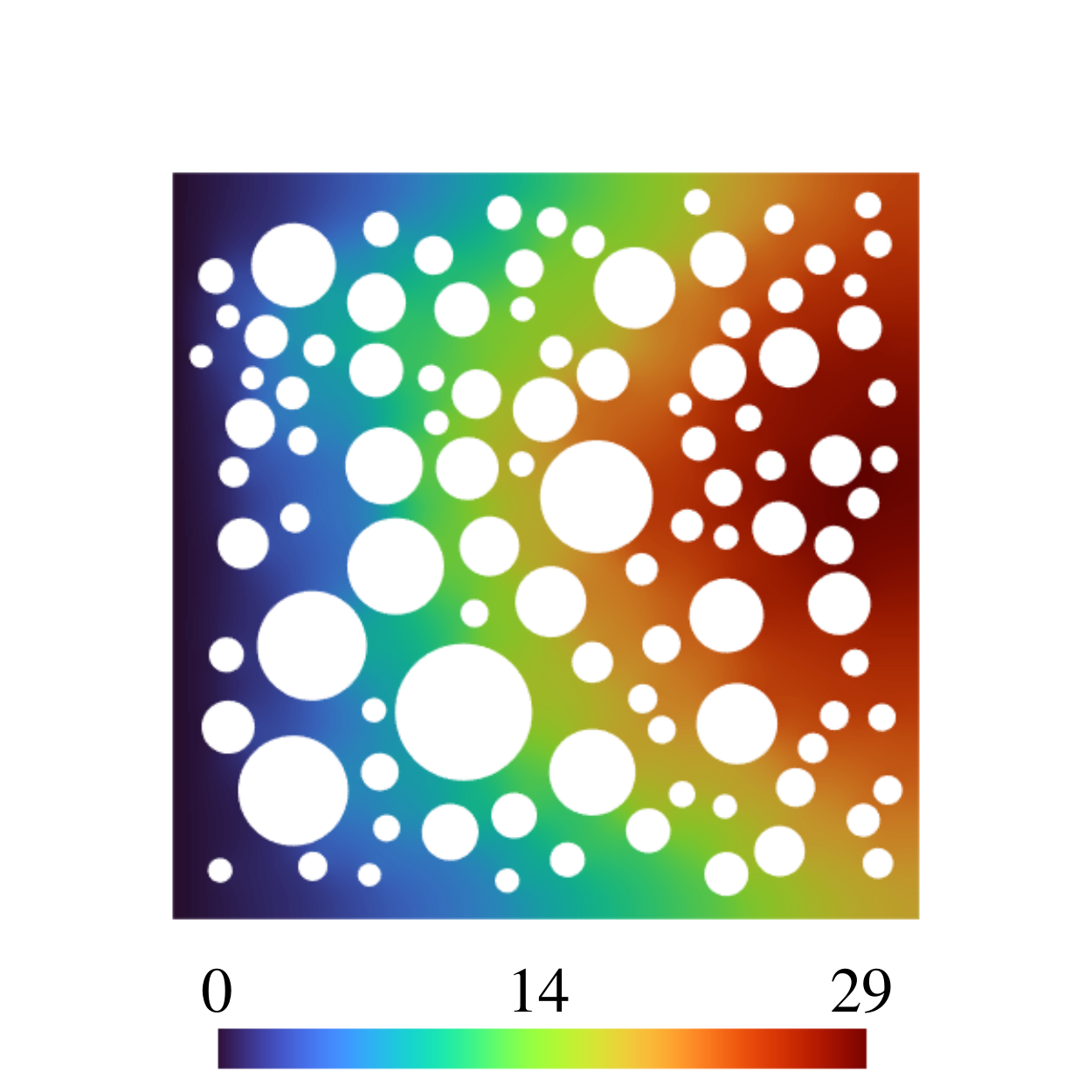}
         \caption{$\theta$ values (Ri = 0)}
     \end{subfigure}
     \centering
     \begin{subfigure}[t]{0.235\textwidth}
         \centering
         \includegraphics[trim={2.7cm 0 3cm 2cm},clip,width=\textwidth]{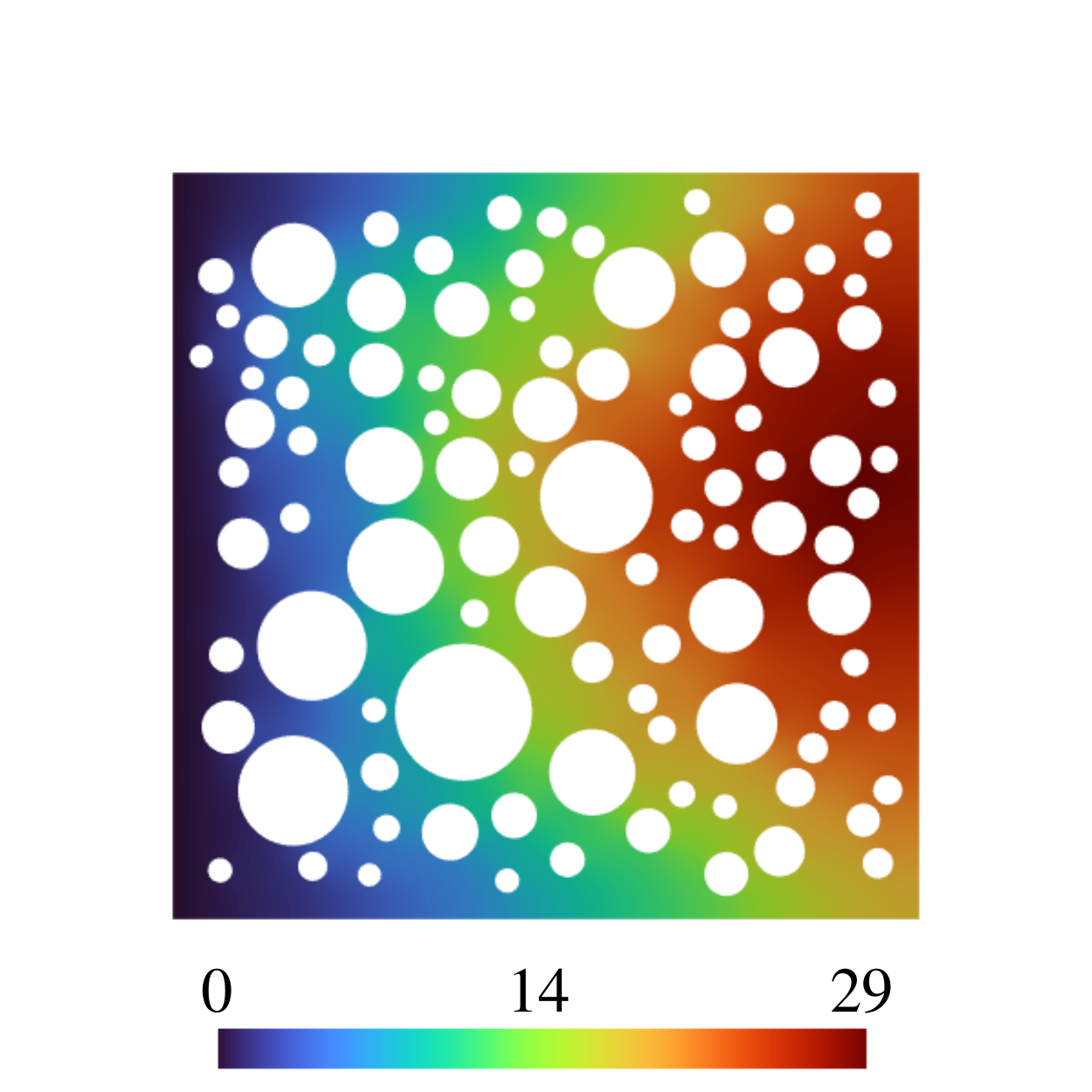}
         \caption{$\theta$ values (Ri = 0.1)}
     \end{subfigure}
     \begin{subfigure}[t]{0.235\textwidth}
         \centering
         \includegraphics[trim={2.7cm 0 3cm 2cm},clip,width=\textwidth]{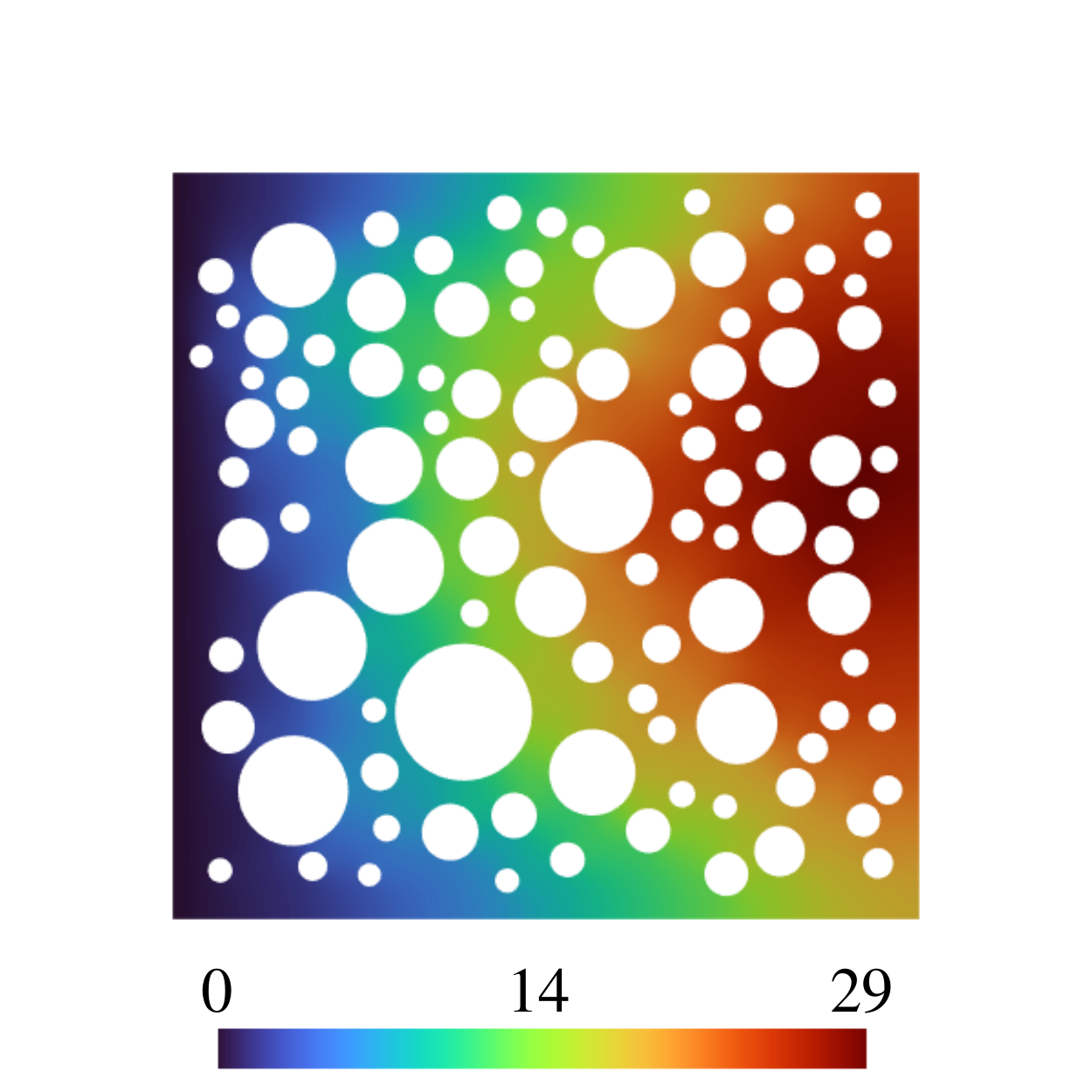}
         \caption{$\theta$ values (Ri = 1)}
     \end{subfigure}
     \begin{subfigure}[t]{0.235\textwidth}
         \centering
         \includegraphics[trim={2.7cm 0 3cm 2cm},clip,width=\textwidth]{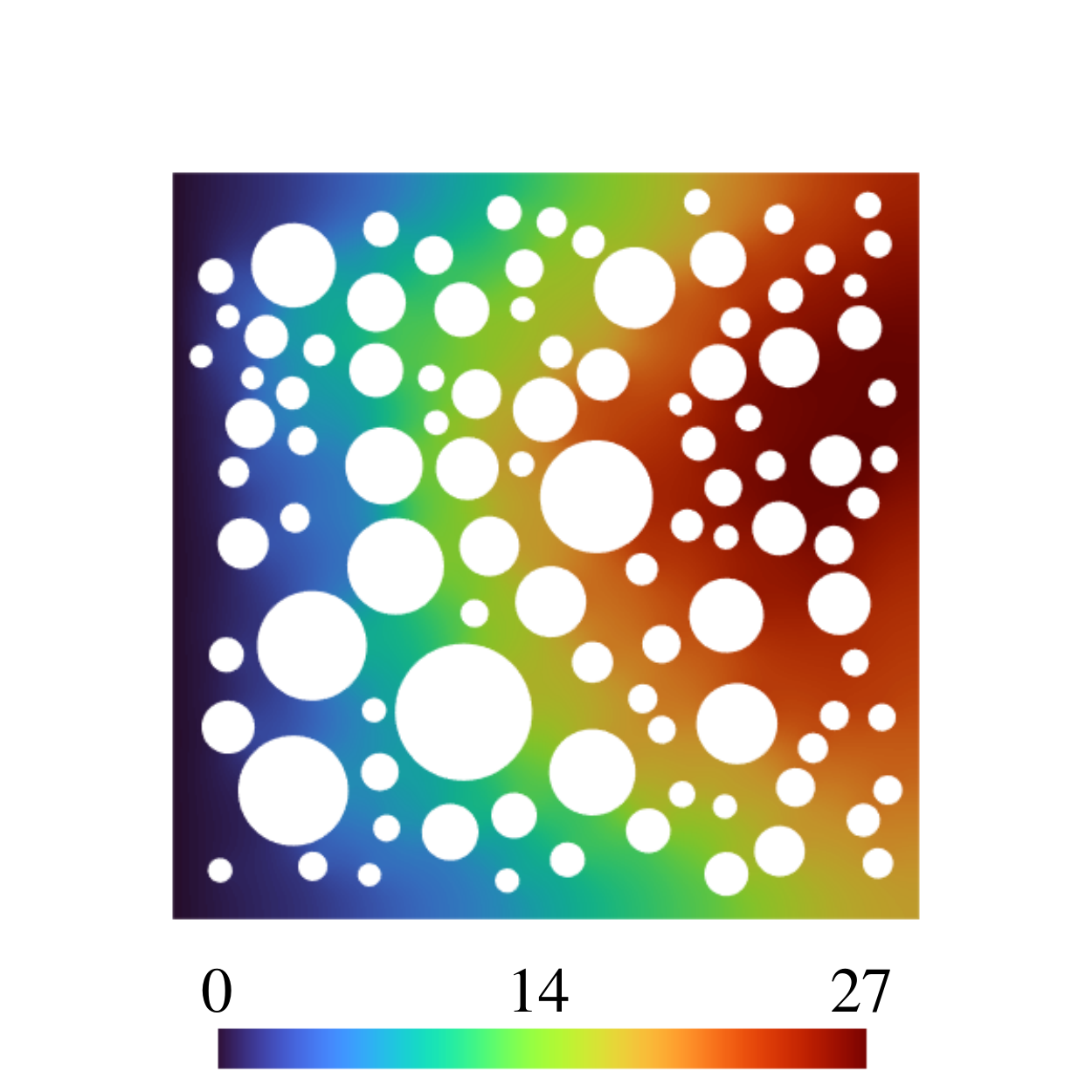}
         \caption{$\theta$ values (Ri = 10)}
     \end{subfigure}
    \caption{Temperature field $\theta$ for $\Ri \in \{0, 0.1, 1, 10\}$ with $\Rey = 10$.}
    \label{fig:temperature_Re_10}
\end{figure}
\begin{figure}[!h]
     \centering
     \begin{subfigure}[t]{0.235\textwidth}
         \centering
         \includegraphics[trim={2.7cm 0 3cm 2cm},clip,width=\textwidth]{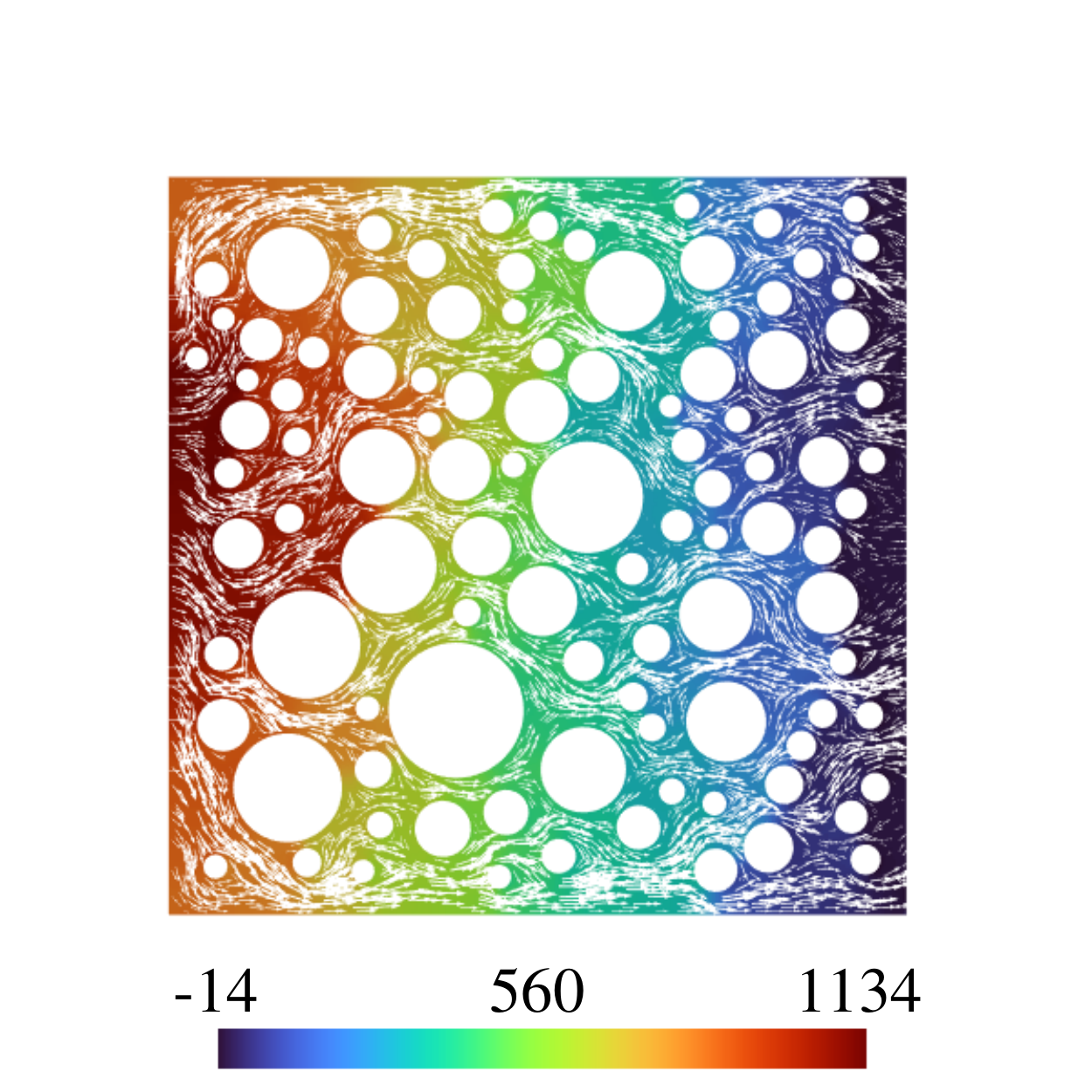}
         \caption{$p$ and $\bu$ (Ri = 0)}
     \end{subfigure}
     \centering
     \begin{subfigure}[t]{0.235\textwidth}
         \centering
         \includegraphics[trim={2.7cm 0 3cm 2cm},clip,width=\textwidth]{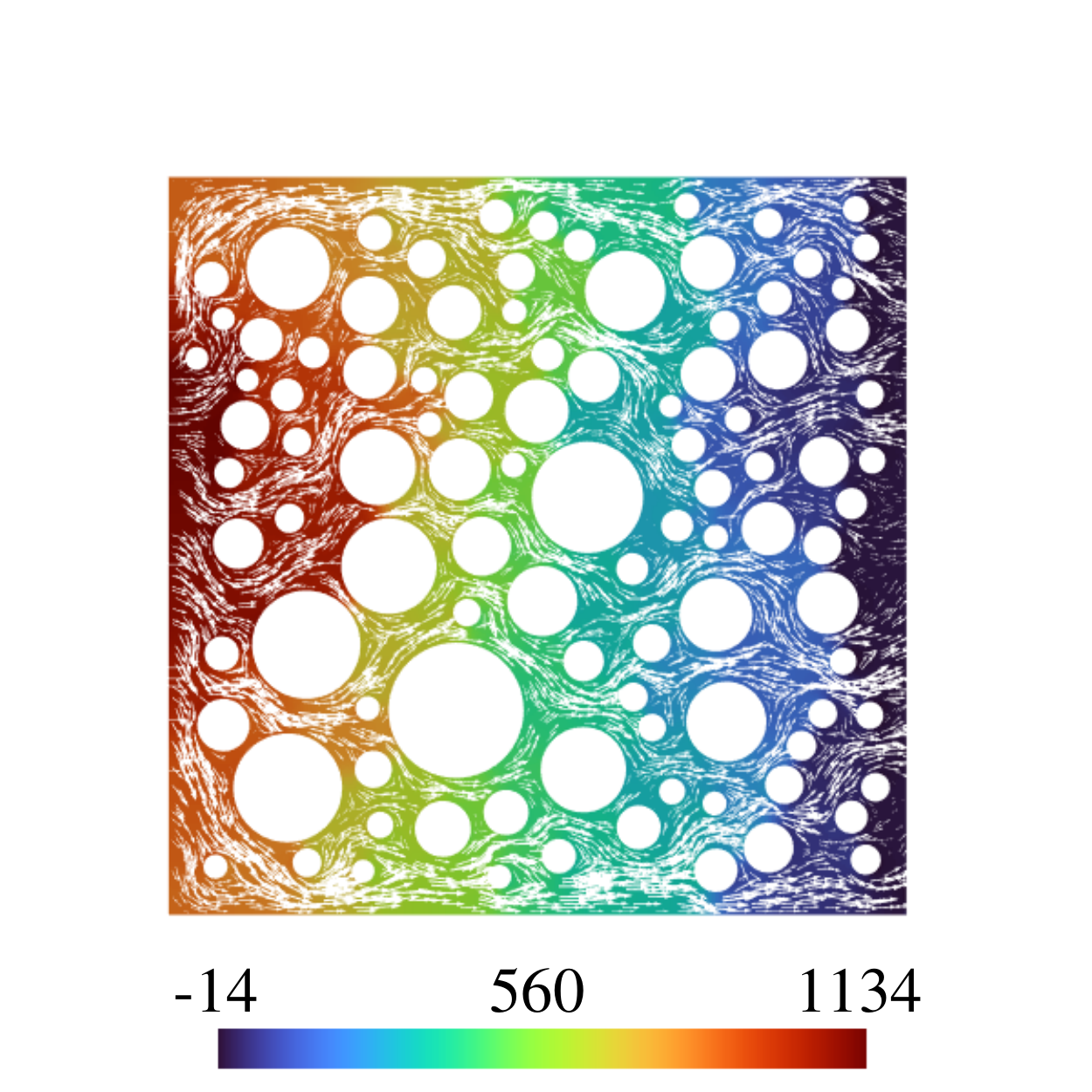}
         \caption{$p$ and $\bu$ (Ri = 0.1)}
     \end{subfigure}
     \begin{subfigure}[t]{0.235\textwidth}
         \centering
         \includegraphics[trim={2.7cm 0 3cm 2cm},clip,width=\textwidth]{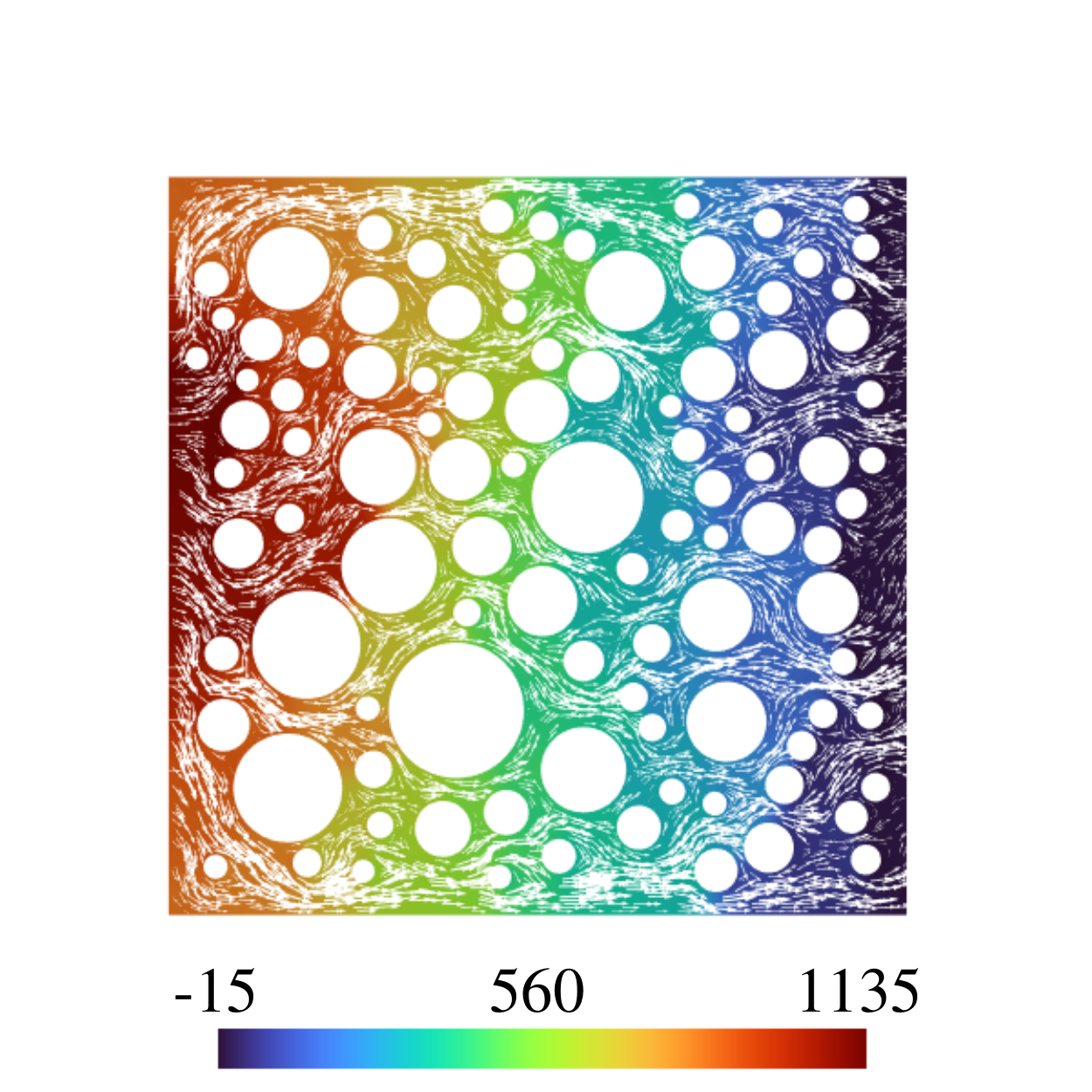}
         \caption{$p$ and $\bu$ (Ri = 1)}
     \end{subfigure}
     \begin{subfigure}[t]{0.235\textwidth}
         \centering
         \includegraphics[trim={2.7cm 0 3cm 2cm},clip,width=\textwidth]{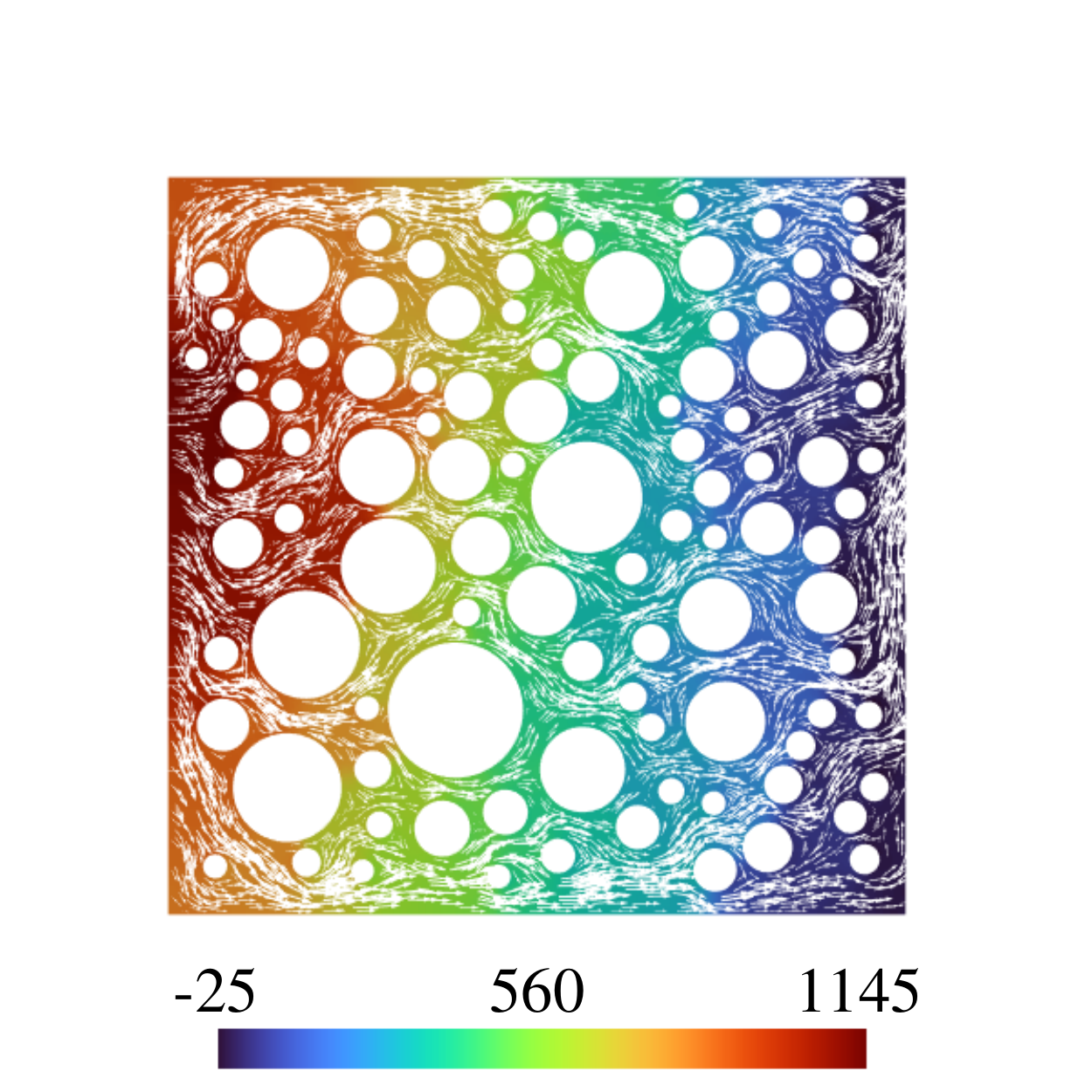}
         \caption{$p$ and $\bu$ (Ri = 10)}
     \end{subfigure}
    \caption{Pressure distribution $p$ and velocity field $\bu$ (arrows) for $\Ri \in \{ 0, 0.1, 1, 10 \}$ with $\Rey = 10$.}
    \label{fig:pressure_Re_10}
\end{figure}

Figure~\ref{fig:temperature_Re_10} shows that the temperature distribution
remains essentially unchanged for the different Richardson numbers at
$\Rey = 10$. As noted previously, at such a low Reynolds number diffusion is
the dominant mode of heat transfer; therefore, variations in $\Ri$ have little
effect on the temperature field. Similarly, the pressure contour plots in
Figure~\ref{fig:pressure_Re_10} indicate that the pressure distribution is also
nearly independent of $\Ri$: the pressure is high at the inlet and low at the
outlet. The velocity vectors exhibit only minimal variation. Although we
observe some upward-directed velocity near the outlet for $\Ri = 10$, diffusion
still dominates the heat transfer, and the temperature field does not show any
significant change despite this slight alteration in the velocity field.

\begin{figure}[!h]
     \centering
     \begin{subfigure}[t]{0.235\textwidth}
         \centering
         \includegraphics[trim={2.7cm 0 3cm 2cm},clip,width=\textwidth]{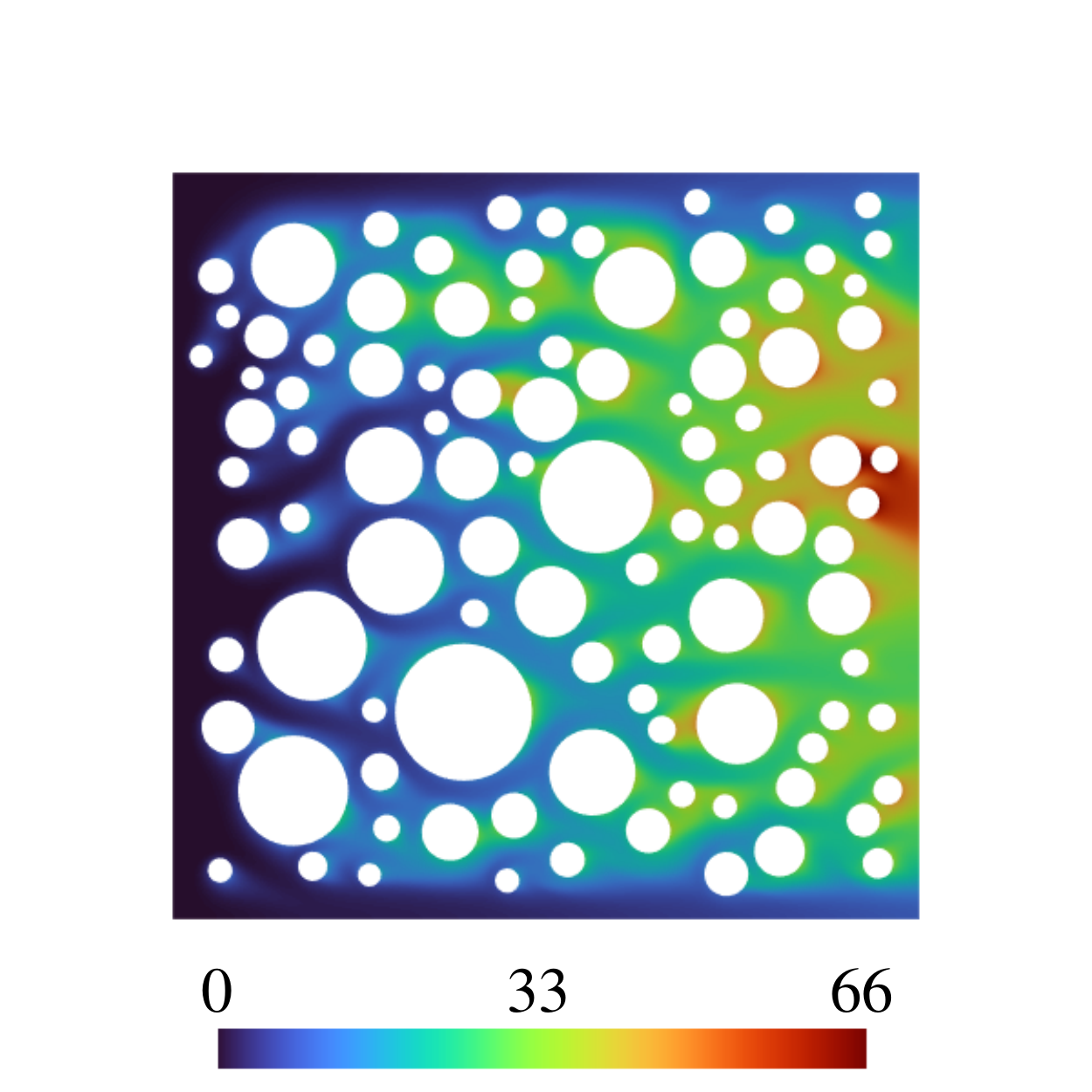}
         \caption{$\theta$ values (Ri = 0)}
     \end{subfigure}
     \centering
     \begin{subfigure}[t]{0.235\textwidth}
         \centering
         \includegraphics[trim={2.7cm 0 3cm 2cm},clip,width=\textwidth]{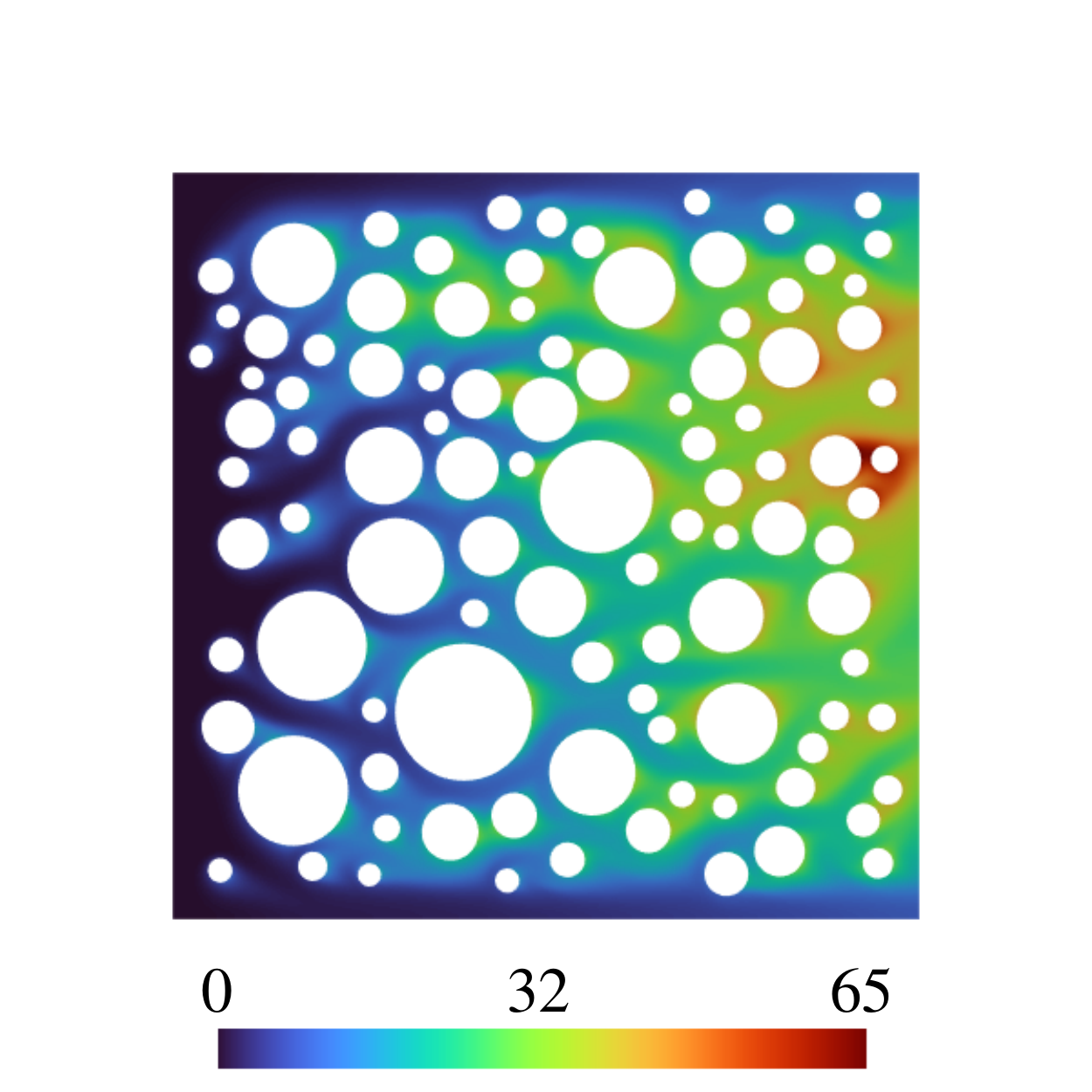}
         \caption{$\theta$ values (Ri = 0.1)}
     \end{subfigure}
     \begin{subfigure}[t]{0.235\textwidth}
         \centering
         \includegraphics[trim={2.7cm 0 3cm 2cm},clip,width=\textwidth]{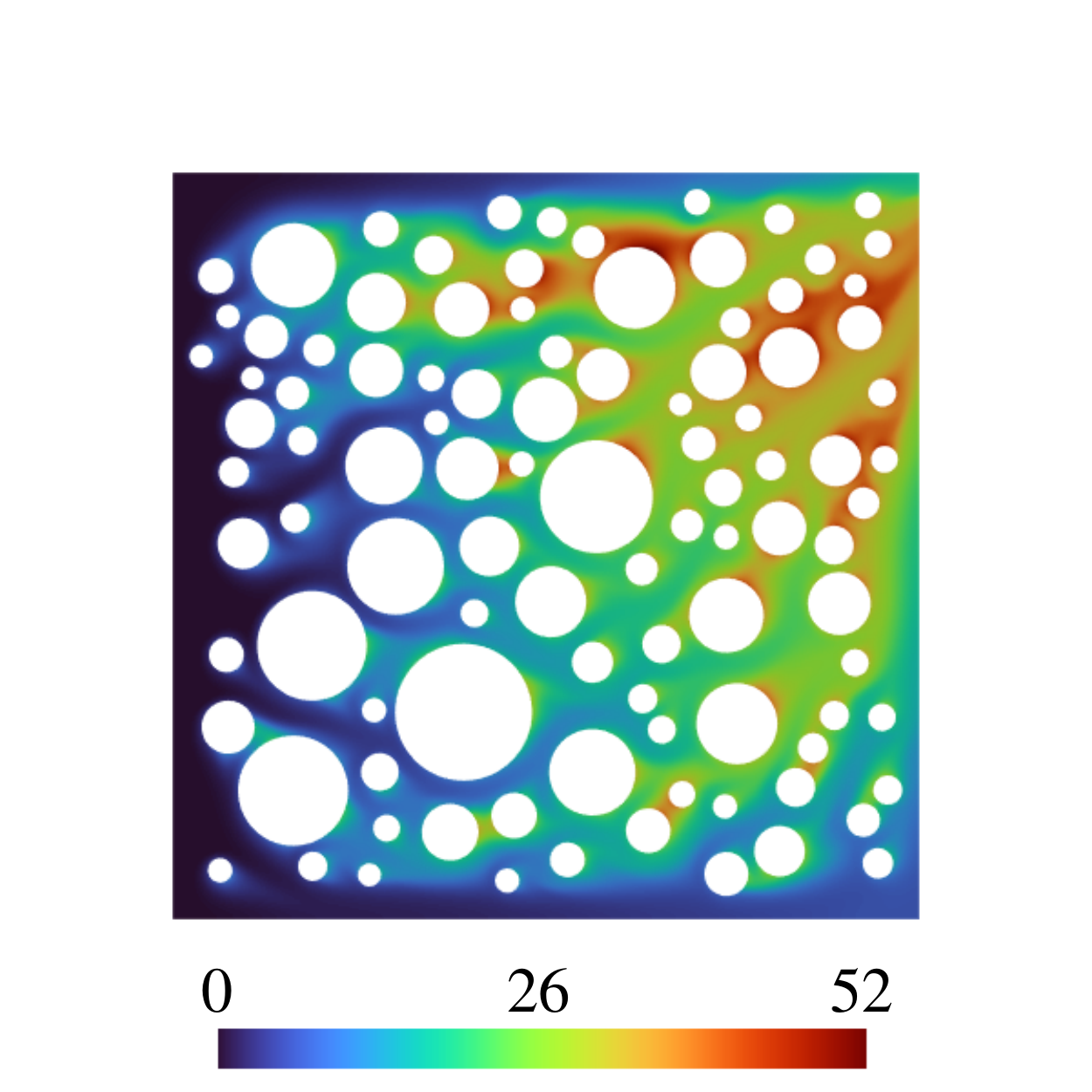}
         \caption{$\theta$ values (Ri = 1)}
     \end{subfigure}
     \begin{subfigure}[t]{0.235\textwidth}
         \centering
         \includegraphics[trim={2.7cm 0 3cm 2cm},clip,width=\textwidth]{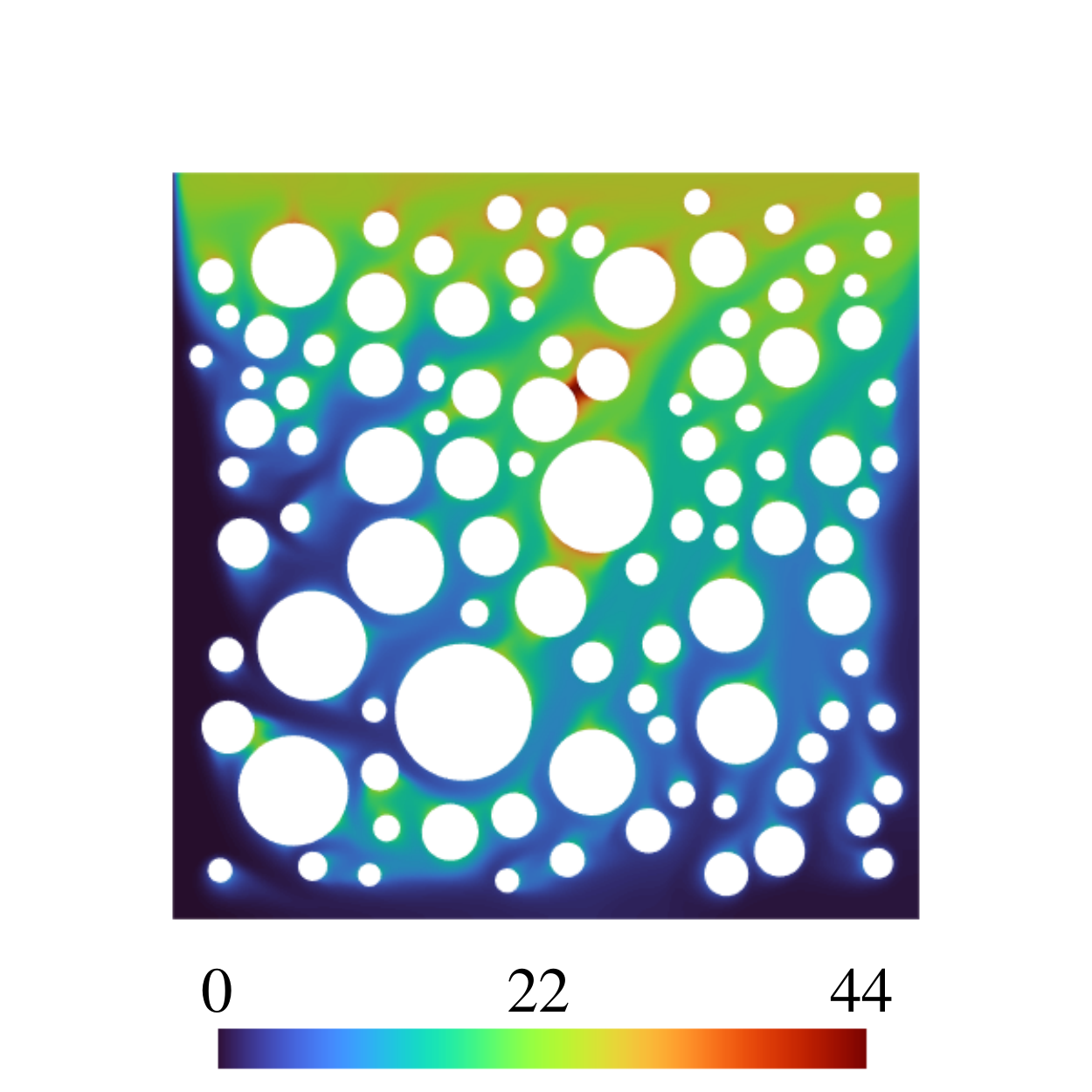}
         \caption{$\theta$ values (Ri = 10)}
     \end{subfigure}
\caption{Temperature field $\theta$ for $\Ri \in \{0, 0.1, 1, 10\}$ with $\Rey = 1000$.}
    \label{fig:temperature_Re_1000}
\end{figure}
\begin{figure}[!h]
     \centering
     \begin{subfigure}[t]{0.235\textwidth}
         \centering
         \includegraphics[trim={2.7cm 0 3cm 2cm},clip,width=\textwidth]{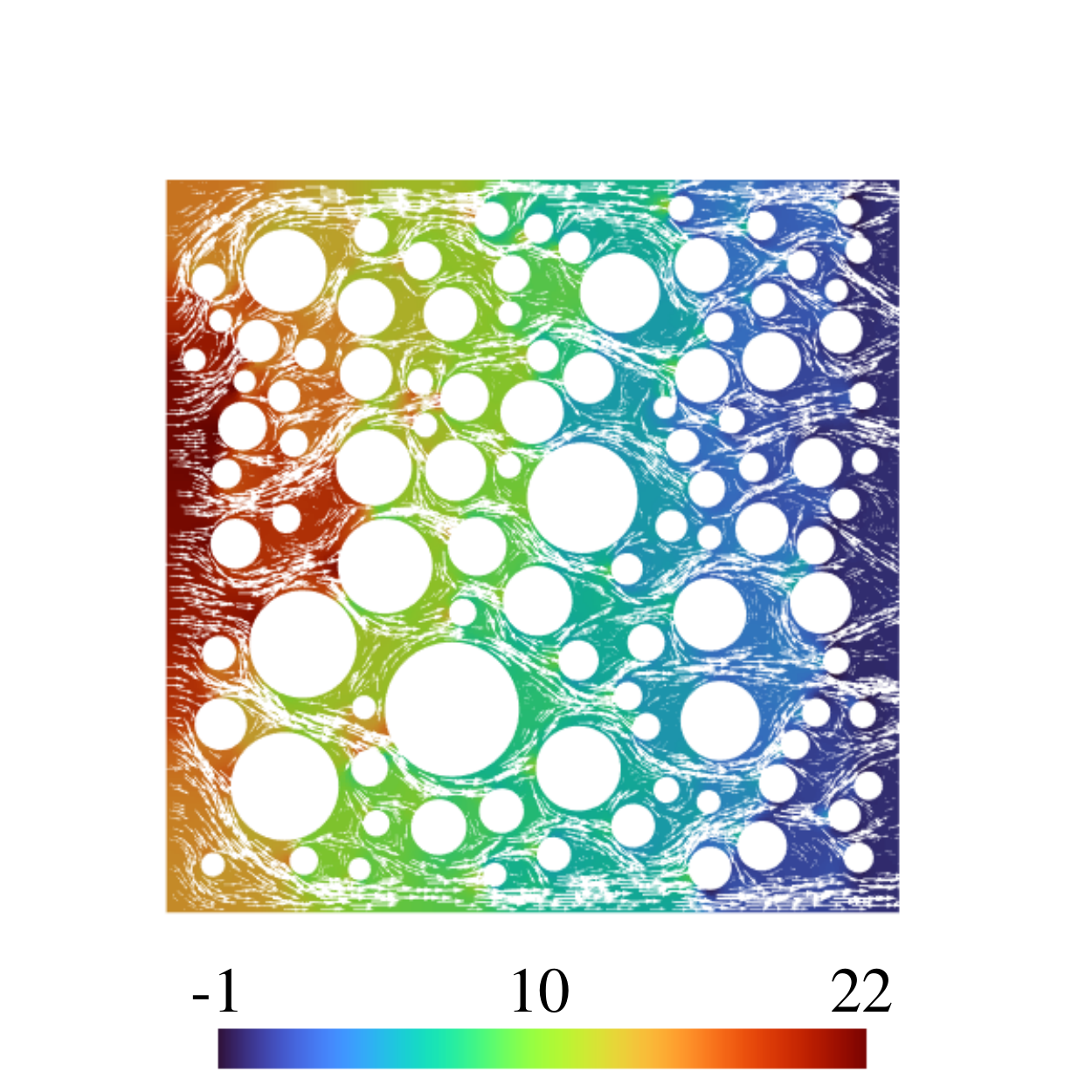}
         \caption{$p$ and $\bu$ (Ri = 0)}
     \end{subfigure}
     \centering
     \begin{subfigure}[t]{0.235\textwidth}
         \centering
         \includegraphics[trim={2.7cm 0 3cm 2cm},clip,width=\textwidth]{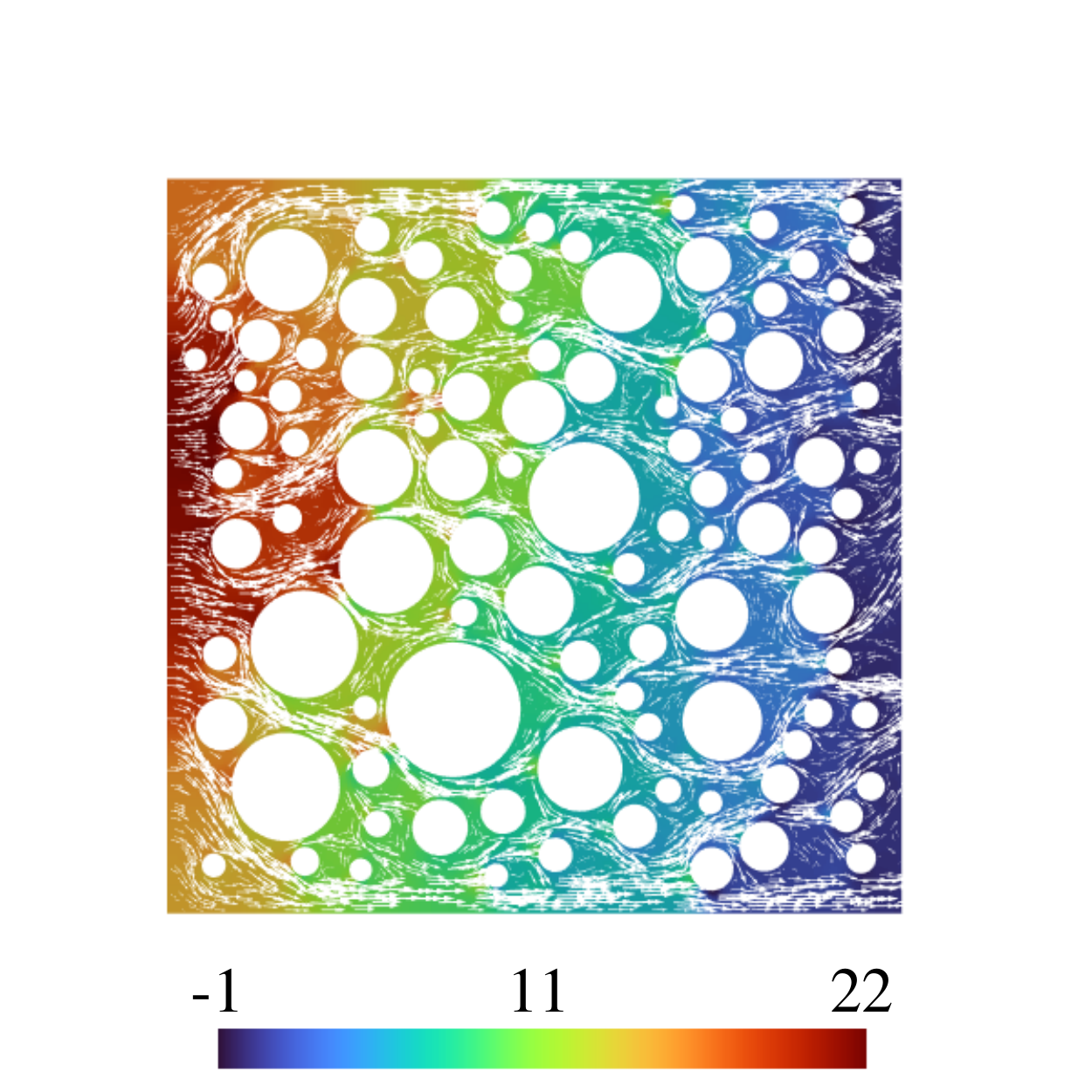}
         \caption{$p$ and $\bu$ (Ri = 0.1)}
     \end{subfigure}
     \begin{subfigure}[t]{0.235\textwidth}
         \centering
         \includegraphics[trim={2.7cm 0 3cm 2cm},clip,width=\textwidth]{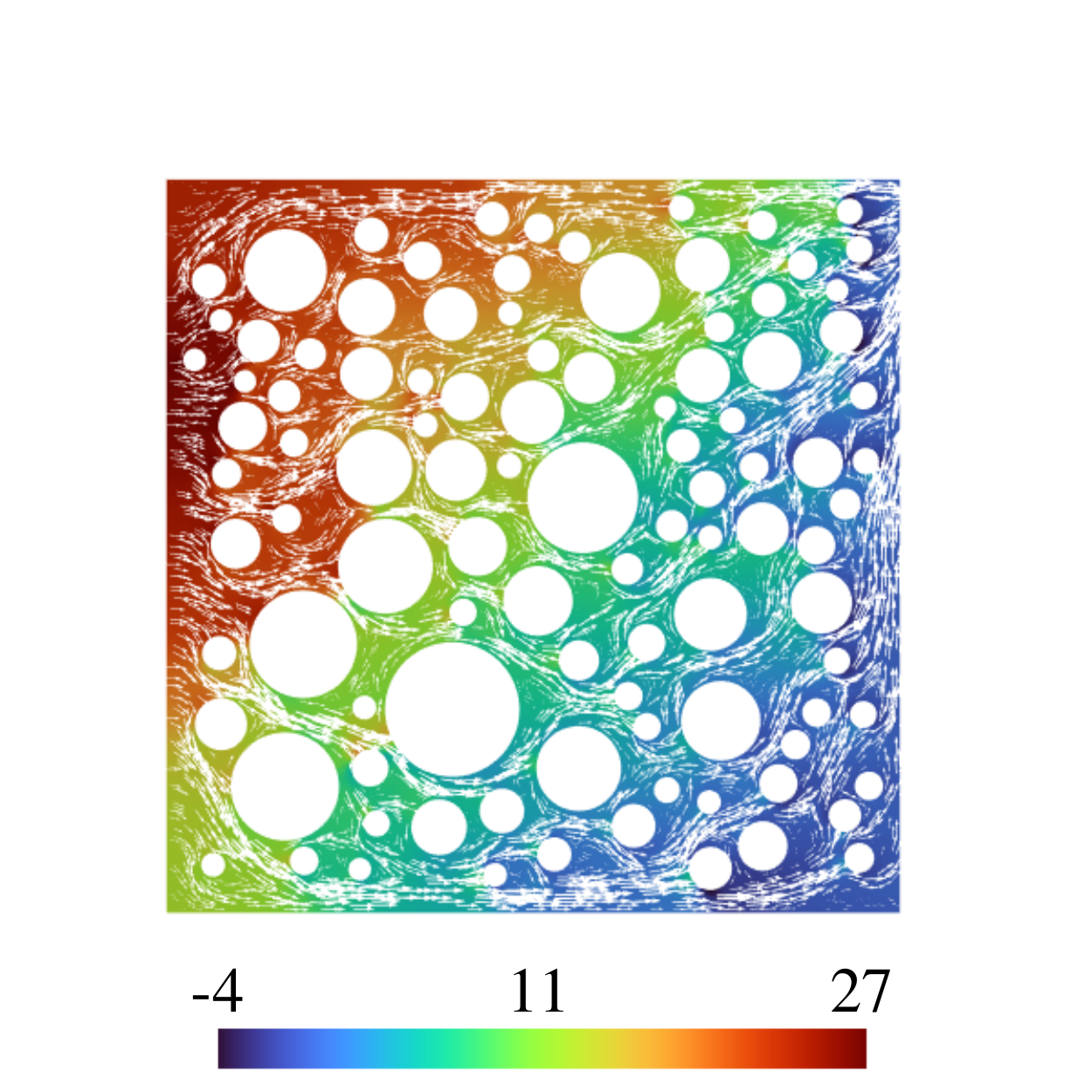}
         \caption{$p$ and $\bu$ (Ri = 1)}
     \end{subfigure}
     \begin{subfigure}[t]{0.235\textwidth}
         \centering
         \includegraphics[trim={2.7cm 0 3cm 2cm},clip,width=\textwidth]{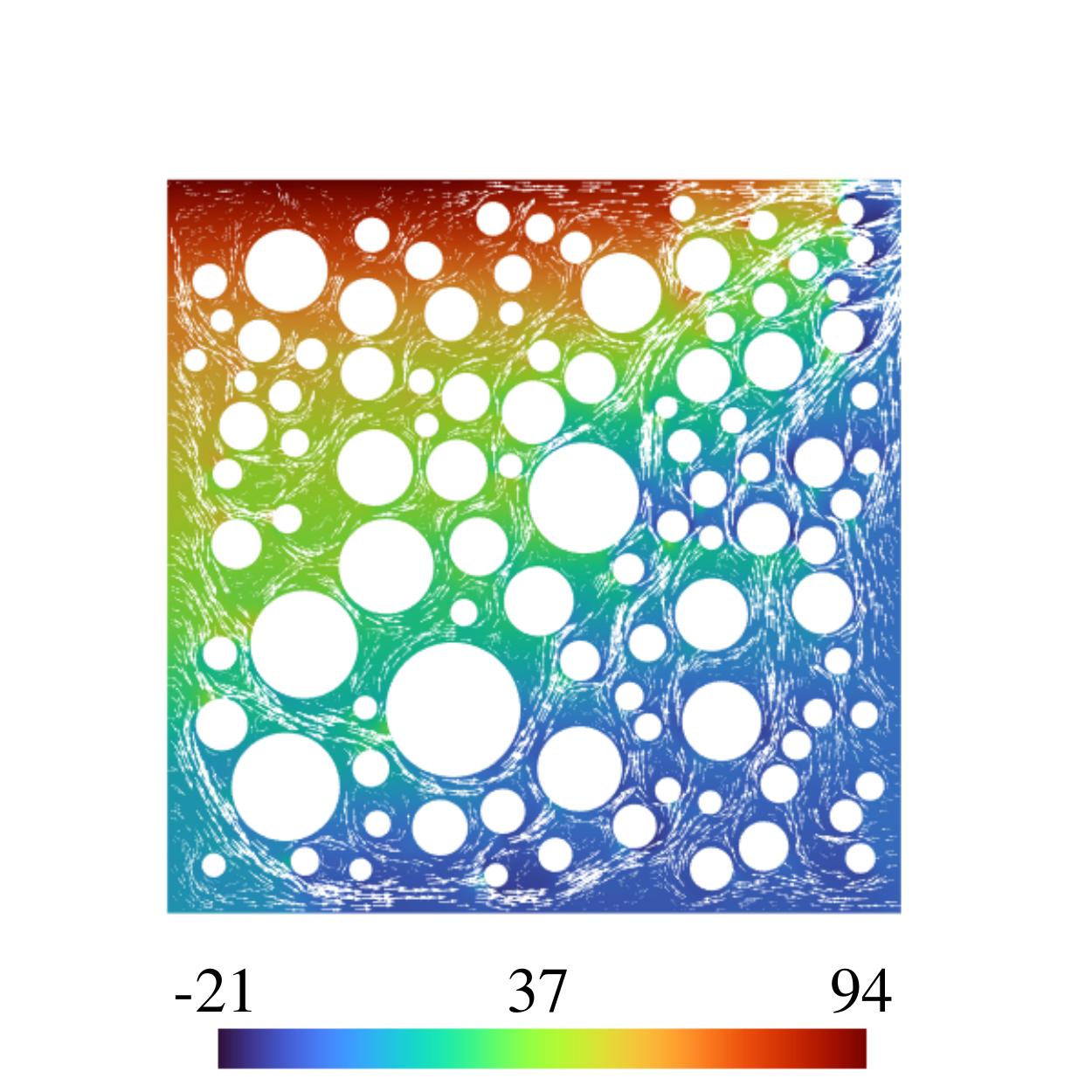}
         \caption{$p$ and $\bu$ (Ri = 10)}
     \end{subfigure}
     \caption{Pressure distribution $p$ and velocity field $\bu$ (arrows) for $\Ri \in \{0, 0.1, 1, 10\}$ with $\Rey = 1000$.}
   \label{fig:pressure_Re_1000}
\end{figure} 

At a high Reynolds number ($\Rey = 1000$), the temperature, pressure, and velocity profiles exhibit pronounced changes as the Richardson number varies. In this regime, advection becomes the dominant mode of heat transfer, so the velocity field has a strong influence on the temperature distribution. When the
Richardson number is zero ($\Ri = 0$), corresponding to the absence of thermal buoyancy effects, the temperature is primarily advected from left to right. However, as $\Ri$ increases, thermal buoyancy becomes more significant, causing the temperature to be advected upward toward the top boundary. This behavior is
clearly visible in Figure~\ref{fig:temperature_Re_1000}. The cold fluid is pushed downward, as seen in the pressure contour plot in Figure~\ref{fig:pressure_Re_1000}, where the pressure field shifts toward the
top boundary. As a result, the hot fluid moves upward and exits the domain near the top-right corner.

\begin{figure}[!h]
    \centering
    \includegraphics[width=0.4\linewidth]{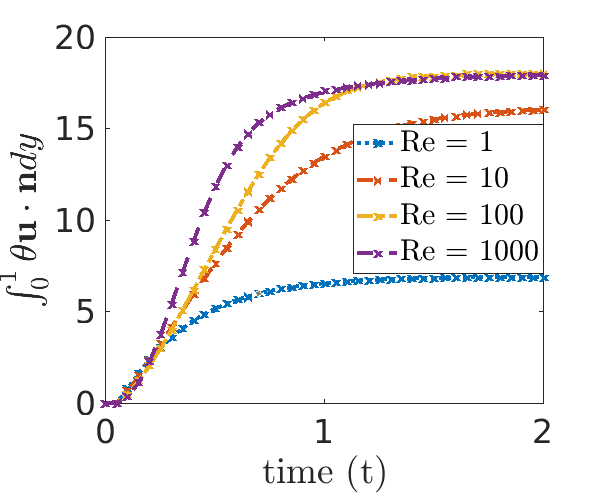}
   \caption{Convective heat flux across the boundary for different Reynolds numbers ($\Rey$) at a fixed Richardson number ($\Ri = 10$).}
    \label{fig:heat_flux_different_Re_Ri_10}
\end{figure}

Additionally, we examine the solution at a fixed Richardson number
$\Ri = 10$ for varying Reynolds numbers. Figure~\ref{fig:heat_flux_different_Re_Ri_10}
shows the convective heat flux values for
$\Rey \in \{1, 10, 100, 1000\}$. As noted previously, at low Reynolds numbers, diffusion dominates the heat transfer process, leading to a minimal convective
heat flux at the outlet. As the Reynolds number increases, however, convection
becomes increasingly dominant and the convective heat flux rises significantly.
The heat flux approaches a peak value, corresponding to the maximum flux that is transferred from the solid matrix into the fluid. For higher Reynolds
numbers, convection continues to dominate, and the system reaches this maximal
heat flux rapidly.

\begin{figure}[!h]
     \centering
     \begin{subfigure}[t]{0.235\textwidth}
         \centering
         \includegraphics[trim={2.7cm 0 3cm 3cm},clip,width=\textwidth]{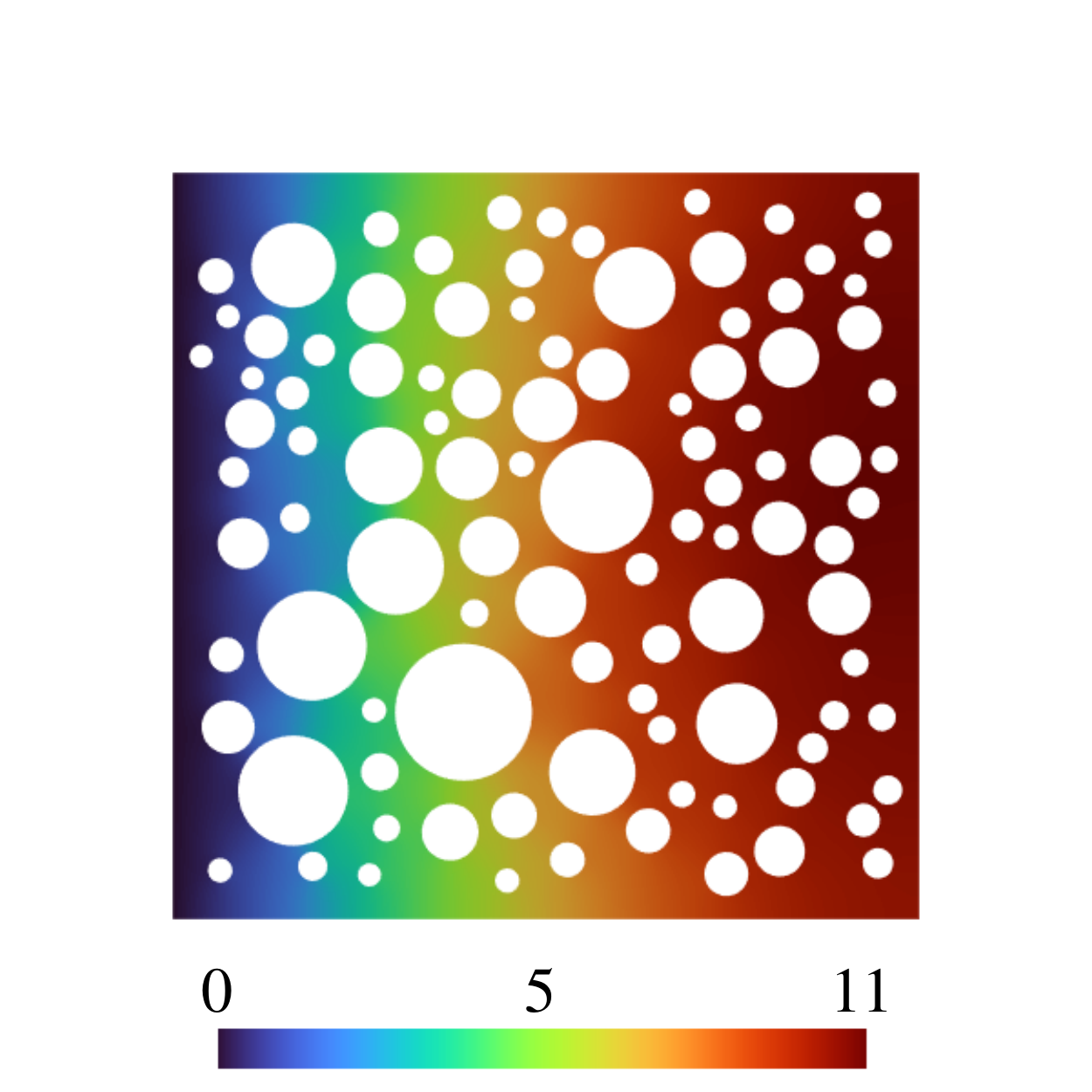}
         \caption{$\theta$ (Re = 1)}
     \end{subfigure}
     \centering
     \begin{subfigure}[t]{0.235\textwidth}
         \centering
         \includegraphics[trim={2.7cm 0 3cm 3cm},clip,width=\textwidth]{figures/Pr_0.710000_Re_10.000000_Ri_10.000000_temperature.pdf}
         \caption{$\theta$ (Re = 10)}
     \end{subfigure}
     \begin{subfigure}[t]{0.235\textwidth}
         \centering
         \includegraphics[trim={2.7cm 0 3cm 3cm},clip,width=\textwidth]{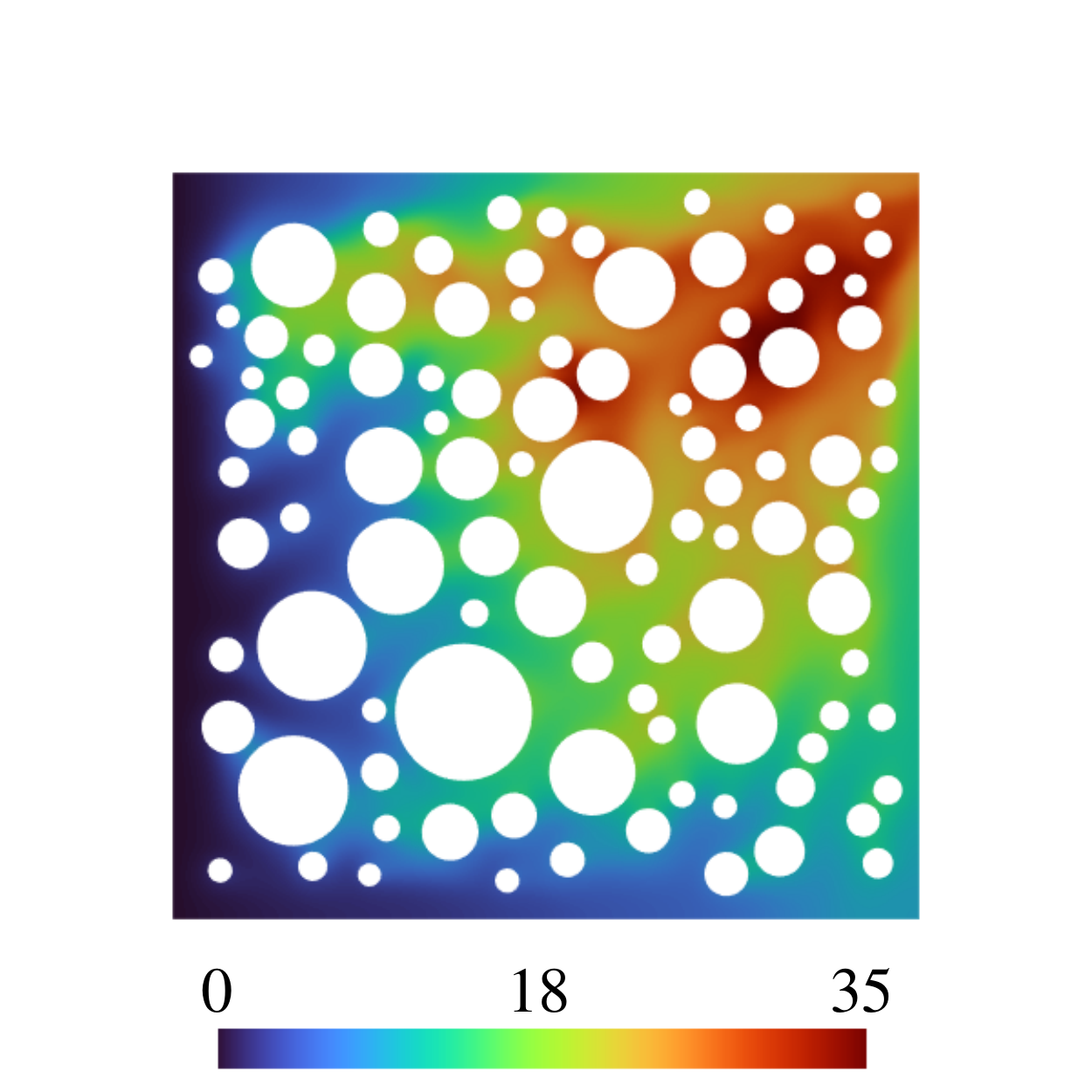}
         \caption{$\theta$  (Re = 100)}
     \end{subfigure}
     \begin{subfigure}[t]{0.235\textwidth}
         \centering
         \includegraphics[trim={2.7cm 0 3cm 3cm},clip,width=\textwidth]{figures/Pr_0.710000_Re_1000.000000_Ri_10.000000_temperature.pdf}
         \caption{$\theta$ (Re = 1000)}
     \end{subfigure}
     \caption{Temperature field $\theta$ for $\Rey \in \{1, 10, 100, 1000\}$ with $\Ri = 10$.}
    \label{fig:temperature_Ri_10}
\end{figure}
\begin{figure}[!h]
     \centering
     \begin{subfigure}[t]{0.235\textwidth}
         \centering
         \includegraphics[trim={2.7cm 0 3cm 3cm},clip,width=\textwidth]{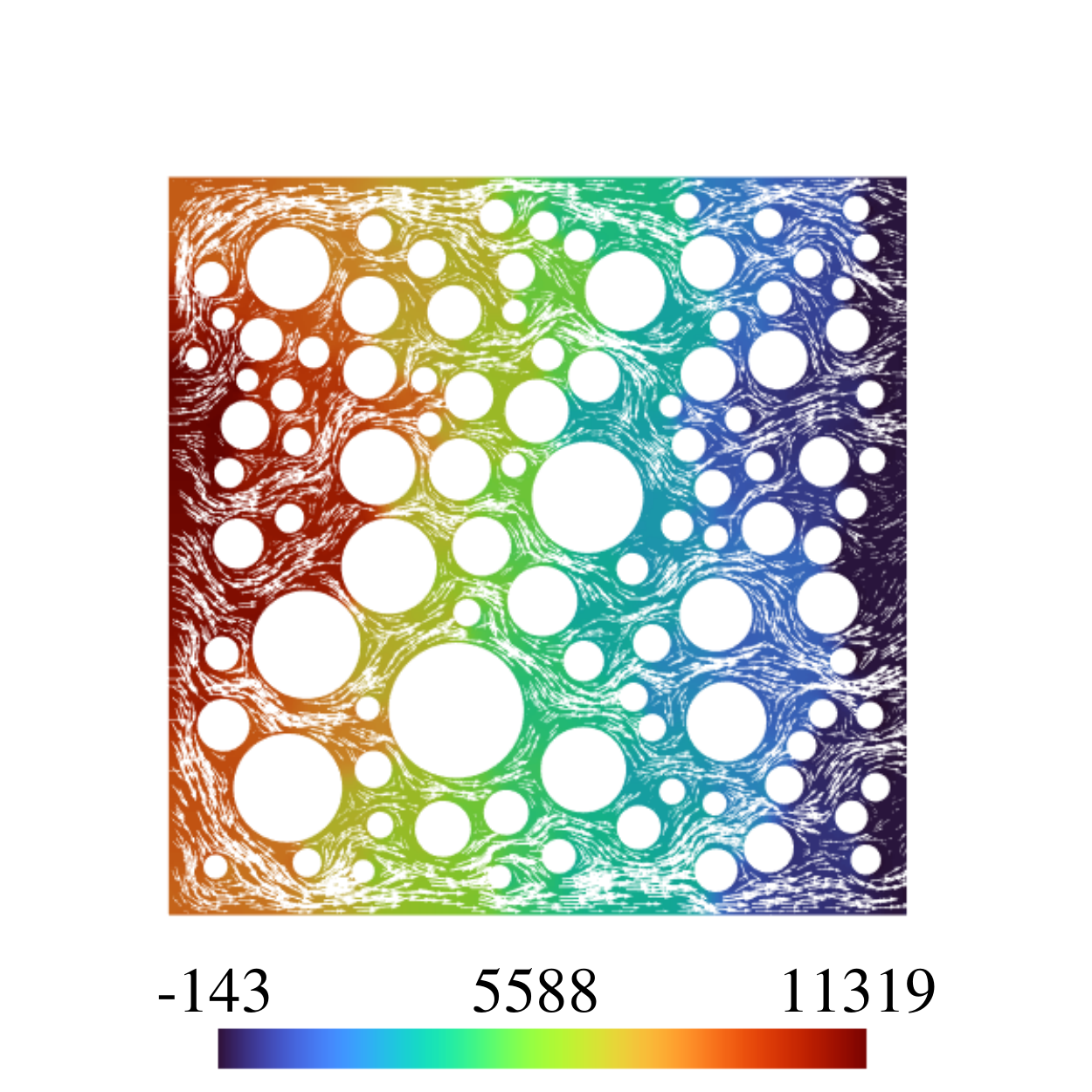}
         \caption{$p$, $\bu$ (Re = 1)}
     \end{subfigure}
     \centering
     \begin{subfigure}[t]{0.235\textwidth}
         \centering
         \includegraphics[trim={2.7cm 0 3cm 3cm},clip,width=\textwidth]{figures/Pr_0.710000_Re_10.000000_Ri_10.000000_velocity.pdf}
         \caption{$p$, $\bu$ (Re = 10)}
     \end{subfigure}
     \begin{subfigure}[t]{0.235\textwidth}
         \centering
         \includegraphics[trim={2.7cm 0 3cm 3cm},clip,width=\textwidth]{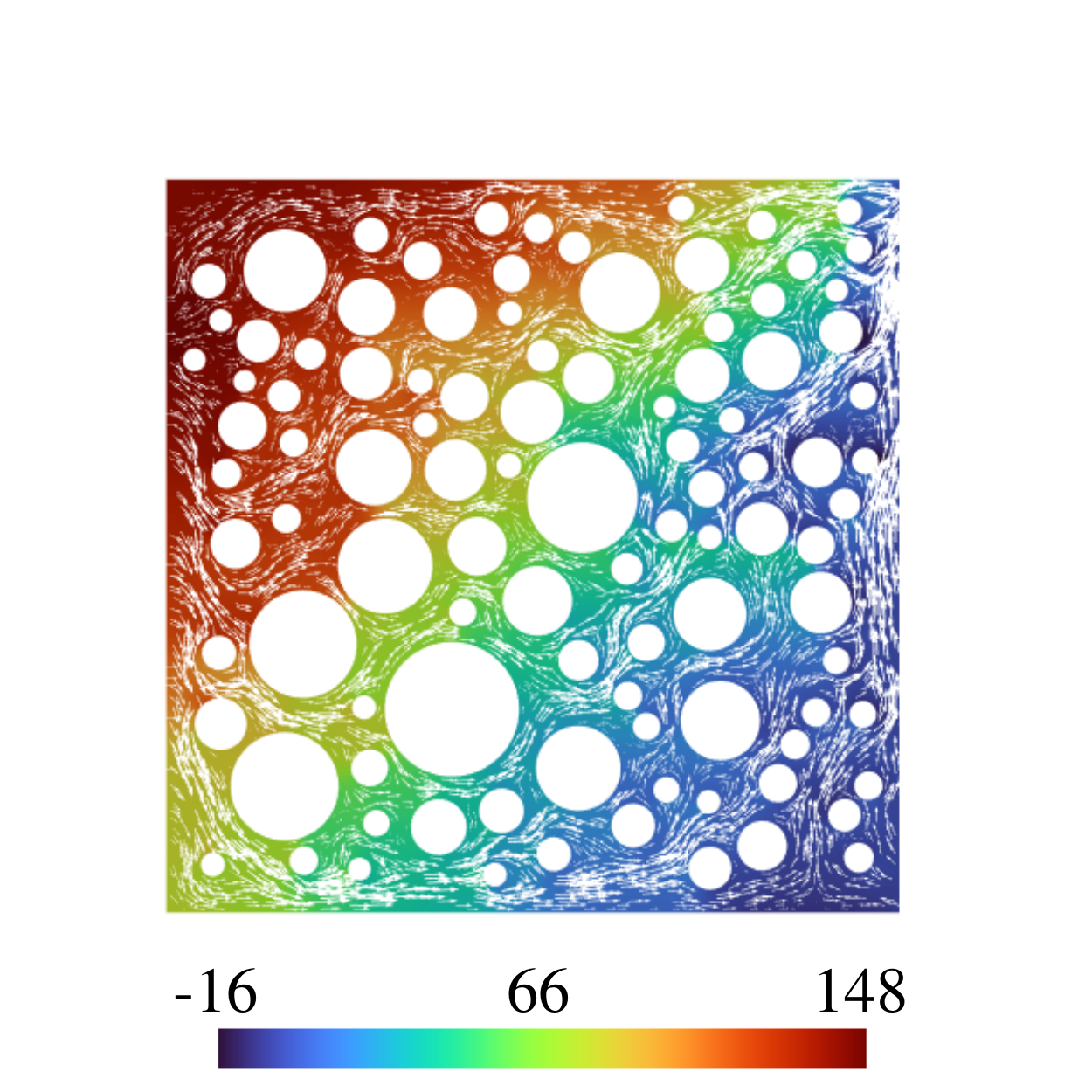}
         \caption{$p$, $\bu$ (Re = 100)}
     \end{subfigure}
     \begin{subfigure}[t]{0.235\textwidth}
         \centering
         \includegraphics[trim={2.7cm 0 3cm 3cm},clip,width=\textwidth]{figures/Pr_0.710000_Re_1000.000000_Ri_10.000000_velocity.pdf}
         \caption{$p$, $\bu$ (Re = 1000)}
     \end{subfigure}
    \caption{Pressure distribution $p$ and velocity field $\bu$ (arrows) for $\Rey \in \{1, 10, 100, 1000\}$ with $\Ri = 10$.}
    \label{fig:pressure_Ri_10}
\end{figure}

Figure~\ref{fig:temperature_Ri_10} illustrates the temperature distribution for
$\Ri = 10$ and $\Rey \in \{1, 10, 100, 1000\}$. As the Reynolds number
increases, the temperature is increasingly transported toward the top boundary.
The pressure contour plots in Figure~\ref{fig:pressure_Ri_10} exhibit a similar
trend: the pressure field shifts toward the top boundary as $\Rey$ increases,
pushing the colder fluid downward.

At low Reynolds numbers, the fluid travels more or less from the left
to the right boundary (the fluid does not move upwards towards the top boundary). This is reflected in the corresponding temperature
distribution, which shows nearly constant values along the vertical direction, indicating that heat is primarily transported by diffusion from inlet to outlet with little vertical advection. As the Reynolds number increases, however, the velocity profile changes and the hot fluid is increasingly advected toward the top boundary.

Finally, this example shows how the flow pattern and temperature distribution vary with different Reynolds and Richardson numbers. At low Reynolds numbers, changes in the Richardson number have minimal impact on the temperature distribution within the porous domain. However, as the Reynolds number increases, the influence of the Richardson number on the temperature distribution becomes more pronounced.

\section{Conclusion}
\label{sec:conclusion}
In this paper, we developed a pressure-robust enriched Galerkin (EG) framework for the incompressible Navier--Stokes and heat equations under the Boussinesq approximation. The method combines an EG velocity--pressure pair with a velocity reconstruction that produces an exactly divergence-free, $H(\mathrm{div})$-conforming field. Using Arbogast--Correa mixed spaces, the resulting scheme is locally mass conservative, inf--sup stable, and robust to pressure gradients and mesh distortion.

Numerical tests confirm these advantages. For manufactured solutions at high Reynolds numbers on distorted meshes, the pressure-robust EG method yields significantly smaller, Reynolds-independent velocity errors compared with standard EG, while maintaining similar pressure accuracy. In the cavity benchmark, it matches established reference values across Rayleigh numbers, and Anderson-accelerated Picard iteration remains effective when classical Picard stagnates. Pore-scale simulations further demonstrate accurate transitions between diffusion- and convection-dominated regimes and clarify how Reynolds and Richardson numbers affect convective heat flux in complex geometries.

Overall, pressure-robust EG discretizations coupled with Anderson-accelerated nonlinear solvers provide a practical and accurate tool for coupled flow and heat transport in complex domains. Future work will consider higher-order and 3D/unstructured extensions, as well as richer physics and multiscale couplings relevant to subsurface energy applications.

\section*{Acknowledgments}
S. Lee was partially supported by the U.S. National Science Foundation under Grant DMS-220840
and the U.S. Department of Energy, Office of Science, Energy Earthshots Initiatives under Award Number DE-SC-0024703.
L. Mu was partially supported by the U.S. National Science Foundation under Grant DMS-2309557.

\bibliographystyle{unsrt}
\bibliography{Navier-Stokes_and_Heat}

\begin{thebibliography}{10}

\bibitem{sha_multidimensional_1982}
W.~T. Sha, C.~I. Yang, T.~T. Kao, and S.~M. Cho.
\newblock Multidimensional numerical modeling of heat exchangers.
\newblock {\em Journal of Heat Transfer}, 104(3):417--425, 08 1982.

\bibitem{stamou2006verification}
A~Stamou and Ioannis Katsiris.
\newblock Verification of a cfd model for indoor airflow and heat transfer.
\newblock {\em Building and Environment}, 41(9):1171--1181, 2006.

\bibitem{sondak2000simulation}
Douglas~L Sondak and Daniel~J Dorney.
\newblock Simulation of coupled unsteady flow and heat conduction in turbine stage.
\newblock {\em Journal of Propulsion and Power}, 16(6):1141--1148, 2000.

\bibitem{olasolo2016enhanced}
P~Olasolo, MC~Ju{\'a}rez, MP~Morales, IA~Liarte, et~al.
\newblock Enhanced geothermal systems ({E}{G}{S}): A review.
\newblock {\em Renewable and Sustainable Energy Reviews}, 56:133--144, 2016.

\bibitem{boussinesq_theorie_1897}
Joseph Boussinesq.
\newblock {\em Th{\'e}orie de l'{\'e}coulmnent tourbillonnant et tumultuex des liquides dans les lits rectilignes {\`a} grande section}, volume~1.
\newblock Gauthier-Villars, 1897.

\bibitem{pollock2021acceleration}
Sara Pollock, Leo~G Rebholz, and Mengying Xiao.
\newblock Acceleration of nonlinear solvers for natural convection problems.
\newblock {\em Journal of Numerical Mathematics}, 29(4):323--341, 2021.

\bibitem{de1983natural}
Graham de~Vahl~Davis.
\newblock Natural convection of air in a square cavity: a bench mark numerical solution.
\newblock {\em International Journal for numerical methods in fluids}, 3(3):249--264, 1983.

\bibitem{kronbichler2012high}
Martin Kronbichler, Timo Heister, and Wolfgang Bangerth.
\newblock High accuracy mantle convection simulation through modern numerical methods.
\newblock {\em Geophysical Journal International}, 191(1):12--29, 2012.

\bibitem{mcwilliams1996modeling}
James~C McWilliams.
\newblock Modeling the oceanic general circulation.
\newblock {\em Annual Review of Fluid Mechanics}, 28(1):215--248, 1996.

\bibitem{babuska_finite_1973}
Ivo Babu{\v{s}}ka.
\newblock The finite element method with lagrangian multipliers.
\newblock {\em Numerische Mathematik}, 20(3):179--192, 1973.

\bibitem{brezzi_existence_1974}
Franco Brezzi.
\newblock On the existence, uniqueness and approximation of saddle-point problems arising from lagrangian multipliers.
\newblock {\em Publications des s{\'e}minaires de math{\'e}matiques et informatique de Rennes}, (S4):1--26, 1974.

\bibitem{su2025pressure}
Shuai Su, Xiurong Yan, and Qian Zhang.
\newblock A pressure-robust and parameter-free enriched galerkin method for the navier-stokes equations of rotational form.
\newblock {\em arXiv preprint arXiv:2511.11330}, 2025.

\bibitem{yi2022enriched}
Son-Young Yi, Xiaozhe Hu, Sanghyun Lee, and James~H Adler.
\newblock An enriched galerkin method for the stokes equations.
\newblock {\em Computers \& Mathematics with Applications}, 120:115--131, 2022.

\bibitem{hu2024pressure}
Xiaozhe Hu, Seulip Lee, Lin Mu, and Son-Young Yi.
\newblock Pressure-robust enriched galerkin methods for the stokes equations.
\newblock {\em Journal of Computational and Applied Mathematics}, 436:115449, 2024.

\bibitem{yi2022locking}
Son-Young Yi, Sanghyun Lee, and Ludmil Zikatanov.
\newblock Locking-free enriched galerkin method for linear elasticity.
\newblock {\em SIAM Journal on Numerical Analysis}, 60(1):52--75, 2022.

\bibitem{lee2023locking}
Sanghyun Lee and Son-Young Yi.
\newblock Locking-free and locally-conservative enriched galerkin method for poroelasticity.
\newblock {\em Journal of Scientific Computing}, 94(1):26, 2023.

\bibitem{arbogast2016two}
Todd Arbogast and Maicon~R Correa.
\newblock Two families of h (div) mixed finite elements on quadrilaterals of minimal dimension.
\newblock {\em SIAM Journal on Numerical Analysis}, 54(6):3332--3356, 2016.

\bibitem{lee2016locally}
Sanghyun Lee, Young-Ju Lee, and Mary~F Wheeler.
\newblock A locally conservative enriched galerkin approximation and efficient solver for elliptic and parabolic problems.
\newblock {\em SIAM Journal on Scientific Computing}, 38(3):A1404--A1429, 2016.

\bibitem{choo2018enriched}
Jinhyun Choo and Sanghyun Lee.
\newblock Enriched galerkin finite elements for coupled poromechanics with local mass conservation.
\newblock {\em Computer Methods in Applied Mechanics and Engineering}, 341:311--332, 2018.

\bibitem{anderson1965iterative}
Donald~G Anderson.
\newblock Iterative procedures for nonlinear integral equations.
\newblock {\em Journal of the ACM (JACM)}, 12(4):547--560, 1965.

\bibitem{evans2020proof}
Claire Evans, Sara Pollock, Leo~G Rebholz, and Mengying Xiao.
\newblock A proof that anderson acceleration improves the convergence rate in linearly converging fixed-point methods (but not in those converging quadratically).
\newblock {\em SIAM Journal on Numerical Analysis}, 58(1):788--810, 2020.

\bibitem{li2023accelerating}
Xuejian Li, Elizabeth~V Hawkins, Leo~G Rebholz, and Duygu Vargun.
\newblock Accelerating and enabling convergence of nonlinear solvers for navier--stokes equations by continuous data assimilation.
\newblock {\em Computer Methods in Applied Mechanics and Engineering}, 416:116313, 2023.

\bibitem{pollock2019anderson}
Sara Pollock, Leo~G Rebholz, and Mengying Xiao.
\newblock Anderson-accelerated convergence of picard iterations for incompressible navier--stokes equations.
\newblock {\em SIAM Journal on Numerical Analysis}, 57(2):615--637, 2019.

\bibitem{walker2011anderson}
Homer~F Walker and Peng Ni.
\newblock Anderson acceleration for fixed-point iterations.
\newblock {\em SIAM Journal on Numerical Analysis}, 49(4):1715--1735, 2011.

\bibitem{pollock_filtering_2023}
Sara Pollock and Leo~G Rebholz.
\newblock Filtering for anderson acceleration.
\newblock {\em SIAM Journal on Scientific Computing}, 45(4):A1571--A1590, 2023.

\bibitem{raviart2006mixed}
Pierre-Arnaud Raviart and Jean-Marie Thomas.
\newblock A mixed finite element method for 2-nd order elliptic problems.
\newblock In {\em Mathematical Aspects of Finite Element Methods: Proceedings of the Conference Held in Rome, December 10--12, 1975}, pages 292--315. Springer, 2006.

\bibitem{brezzi1985two}
Franco Brezzi, Jim Douglas~Jr, and L~Donatella Marini.
\newblock Two families of mixed finite elements for second order elliptic problems.
\newblock {\em Numerische Mathematik}, 47(2):217--235, 1985.

\bibitem{deal9402022}
Daniel Arndt, Wolfgang Bangerth, Marco Feder, Marc Fehling, Rene Gassm{\"o}ller, Timo Heister, Luca Heltai, Martin Kronbichler, Matthias Maier, Peter Munch, Jean-Paul Pelteret, Simon Sticko, Bruno Turcksin, and David Wells.
\newblock The \texttt{deal.II} library, version 9.4.
\newblock {\em Journal of Numerical Mathematics}, 30(3):231--246, 2022.

\bibitem{kuznik2007double}
F~Kuznik, J~Vareilles, G~Rusaouen, and G~Krauss.
\newblock A double-population lattice boltzmann method with non-uniform mesh for the simulation of natural convection in a square cavity.
\newblock {\em International Journal of Heat and Fluid Flow}, 28(5):862--870, 2007.

\bibitem{choi2011comparative}
Seok-Ki Choi and Seong-O Kim.
\newblock Comparative analysis of thermal models in the lattice boltzmann method for the simulation of natural convection in a square cavity.
\newblock {\em Numerical Heat Transfer, Part B: Fundamentals}, 60(2):135--145, 2011.

\end{thebibliography}

\end{document}